\documentclass[twocolumn]{aastex631}

\usepackage{ae,aecompl}

\usepackage{url}
\usepackage{graphicx}
\usepackage[caption=false]{subfig}

\usepackage{float}     
\usepackage{hyperref} 
\usepackage{array}  
\usepackage{xcolor} 
\usepackage[normalem]{ulem}

\usepackage{lipsum} 

\begin{document}

\title{Spectroscopic and photometric confirmation of 3 globular and 14 intermediate-age clusters in the Irr\,II galaxy NGC\,3077}

\author[0000-0002-4437-347X]{P. A. Ovando$^\star$}
\correspondingauthor{\href{mailto:kpa.astro@gmail.com}{kpa.astro@gmail.com$^\star$}}
\affiliation{Instituto Nacional de Astrofísica, Óptica y Electrónica, Luis Enrique Erro 1,Tonantzintla 72840,
	Puebla, Mexico.}


\author[0000-0002-4677-0516]{Y. D. Mayya}
\affiliation{Instituto Nacional de Astrofísica, Óptica y Electrónica, Luis Enrique Erro 1,Tonantzintla 72840,
	Puebla, Mexico.}

\author[0000-0003-0961-3105]{L. H. Rodríguez-Merino}
\affiliation{Instituto Nacional de Astrofísica, Óptica y Electrónica, Luis Enrique Erro 1,Tonantzintla 72840,
	Puebla, Mexico.}

 \author[0000-0003-2127-2841]{L. Lomelí-Nuñez}
\affiliation{Valongo Observatory, Federal University of Rio de Janeiro, Ladeira Pedro Antonio 43, Saude Rio de Janeiro, RJ, 20080-090, Brazil.}

 \author[0000-0002-1046-1500]{B. Cuevas-Otahola}
\affiliation{Departamento de Matemáticas-FCE, Benemérita Universidad Autónoma de Puebla, Puebla, 72570, Mexico.}

\author[0000-0003-1327-0838]{D. Rosa-González}
\affiliation{Instituto Nacional de Astrofísica, Óptica y Electrónica, Luis Enrique Erro 1,Tonantzintla 72840,
	Puebla, Mexico.}

\author{L. Carrasco}
\affiliation{Instituto Nacional de Astrofísica, Óptica y Electrónica, Luis Enrique Erro 1,Tonantzintla 72840,
	Puebla, Mexico.}







\begin{abstract}

We present the results from spectroscopic and photometric analysis of 17 globular cluster (GC) candidates in the Irr II galaxy NGC~3077. The GC candidates were selected on the Hubble Space Telescope (HST) images and were cleaned of foreground Galactic stars using the GAIA parameters. We carried out aperture photometry using the multi-band archival images from SDSS, and 2MASS of all candidates, and low resolution (R= 1000) spectroscopic observations of 12 GC candidates and three suspected foreground stars using the OSIRIS/MOS mode at the Gran Telescopio Canarias (GTC). 
Age, metallicity and extinction values were determined both using spectroscopic and photometric data, independently. 
We { find} 
three of the 17 candidates { are} 
old (age $>$ 10~Gyr), metal-poor ([Fe/H] $< -1.0$~dex), massive
(mass $> 10^5 M_\odot$) GCs with characteristics similar to the classical GCs in the Milky Way. 
The rest are intermediate-age clusters (IACs) with typical ages of 3 to 4~Gyr, and in general metal-rich clusters. The radial velocities of both  populations are within 100~km\,s$^{-1}$ of the recessional velocity of the { host galaxy}. 
A relatively large population of IACs and low value of GC specific frequency ($S_{\rm N}$ = 0.7) suggest that the pre-interaction galaxy was actively forming stars and star clusters, and is unlikely to be a dE as suggested 
in some previous 
works.
\end{abstract}
%

\section{Introduction} \label{Int}

Globular clusters (GCs) are relics of early star formation in the Universe, hence their ages and metallicities, two parameters heavily dependent on the epoch and physical properties of their formation, put strong constrains on the formation history of galaxies  
(e.g. \citealt{Searle1978}; \citealt{Kauffmann1993}; \citealt{Ashman1998}; \citealt{Brodie2006}).
{ Based on cosmological N-body simulations, \cite{Bekki2008} found that around $\sim$90 per cent of GCs currently in haloes of galaxies are formed in low-mass galaxies at redshifts greater than 3.}
{ In recent years, simulations that
incorporate formation and evolution
of star clusters { in cosmological galaxy formation models have 
reproduced observational properties of GC populations in spiral galaxies such as the Milky Way and M\,31} 
(e.g. E-MOSAICS; \citealt{Pfeffer2018}).}
The majority of GCs are older than 10~Gyr, especially those formed in massive galaxies (\citealt{VandenBerg2013}).
{ 
With the advent of the James Webb Space Telescope
(JWST), it has been possible to directly detect  GC-like objects forming at times  as early as 460~Myr of the Big Bang (\citealt{Adamo2024}).}
{Meanwhile, the metallicity distribution function in several massive early-type galaxies displays bimodality, and some even trimodality \citep[e.g. M31;][]{Caldwell2016, Wang2019}.
However, there exist cases where the metallicity distribution is consistent with a single peak  such as in NGC~5128 \citep{Zepf1993} and NGC~1399 \citep{Blakeslee2012}. 
}

{
The bimodal metallicity distribution can be understood by the two-phase galaxy formation  scenario \citep{Oser2010, Forbes2011, Usher2012, Beasley2018, Alamo2021}. In this model, the first phase consists of rapid {\it in situ} star formation in which most of the metal-rich population is formed in the central $\sim$10\% of the present day radius of galaxies. Galaxies grow in mass and radius in the second phase by the accretion of metal-poor dwarf galaxies. Metal-rich population can be formed in the host galaxy during this second phase, if the accreting galaxy is gas-rich. GCs are expected to form in each of these two phases of galaxy formation, with metal-rich and metal-poor GCs located in the inner and outer parts of the galaxies, respectively \citep{Pfeffer2018, Kruijssen2019}. The relative number of {\it in situ} and accreted GCs in the two phases depends on the present-day galaxy mass, with massive galaxies having a higher fraction of accreted material. In this scenario, low-mass galaxies (mass $\leq 10^{10}$~M$_\odot$) are expected to have a higher fraction of old metal-rich GCs as compared to the massive galaxies. On the other hand, GCs in genuine dwarf galaxies are expected to be metal-poor \citep{Kauffmann1993}. Studies of age and metallicities of GC systems in low-mass galaxies are required to test these predictions of the cosmological simulations. We summarize below some of the existing attempts in this direction. 
}
\citet{Georgiev2008} carried out a photometric study of GCs in 19 Magellanic type dwarf irregular galaxies, the majority of { which} 
have colors similar to the old metal-poor GCs found in the haloes of giant local galaxies. 
{On the other hand, \citet{Parisi2014} found a higher fraction of younger GCs (3 $<$ Age/Gyr $<10$) in the Small Magellanic Cloud (SMC).} 
\citet{Strader2012} found an unusual abundance of luminous red star clusters in the dwarf Magallanic irregular galaxy NGC 4449. Some of the clusters have properties of the old GCs. They also found two clusters with unexpected abundance properties, the clusters were neither metal-poor ([Fe/H]= $-$1.0 dex) nor $\alpha$-elements enhanced ([Mg/Fe] $\sim-0.1$ to $-0.2$). { These unusual properties are confirmed by}
\citet{Annibali2018} who found the old GCs in the NGC 4449 to be of intermediate metallicities ($-1.2 \lesssim$ [Fe/H] $\lesssim -0.7$~dex), with  subsolar [$\alpha$/Fe] ratios ($-0.8  \lesssim$ [$\alpha$/Fe] $\lesssim 0.1$; peak at $\sim -0.4$). 
{These studies illustrate the diversity in the age and metallicity properties of GC systems in dwarf galaxies, highlighting the importance of detailed studies of star cluster systems in more { low-mass} galaxies}. 

Color-magnitude diagram (CMD) of the constituent stars with the best fitting of an isochrone is the most widely recognized and direct technique to estimate the age of a star cluster (e.g., \citealt{Hodge1999,Bastian2016}). However, star clusters located in galaxies beyond the Local Group cannot be resolved even using images obtained with the Hubble Space Telescope (HST). Therefore, most of extagalatic star cluster analysis depends on integrated-light measurements to determine ages and/or metallicities (e.g. \citealt{Nantais2010,Mayya2013,Asad2014,Annibali2018,Luis2024}). 

\begin{table}
\centering
\caption{Basic properties of NGC 3077.}
\label{tab:0}
\resizebox{\columnwidth}{!}
{
\begin{tabular}{lll} 

\hline
Parameter & Value  & References \\
  \hline
\hline
 Morphology Type  & Irr II (I0) & (a,b) \\ 
 Distance         & 3.63 Mpc    & (c) \\
 Angular size     & 5.4$\times$4.5 arcmin & (d)  \\
 Linear size      & 5.6 kpc &   \\
 $R_{25}$ & 2.68 arcmin   &  (d)   \\
 Radial velocity ($V_{\rm hel}$)& 14.09 km s$^{-1}$& (d)\\
 Total Mass            & $2 \times 10^{10}M\odot$ & (e) \\
  $(B-I)$         & 1.692 mag & (f)  \\
 $M_V$            & $-$16.202 mag & (f)  \\
  $H_I$ Mass       & 6.9$\times10^8 M_\odot$ & (e) \\
 $H_2$ Mass       & $1.6\times10^6 M_\odot$ &  (g)\\
 $E(B-V)_{\rm MW}$    & 0.07 mag & (h)   \\
 $E({\rm F475W}-{\rm F814W})$ & 0.12 mag & (h,i)\\
 Metallicity (Z)  &  0.02 &  (j)\\
 \hline  
   \end{tabular}
   }
   \footnotesize{\\References:
   (a) \citet{Holmberg1950}, (b) \citet{Vaucouleurs1959}, (c) \citet{Freedman1994}, (d) \citet{deVaucouleurs1991}, (e) \citet{yun1999}, (f) \citet{Wisniewski1968}, (g) \citet{Walter2002}, (h) \citet{Schlegel1998}, (i) using \citet{Cardelli1989} extinction curve, (j) \citet{Calzetti2004}.}
   \end{table}
   
The well-known M81 group { consists} of more than thirty galaxies and is located 3.63 Mpc away (\citealt{Freedman1994}). The three brightest members 
{ of the group, namely} M81, M82 and NGC~3077 (M81 triplet), 
were involved in a tidal interaction \citep{vanderhulst1979,Yun1994}. This interaction occurred 
around 300~Myr ago (\citealt{yun1999}) and 
has triggered a large-scale star formation in the disk of M82 \citep{Mayya2006,Lino2011}, and is also likely responsible for the nuclear starburst activities in NGC~3077. It is a dusty late-type galaxy, currently witnessing intense star-formation in the central 500~pc (Table \ref{tab:0} displays the basic properties of NGC~3077). Morphologically, it has been classified as Irregular II (Irr II) galaxy \citep[see][]{Holmberg1950, Vaucouleurs1959,Krienke1974}. { A total mass of $2\times10^{10}$~M$_\odot$ \citep{yun1999} puts this galaxy in the upper fringe of what can be considered as a low-mass galaxy.}

\defcitealias{Davidge2004}{D04}

{ The central starburst region of NGC\,3077 is host to a rich population of young Super Star Clusters \citep[SSCs;][]{Harris2004, Notni2004}. }
\citet[][hereafter \citetalias{Davidge2004}]{Davidge2004} used ground-based wide-field near-IR (NIR) images to identify 12 GC candidates located in the periphery of NGC~3077. It is still unclear if all these candidates belong to this galaxy or are part of other members of the M81 group. More recently, \cite{Chies2022} have elaborated a new catalogue of 642 GC candidates located in the M81 triplet. 
However, they did not recover a significant population of GC candidates in NGC~3077. Similarly, \cite{Pan2022} used multi-wavelength data to select GC candidates in the outer halo of the M81 group, finding three candidates associated to NGC~3077. { In spite of all these studies, there does not exist a consensus on the total number of GCs in this galaxy.}

According to \cite{Harris1981} the number of GCs ($N_{\rm GC}$) in a galaxy normalised by its luminosity is given by the specific frequency equation, $S_{\rm N}=N_{\rm GC}10^{0.4(M_V+15)}$. The specific frequency associated to irregular galaxies (e.g. SMC and LMC) is $<S_{\rm N}>$= 0.5 \citep{Harris1991}. In case of NGC~3077, based on the observed absolute visual magnitude of $M_V\sim-16$ mag (see Table~\ref{tab:0}), and on the estimated specific frequency for irregular galaxies, the expected number of GCs is $N_{\rm GC}\sim$ 2.
%
%

In this paper, we { present the results of a spectroscopic study of 15 GC candidates in NGC\,3077, which is  the first of its kind for this galaxy. The GC candidates were selected from a list of 21 objects obtained in a search on } 
the Advanced Camera for Surveys 
(ACS) images of the HST  observations. 
We further carried out multi-band photometric analysis of all the candidates, with the goal of obtaining a complete census of the GCs in this low-mass galaxy.
The work is organized as follows: in Section \ref{DataO}, we { discuss in detail the criteria used to select a sample of GC candidates and describe new and existing observational data used in this work.}
In Section \ref{Pst}, we {  implement tests to reject foreground stars and non GC-like objects from our catalog.}
{ The results from} photometric/spectroscopic analysis { are presented} in Section
\ref{Analy}. Results are discussed in Section \ref{Discu}. 
Finally, in Section \ref{Conc}, we give a summary and the conclusions.

\section{Sample of GC candidates, photometric data and spectroscopic observations}
\label{DataO}
\defcitealias{Girardi2002}{G02}
\defcitealias{Cardelli1989}{C89}
\begin{figure*}
    \centering
    \includegraphics[{trim= 0 0 0 0},clip,height=6.7cm] {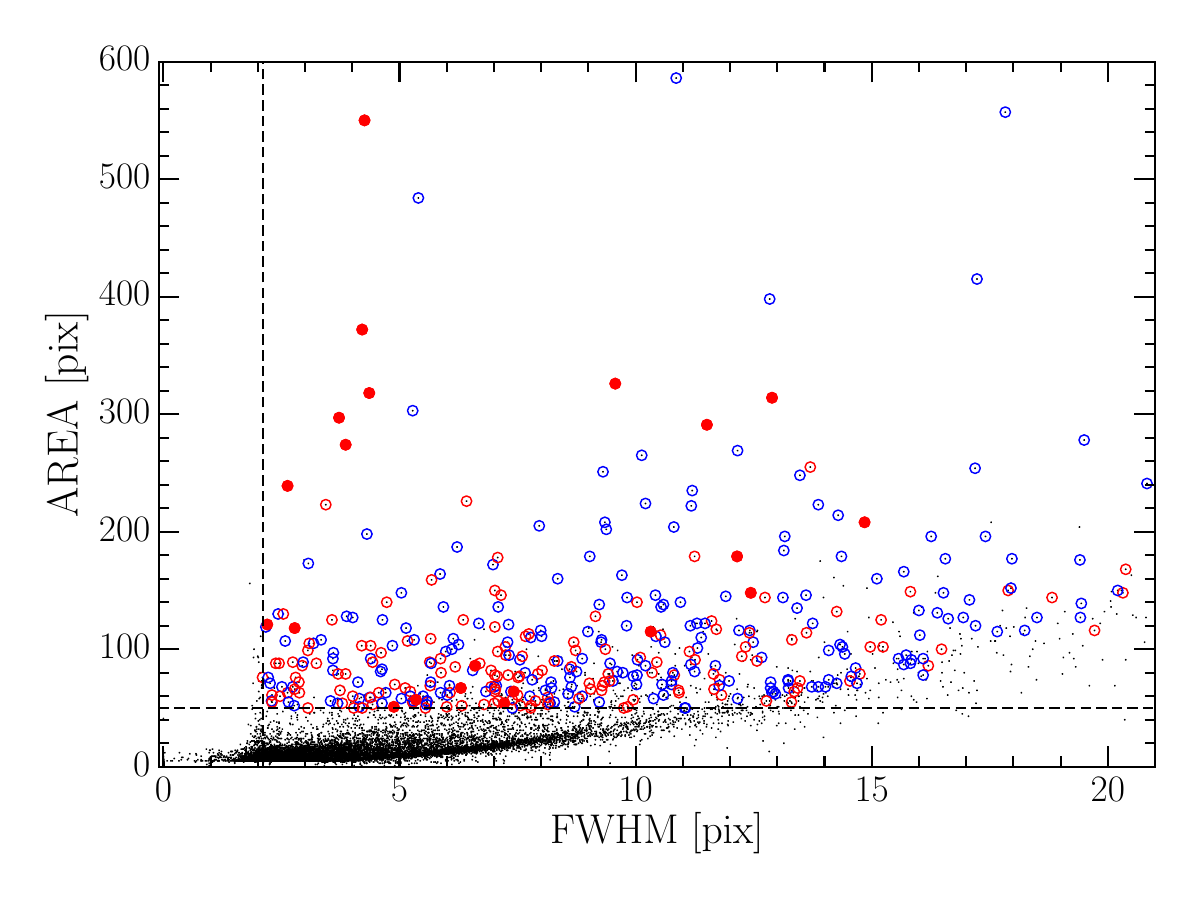}
    \includegraphics[{trim= 0 0 0 0},clip,height=6.7cm] {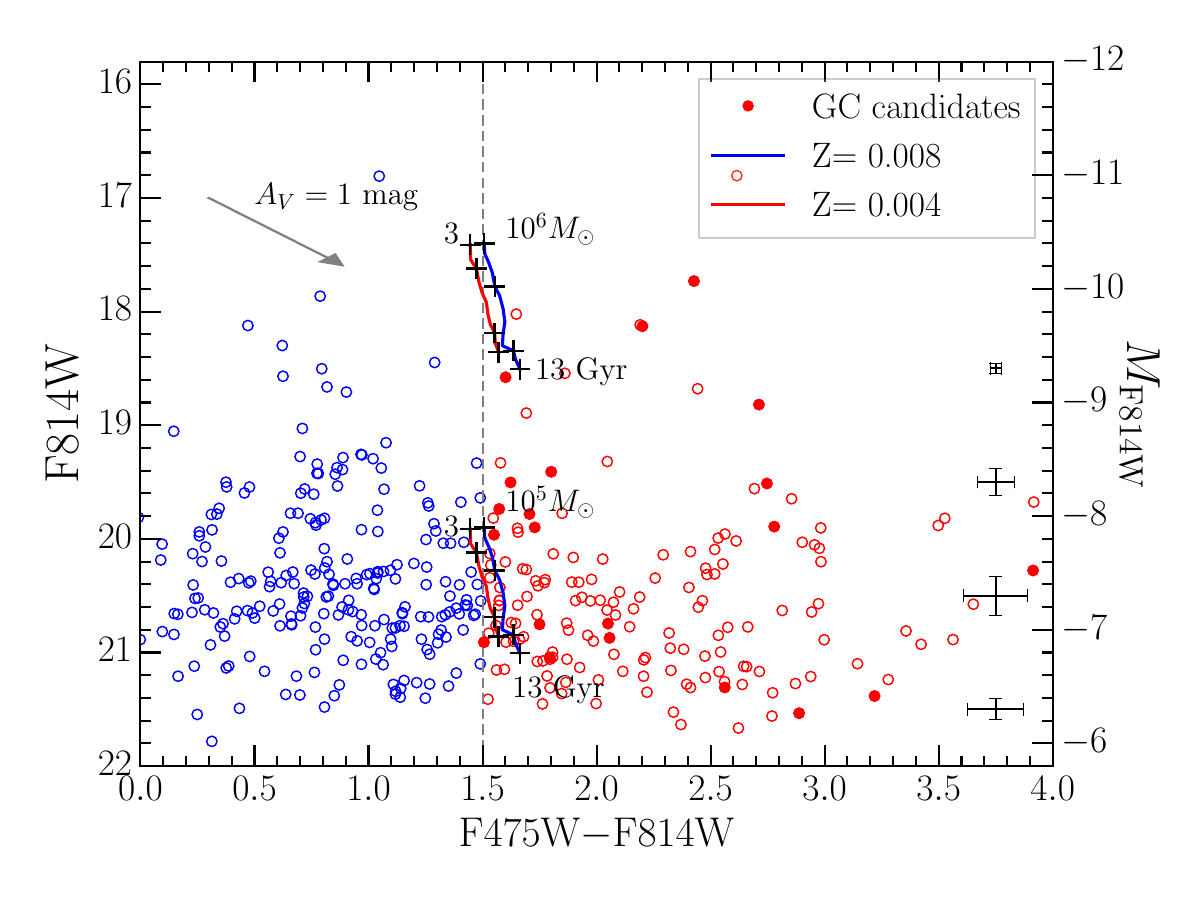}
    \caption{ Illustration of selection criteria used in this work to select GC candidates.
    {\it Left panel:}
    All {\sc SExtractor}-selected sources (gray dots) plotted along with the  clusters candidates (circles) in  {\sc area} vs {\sc fwhm} 
plane. Clusters are defined as those objects 
having {\sc fwhm}$\geq$ 2.1~pix,
{\sc area}$>$ 50~pix (dashed lines), F814W$\leq$ 22 mag 
and {\sc ellipticity}$\leq$ 0.66 (see text). 
{\it Right panel:}  
GC candidates (solid red circles) are separated from SSC candidates (blue circles) using the color cut F475W$-$F814W= 1.5~mag (vertical dashed line). 
Locus of photometric evolution of clusters between 3 and 13~Gyr using \citetalias{Girardi2002} SSPs are shown for two metallicities (Z= 0.004 and 0.008) and two cluster masses ($10^5$~M$_\odot$ and $10^6$~M$_\odot$). 
The absolute magnitude of clusters is shown along the right axis. The reddening vector corresponding to $A_V=1$~mag is shown. Open red circles are the red clusters located in the central region of radius= 25~arcsec, which are most likely reddened SSCs (see text and Figure~\ref{fig:Pos}).
}
    \label{fig:FWHM}
\end{figure*}

\begin{figure*}
    \centering
    \includegraphics[{trim= 0 0 0 0},clip,width=16.5cm] {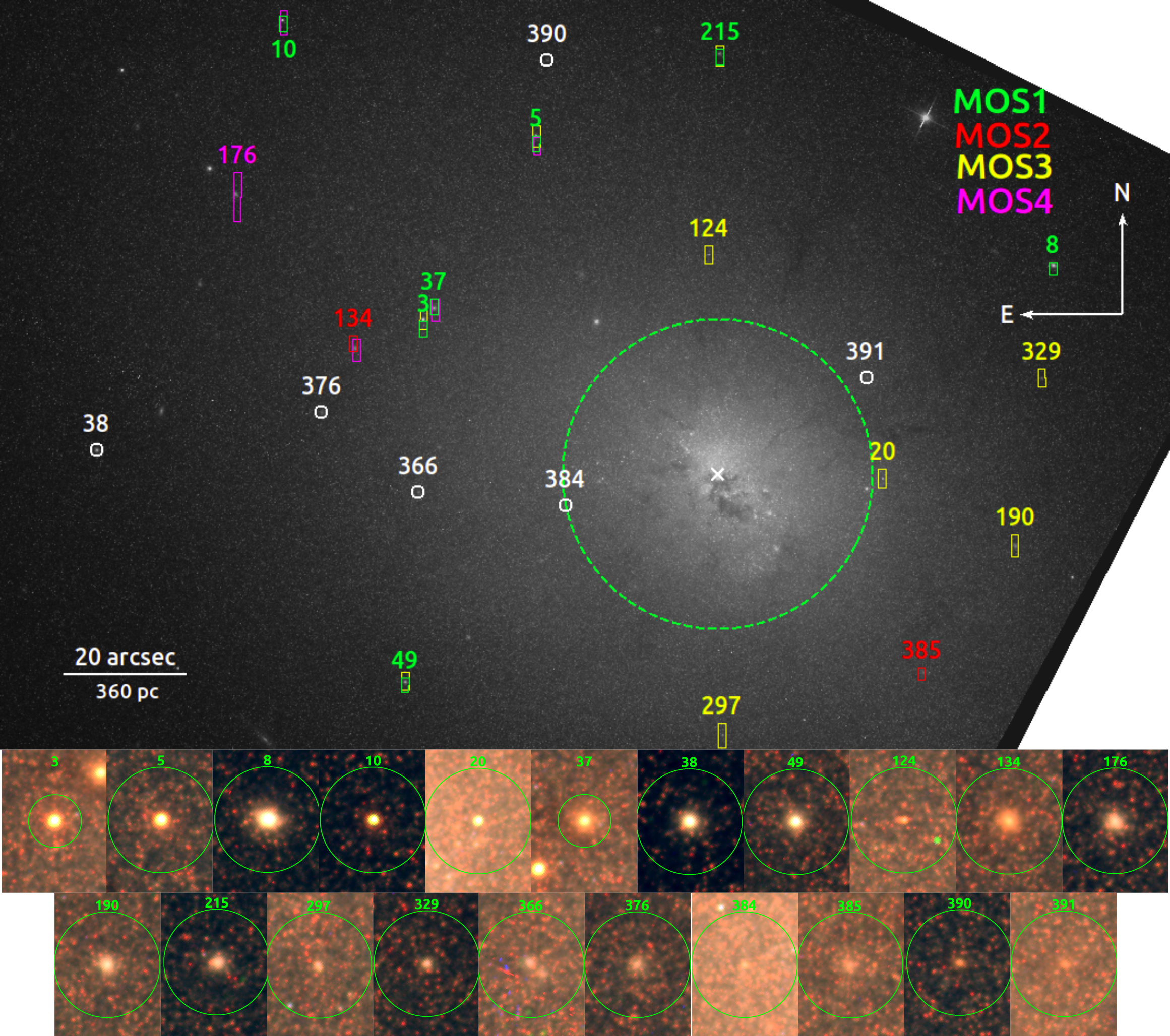}
    \caption{{\it Top panel:}
    GC candidates are displayed on the image of NGC~3077 (F814W band). Objects inside a rectangle were observed with MOS, and the sizes of the rectangles correspond to the sizes of the slitlet, the color indicates the observing run (see the top-right corner). GC candidates not observed with MOS are located inside white circles. The central dusty starburst zone is enclosed with a green circle with a radius of {25 arcsec}. Center at $\alpha$= 10h03m19.07s and $\delta= +68^{\circ}44\arcmin02\arcsec.1$ is indicated with white cross.
    {\it Bottom panels}: The RGB (F475W, F606W and F814W) stamps of the GC candidates are displayed, each circle has a radius of {2 arcsec} (36 pc), except ID 3 and 37, both have a radius of 1 arcsec to avoid reciprocal contamination.
    }
    \label{fig:Pos}
\end{figure*}

\begin{table}
\centering
\setlength{\tabcolsep}{2.pt}
\caption{HST and multi-band images used in this work.}
\label{tab:MB}
\resizebox{\columnwidth}{!}
{
\begin{tabular}{cccccc} 
\hline
 Instrum. & Band&Central $\lambda$ & Scale &   ZP  & Exp. Time \\
 (1) & (2) & (3) & (4) & (5) & (6) \\
\hline \hline
        HST/ACS & F814W & 8037 & 0.05 &25.512  &1622 \\
        HST/ACS & F606W & 5961 & 0.05 &26.399  &1596  \\
        HST/ACS & F475W & 4818 & 0.05 &26.144  &1570  \\			
		HST/WFPC2&F300W & 3266 & 0.10 &19.433 &2400  \\ 
			 SDSS  & $u$& 3551 & 0.4&21.589 & 53.9\\
			 SDSS  & $g$& 4686 & 0.4&22.596 & 53.9\\
			 SDSS  & $r$& 6165 & 0.4&22.350  & 53.9\\
			 SDSS & $i$& 7481 & 0.4&22.138 & 53.9\\
			 SDSS & $z$& 8931 & 0.4&21.978  & 53.9\\
			 2MASS & $J$& 12350 & 1.0&20.952 & 7.8\\
			 2MASS & $H$& 16620 & 1.0&20.728 & 7.8\\
			 2MASS & $K_s$& 21590 & 1.0&20.085 & 7.8\\
			\hline
\end{tabular}
}
\footnotesize{\\Notes: (1) telescope/camera,
(2) filter/band, (3) central wavelength
in [\AA], (4) scale of plate in 
[arcsec pix$^{-1}$], (5) zero point in VEGAMAG system, (6) exposure time in [s].}
\end{table}

\subsection{The sample of GC candidates and photometric data}
\label{Sample}

\begin{table*}
\centering
\caption{{ Parameters of the GC candidates.}}
\label{tab:1}
\resizebox{\textwidth}{!}
{
\begin{tabular}{lccccccclll} 
\hline
ID & R.A. & Dec & F814W & F475W$-$F814W &F300W    & FWHM  & $R_{G}$ &{ Obj type} &{ Confirmation}&Others IDs \\
 (1)  & (2)  & (3) & (4) & (5)          &  (6) & (7) & (8) & (9) &(10) & (11) \\
\hline \hline

3*                 &  10h03m27.84s &+68$^\circ$44$\arcmin$27$\arcsec$.21& 17.731$\pm$0.001 & 2.427$\pm$0.002 &23.633$\pm$0.441  & 3.86  & 949.0     & S   &GAIA &11(D),100(N) \\
  5*                 &  10h03m24.51s &+68$^\circ$44$\arcmin$56$\arcsec$.81& 18.129$\pm$0.001 & 2.201$\pm$0.002 &23.392$\pm$0.204  & 3.21  & 1094.5    & S   &GAIA &15(D),107(N) \\
  8                  & 10h03m09.17s  &+68$^\circ$44$\arcmin$35$\arcsec$.93 & 18.578$\pm$0.001 & 1.601$\pm$0.002 &\null             & 4.26  & 1118.7   & GC  &Spectra&27(D),74(N) \\
  10*                &  10h03m32.02s &+68$^\circ$45$\arcmin$15$\arcsec$.56& 18.818$\pm$0.001 & 2.712$\pm$0.004 &\null             & 3.13  & 1790.4    & S   &GAIA  &8(D),114(N) \\
  20                 & 10h03m14.23s  &+68$^\circ$44$\arcmin$01$\arcsec$.53 & 19.409$\pm$0.005 & 1.801$\pm$0.006 &23.773$\pm$0.255  & 2.20  & 463.5    & GC  &Spectra &\null \\
  37                 & 10h03m27.53s  &+68$^\circ$44$\arcmin$29$\arcsec$.03 & 19.513$\pm$0.004 & 2.747$\pm$0.010 &24.348$\pm$0.696  & 4.21  & 938.2    &IAC  &SED  &12(D),101(N) \\
  38                 & 10h03m37.52s  &+68$^\circ$44$\arcmin$06$\arcsec$.15 & 19.503$\pm$0.003 & 1.623$\pm$0.004 &\null             & 4.36  & 1768.1   & GC  &SED  &1(D),104(N) \\
  49                 & 10h03m28.38s  &+68$^\circ$43$\arcmin$28$\arcsec$.72 & 19.737$\pm$0.004 & 1.573$\pm$0.005 &22.909$\pm$0.184  & 4.12  & 1068.0   &IAC  &Spectra   &48(N) \\
  124$^\dagger$      & 10h03m19.38s  &+68$^\circ$44$\arcmin$37$\arcsec$.62  & 20.278$\pm$0.009 & 3.914$\pm$0.043 &\null             & 3.15  & 625.8   &Blended &Visual &\null \\
  134                & 10h03m29.86s  &+68$^\circ$44$\arcmin$22$\arcsec$.55 & 19.892$\pm$0.011 & 2.779$\pm$0.014 &23.809$\pm$0.725  & 7.66  & 1094.2   &IAC  &Spectra  &9(D),102(N) \\
  176                & 10h03m33.41s  &+68$^\circ$44$\arcmin$47$\arcsec$.47 & 19.781$\pm$0.015 & 1.706$\pm$0.009 &23.341$\pm$0.886  & 7.02  & 1587.3   &IAC  &Spectra  &111(N) \\
  190                & 10h03m10.30s  &+68$^\circ$43$\arcmin$50$\arcsec$.68 & 19.964$\pm$0.019 & 1.550$\pm$0.010 &\null             & 7.11  & 863.7    &IAC  &Spectra &\null \\
  215                & 10h03m19.06s  &+68$^\circ$45$\arcmin$10$\arcsec$.05 & 19.899$\pm$0.015 & 1.729$\pm$0.009 &\null             & 6.60  & 1195.8   &IAC  &Spectra   &\null \\
  297                & 10h03m18.97s  &+68$^\circ$43$\arcmin$20$\arcsec$.23  & 20.871$\pm$0.022 & 2.057$\pm$0.020 &\null             &5.35   &  736.9  &IAC  &SED  &\null \\
  329                & 10h03m09.51s  &+68$^\circ$44$\arcmin$17$\arcsec$.72  & 21.042$\pm$0.015 & 1.794$\pm$0.014 &\null             &4.23   &  955.8  &IAC  &SED  &\null \\
  366                & 10h03m27.99s  &+68$^\circ$43$\arcmin$59$\arcsec$.31  & 20.910$\pm$0.050 & 1.506$\pm$0.026 &\null             &6.21   &  855.8  &IAC  &SED  &88(N) \\
  376                & 10h03m30.85s  &+68$^\circ$44$\arcmin$12$\arcsec$.28  & 20.752$\pm$0.039 & 1.750$\pm$0.021 &24.010$\pm$1.389  &6.91   & 1142.3  &IAC  &SED  &97(N) \\
  384                & 10h03m23.63s  &+68$^\circ$43$\arcmin$57$\arcsec$.11  & 21.308$\pm$0.056 & 2.562$\pm$0.064 &\null             &4.25   & 445.4   &IAC  &SED  &\null \\
  385                & 10h03m13.03s  &+68$^\circ$43$\arcmin$30$\arcsec$.28  & 20.745$\pm$0.046 & 2.050$\pm$0.027 &\null             &7.24   &  805.4  &IAC  &Spectra &20(N) \\
  390                & 10h03m24.16s  &+68$^\circ$45$\arcmin$09$\arcsec$.00  & 21.534$\pm$0.018 & 2.888$\pm$0.047 &\null             &3.73   &  1274.0 &IAC  &SED  &\null \\
  391                & 10h03m14.71s  &+68$^\circ$44$\arcmin$17$\arcsec$.82  & 21.383$\pm$0.028 & 3.219$\pm$0.070 &\null             &5.44   &  501.1  &IAC  &SED  &\null \\  
  
\hline
\end{tabular}
}
\footnotesize{\\
Notes: (1): Identifier; ID numbers followed
by { a symbol are rejected objects either because they are identified as bright foreground stars (``*'') or blended stars ($^\dagger$}; see Section \ref{Pst}),  
(2--3) coordinates in the GAIA system (J2000),
(4--6) magnitudes and color in VEGAMAG system, 
(7) FWHM in [pix] from F814W HST/ACS,  
 (8) galactocentric radius in [pc]
 (center at $\alpha=$ 10h03m19.07s and $\delta= +68^{\circ}44\arcmin02\arcsec.1$), 
 { (9) 
 object type: GC, S and IAC for globular cluster, 
 star and  intermediate-age cluster, respectively,}
 { (10) confirmation of object type
 (see text),}
 (11) others IDs: (D)
for \citet{Davidge2004} and
(N) for \citet{Notni2004}. 
		}
\end{table*} 



The sample of GC candidates was defined on the F435W, F606W and F814W images taken using the ACS/WFC camera onboard the HST (proposal ID 10915, PI: Julianne Dalcaton) 
{ on 2006 September 21.} 
The first three rows of Table~\ref{tab:MB} gives the details of these { observations}. 
The central $\sim$100 arcsec$\times$100 arcsec of the galaxy has been { earlier} observed { on 2001 May 22} with the Wide Field Planetary Camera 2 (WFPC2) in F300W (proposal ID 9144, PI: Daniela Calzetti; { \citealt{Harris2004}}). { F300W being the bluest of the HST bands available for this galaxy, we used this image to carryout photometry on the selected GC candidates in the limited Field of View of the WFPC2 observations.}

We used  the {\sc SExtractor} program \citep{Bertin1996} to select sources on the HST/ACS images. 
%
{ During the {\sc SExtractor} run we saved  a number of geometric and photometric parameters that allowed us to distinguish genuine GC candidates from stars and fake sources. The most important of them are: Full Width at Half Maximum ({\sc fwhm}), {\sc area}, which is the number of contiguous pixels above the detection threshold, {\sc ellipticity}, 
as it is measured at the isophote corresponding to the detection
threshold on the background subtracted image
enclosing the 
{\sc area}, and magnitudes in aperture of 0.2~arcsec radius in all the filters.}
A source is considered a GC candidate if it satisfies the following criteria: {\sc fwhm} $\geq$ 2.1 pixels, {\sc area} $>$ 50~pixels, { {\sc ellipticity}$<$ 0.66
(not elongated), F814W $\leq$ 22 mag} 
and 
F475W$-$F814W $\geq$ 1.5 mag. 
{ 
The {\sc fwhm} of 2.1~pixels corresponds to the Point Spread Function (PSF) of the ACS images and hence the first criterion separates stars from extended objects. This criterion is similar to the concentration index (CI) used in some studies (e.g.
\citealt{Whitmore2014}; see also 
\citealt{Lomeli2022}
 for a comparison between the {\sc fwhm} and CI-based selections). The subsequent selection criteria involving {\sc area, ellipticity} and F814W magnitude allow us to eliminate vast number of false clusters formed from faint superposed  stars. These  criteria have been successfully used earlier to select SSCs on the HST/ACS images by \citet{Mayya2008}, \citet{Mayra2010} and more recently \citet{Lomeli2022}. The last criterion, namely the color-cut,  corresponds to the color of metal-poor (Z $\lesssim$ 0.008) Simple Stellar Populations (SSPs) models
\citep[][hereafter \citetalias{Girardi2002}\footnote{\url{http://stev.oapd.inaf.it/cgi-bin/cmd_3.1}}]{Girardi2002} older than 3~Gyr, and hence is expected to reject relatively blue, and hence, young clusters from our sample.
}


Given the large number of reddened young clusters in the dusty central starburst region \citep{Calzetti00}, we restricted our search of GC candidates to 
areas outside {25 arcsec} (450 pc) from the galactic center  ($\alpha$= 10h03m19.07s, $\delta=+68^{\circ}44\arcmin02\arcsec.1$). 
%
{ This resulted in a sample of }
21 GC candidates. 
{ In Figure~\ref{fig:FWHM}, we illustrate the selection criteria. In the left panel we show all the {\sc Sextractor} sources (points) in {\sc area} vs {\sc fwhm} diagram with the 21  selected GC candidates identified by  filled red circles. The open red circles are red clusters located in the central region of radius= 25~arcsec, whereas the blue circles are bluer clusters, which are  candidates to SSCs, and are not part of the discussion in this study.
In the right panel, we show all the candidates that passed the first four criteria that defines a star cluster in a CMD. The SSP evolutionary locus corresponding to ages $>$3~Gyr are shown for two metallicities (Z= 0.004 and 0.008) and two cluster masses ($10^5$~M$_\odot$ and $10^6$~M$_\odot$). The color-cut  (F475W$-$F814W $\geq$ 1.5~mag)  corresponding to 3~Gyr population at Z= 0.008 separates blue SSC candidates from red GC candidates.
The reddening vector for $A_V=$ 1~mag using the \citet[hereafter \citetalias{Cardelli1989}]{Cardelli1989} 
extinction curve with $R_V= 3.1$ is shown by the arrow.}

Figure~\ref{fig:Pos} shows the locations of the GC candidates in the galaxy. The RGB (F814W, F606W and F475W) stamps of the GC  candidates are displayed in the bottom part of this figure. 
{\sc SExtractor} has provided the coordinates of the GC candidates, with this information at hand 
we have employed the {\it phot} task of {\sc iraf} \citep{Tody1986} 
to perform the circular aperture photometry using a radius of 0.2~arcsec. { Local sky values are subtracted from each aperture magnitude using the median sky in an annular region of inner radius 0.8~arcsec and width 0.2~arcsec around each object.} 

{ GCs are marginally resolved objects on the HST/ACS images, which implies the flux lying outside a fixed aperture radius of 0.2~arcsec used to measure the magnitude depends on the size of the cluster, or in our case its proxy, the {\sc fwhm}. This correction can be calculated using the growth curves for 
isolated clusters in the same image or using simulated clusters as in \citet{Lomeli2022}. Using the isolated blue and red clusters in the image, we found that growth curves reach asymptotic values at radius= 15~pixel.
Based on these measured magnitudes at 15~pixels, we defined a FWHM-dependent aperture correction $\Delta m= 0.096\times\text{\normalfont FWHM}-0.048$, which we applied to both the F475W and F814W aperture magnitudes.
}

{ We calculated errors on each measured magnitude assuming} Poisson noise from the object and the sky. The errors on the magnitudes are added in quadrature to calculate the error on colors. 
{ GAIA Data Release 3 (DR3: \citealt{GAIA2016}, \citealt{GAIA2022}) was used to bring all images to a single astrometric system using the coordinates of field stars in GAIA.}
Table~\ref{tab:1} shows the parameters measured of the GC candidates; coordinates (Columns 2 and 3), photometric data (Columns 4, 5 and 6), FWHM measured by IRAF 
(Column 7), the galactocentric distance is measured in the plane of the sky (Column 8), object type { and the method used to determine it} (based on results of the next sections) 
(Columns 9 { and 10}), and cross identification with previous studies { when available} (Column 11).

\begin{table*}
\centering
\caption{Multi-band photometry of GC candidates.}
\label{tab:pMB}
\resizebox{2\columnwidth}{!}{
\begin{tabular}{lccccccccc} 
\hline
ID   &  $i$               &  $H$               &  $u-i$                   &  $g-i$            &  $r-i$            &  $i-z$               &  $i-K_s$              &  $H-K_s$           &  $J-H$            \\
   (1)&(2)&(3)&(4)&(5)&(6)&(7)&(8)&(9)&(10) \\    
  \hline
\hline

3*    &  17.591$\pm$0.027  &  15.650$\pm$0.094  &  3.859$\pm$0.370$^u$     &  2.709$\pm$0.083  &  0.972$\pm$0.044  &  0.610$\pm$0.037     &  1.686$\pm$0.244      &  $-$0.255$\pm$0.261  &  0.668$\pm$0.156  \\
5*    &  17.957$\pm$0.028  &  16.570$\pm$0.101  &  3.207$\pm$0.353$^u$     &  2.507$\pm$0.078  &  0.837$\pm$0.046  &  0.536$\pm$0.052     &  0.794$\pm$0.283      &  $-$0.593$\pm$0.300  &  0.837$\pm$0.149  \\
8    &  18.373$\pm$0.034  &  17.142$\pm$0.155  &  2.058$\pm$0.199         &  1.444$\pm$0.056  &  0.459$\pm$0.048  &  0.493$\pm$0.069     &  2.052$\pm$0.134      &  0.821$\pm$0.202   &  1.148$\pm$0.281  \\
10*   &  19.003$\pm$0.049  &  17.287$\pm$0.184  &  2.161$\pm$0.355$^u$     &  2.588$\pm$0.136  &  0.943$\pm$0.081  &  0.708$\pm$0.096     &  1.664$\pm$0.248      &  $-$0.052$\pm$0.305  &  0.872$\pm$0.263  \\
20   & 17.525$\pm$0.100   & 16.036$\pm$0.433   & 2.210$\pm$0.208          & 1.768$\pm$0.126   & 0.466$\pm$0.127   & 0.262$\pm$0.221      & 1.842$\pm$0.275       & 0.353$\pm$0.503    & 0.760$\pm$0.587 \\
37   &  19.102$\pm$0.075  &  16.472$\pm$0.192  &  2.348$\pm$0.377$^u$     &  2.744$\pm$0.245  &  0.883$\pm$0.116  &  0.616$\pm$0.110     &  2.524$\pm$0.378$^K$  &  $-$0.106$\pm$0.417  &  0.330$\pm$0.249  \\
38   &  19.338$\pm$0.062  &  $---$             &  1.725$\pm$0.363         &  1.474$\pm$0.099  &  0.429$\pm$0.086  &  0.353$\pm$0.161     &  1.760$\pm$0.357$^K$  &  $---$             &  $---$            \\
49   &  19.589$\pm$0.111  &  $---$             &  1.575$\pm$0.369$^u$     &  1.236$\pm$0.147  &  0.280$\pm$0.139  &  0.385$\pm$0.242     &  2.011$\pm$0.369$^K$  &  $---$             &  $---$            \\
124$^\dagger$  &  20.132$\pm$0.231  &  $---$             &  1.032$\pm$0.421$^u$     &  2.367$\pm$0.569  &  1.085$\pm$0.424  &  0.376$\pm$0.462     &  2.554$\pm$0.421$^K$  &  $---$             &  $---$            \\
134  &  19.417$\pm$0.097  &  $---$             &  1.747$\pm$0.365$^u$     &  2.470$\pm$0.242  &  0.709$\pm$0.145  &  0.675$\pm$0.173     &  1.839$\pm$0.365$^K$  &  $---$             &  $---$            \\
176  &  19.758$\pm$0.098  &  $---$             &  1.406$\pm$0.365$^u$     &  1.403$\pm$0.143  &  0.537$\pm$0.134  &  0.358$\pm$0.260     &  2.180$\pm$0.365$^K$  &  $---$             &  $---$            \\
190  &  19.830$\pm$0.162  &  $---$             &  1.328$\pm$0.414         &  1.160$\pm$0.206  &  0.427$\pm$0.216  &  0.736$\pm$0.269     &  2.252$\pm$0.387$^K$  &  $---$             &  $---$            \\
215  &  20.290$\pm$0.161  &  $---$             &  0.874$\pm$0.387$^u$     &  1.387$\pm$0.222  &  0.989$\pm$0.267  &  0.553$\pm$0.358     &  2.712$\pm$0.387$^K$  &  $---$             &  $---$            \\
297  &  21.685$\pm$0.351  &  $---$             &  $-$0.521$\pm$0.497$^u$  &  0.929$\pm$0.727  &  0.803$\pm$0.505  &  1.677$\pm$0.487     &  4.107$\pm$0.497$^K$  &  $---$             &  $---$            \\
329  &  20.517$\pm$0.215  &  $---$             &  0.647$\pm$0.412$^u$     &  2.586$\pm$0.582  &  1.386$\pm$0.501  &  0.854$\pm$0.359     &  2.939$\pm$0.412$^K$  &  $---$             &  $---$            \\
366  &  19.909$\pm$0.223  &  $---$             &  1.255$\pm$0.417$^u$     &  0.692$\pm$0.250  &  0.218$\pm$0.276  &  0.522$\pm$0.358     &  2.331$\pm$0.417$^K$  &  $---$             &  $---$            \\
376  &  20.519$\pm$0.248  &  $---$             &  0.645$\pm$0.431$^u$     &  1.762$\pm$0.404  &  0.587$\pm$0.347  &  0.496$\pm$0.419$^z$ &  2.941$\pm$0.431$^K$  &  $---$             &  $---$            \\
384  &  19.314$\pm$0.173  &  $---$             &  1.850$\pm$0.392$^u$     &  0.788$\pm$0.219  &  0.260$\pm$0.230  &  $-$0.694$\pm$0.380  &  1.736$\pm$0.392$^K$  &  $---$             &  $---$            \\
385  &  20.036$\pm$0.184  &  $---$             &  1.128$\pm$0.397$^u$     &  1.552$\pm$0.291  &  0.632$\pm$0.268  &  0.618$\pm$0.336     &  2.458$\pm$0.397$^K$  &  $---$             &  $---$            \\
390  &  21.468$\pm$0.465  &  $---$             &  $-$0.304$\pm$0.583$^u$  &  1.926$\pm$0.605  &  1.022$\pm$0.796  &  1.460$\pm$0.575     &  3.890$\pm$0.583$^K$  &  $---$             &  $---$            \\
391  &  19.354$\pm$0.140  &  $---$             &  1.810$\pm$0.379$^u$     &  1.417$\pm$0.214  &  0.591$\pm$0.215  &  $-$0.180$\pm$0.414  &  1.776$\pm$0.379$^K$  &  $---$             &  $---$            \\
 
 \hline  
   \end{tabular}
   }
 \footnotesize{\\Notes: (2-3) magnitudes in VEGAMAG system, (4-10) colors in VEGAMAG system. 
 Most of the clusters are very faint, therefore in some bands their magnitudes are given by an upper limit (denoted with a super index). We did not write a color value if both magnitudes are upper limits. }
   \end{table*}

\begin{table*}
\centering
\caption{Log of the spectroscopic observations using GTC/OSIRIS}
\label{tab:sp}
  \resizebox{\textwidth}{!}
{
\begin{tabular}{ccccccccrr} 
\hline
Run & PI & Date  & SW & Exp. time & AM & Seeing & Night & STD & Obj \\
(1) & (2) & (3) &(4)& (5) &(6)& (7)&
(8) &(9) & (10) \\
\hline \hline
2015B-MOS1 & L. H. Rodríguez-Merino & 2015-12-17 &  1.2 & 3$\times$1300 & 1.34 & 1.0 & G & G191-B2B & 7 \\
2015B-MOS2 & L. H. Rodríguez-Merino & 2015-12-17 &  1.2 & 3$\times$1300 & 1.30 & 0.9 & G & G191-B2B & 2 \\
2017A-MOS3 & L. H. Rodríguez-Merino & 2017-04-22 &  1.2 & 3$\times$1500 & 1.32 & 1.1 & D & HILT600 & 9 \\
2017A-MOS4 & L. H. Rodríguez-Merino & 2017-04-02 &  1.2 & 3$\times$1500 & 1.30 & 0.9 & D & HILT600 & 5 \\
\hline
\end{tabular}
}
\footnotesize{\\Notes: (1) ID of the Run, (2) principal investigator, (3) observational date, (4) slit-width (arcsec), (5) exposure time (number of exposures $\times$ integration time in seconds), (6) mean air mass of the three exposures, (7) seeing (arcsec), (8) type of night (G= gray or D= dark), (9) standard star, (10) number of objects in each run.}
\end{table*} 

\begin{table}
\centering
\caption{Characteristics of the spectra obtained with GTC}
\label{tab:Psp}
\setlength{\tabcolsep}{1.2pt}
\resizebox{\columnwidth}{!}
{
\begin{tabular}{lrcrclc}
\hline
ID &Velocity & RMS & SNR & Obj type & Runs & Ap. size \\
(1)&(2) & (3)  & (4) & (5)& (6) & (7) \\
  \hline
\hline
 
3*  &$-$163$\pm$26  &  1.2& 17 &2 &MOS 1,3  & 3 \\%
5*  &$-$79$\pm$8    &  1.0& 21 &2 &MOS 1,3,4& 3 \\%
8   &$-$78$\pm$38   &  1.5& 19 &1 &MOS 1    & 4 \\%
10* &$-85\pm$35     &  1.4& 8  &2 &MOS 1,4  & 4 \\%
20  &$-$57$\pm$29   &  1.4& 14 &1 &MOS 3    & 4 \\%
37  &$-$90$\pm$30   &  1.4& 12 &3 &MOS 1,4  & 4 \\%
49  &120$\pm$10     &  2.5& 12 &3 &MOS 1,3  & 4 \\%
124$^\dagger$&68$\pm$12&0.9& 9 &3 &MOS 3    & 4 \\%
134 &46$\pm$31      &  1.7& 12 &3 &MOS 2,4  & 4 \\%
176 &$-$164$\pm$42  &  1.2& 6  &3 &MOS 4    & 4 \\%
190 &71$\pm$23      &  1.2&10  &3 &MOS 3    & 4 \\%
215 &$-$38$\pm$36   &  1.8& 9  &3 &MOS 1,3  & 5 \\%
297 &$-$129$\pm$34  &  1.4& 6  &3 &MOS 3    & 4 \\%
329$^\dagger$&$-51\pm$28     &  0.9& 8  &3 &MOS 3    & 4 \\%
385 &$-$84$\pm$32   &  1.3& 17 &3 &MOS 2    & 4 \\%

 \hline  
  \end{tabular}
  }
  \footnotesize{\\Notes: { (1) Identification numbers. A superscript "*" or "$\dagger$" signs imply they are stars, or the spectra have insufficient quality to measure spectral indices, respectively,} (2) Observed recessional velocity 
  (measuring mainly H and K of CaII, and iron lines)
  in [km s$^{-1}$], (3--4) root mean square (RMS) of the continuum  [$\times10^{-18}$ erg s$^{-1}$ cm$^{-2}$ \AA$^{-1}$] and signal to noise ratio (SNR) of the spectrum, both measured at 4150~\AA~ in a 100 \AA~ window. In case a GC candidate has been observed two or more times, we employed the spectrum with the highest SNR (see Section \ref{SpecO}), (5) same as defined in column~8 of Table~\ref{tab:1}, (6) same as defined in column~1 of Table~\ref{tab:sp}, (7) aperture size for extraction in [pix].
  }
  \end{table}
  \subsection{Multi-band ancillary images data}
\label{MBD}

In order to perform an analysis of the GC candidates using a large interval of the wavelength range, the HST's photometry of the GC candidates were complemented with ground-based archival images from the Sloan Digital Sky Survey (SDSS) \citep{Abazajian2003} and Two Micron All Sky Survey (2MASS) \citep{Skrutskie2006} missions (see Table~\ref{tab:MB}).
We performed aperture photometry using {\it phot} of IRAF on images from the five SDSS filters ($ugriz$) and the three 2MASS bands ($JHK_s$). The relative isolation of the selected objects allowed us to obtain magnitudes using an aperture with a {2 arcsec} radius, except for GCs candidates ID 3 and 37 which were calculated using {1 arcsec} radius (see Figure~\ref{fig:Pos}). 

%

Table~\ref{tab:pMB} lists the calculated photometry of the twenty-one GC candidates. Due to the low sensitivity of the 2MASS detectors, photometry of only the brightest clusters was measured.

\subsection{Spectroscopic observations}
\label{SpecO}

Spectroscopic observations of a limited sample of GC candidates were obtained using the Multi-Object Spectra 
(MOS) mode of the Optical System for Imaging and Low-Intermediate-Resolution Integrated Spectroscopy (OSIRIS) instrument on the Gran Telescopio Canarias (GTC). 
A complete log of observations is shown in Table~\ref{tab:sp}. All observations were carried out using the R1000B grism, covering a spectral range of $\sim$3700 $-$ 7500 \AA\ with a spectral resolution of about 5.3~\AA\ and with a sampling of 2.1~\AA~pixel$^{-1}$. The spatial scale covered by the spectra is  $\sim$0.25 arcsec pixel$^{-1}$ at the default binning mode (2$\times$2). There were clear skies, seeing was {$\sim$1 arcsec} during the two runs. And cirrus clouds reported only for the 2017A-MOS4 run.

The top panel of Figure~\ref{fig:Pos} displays the GC candidates observed (colored rectangles) with the MOS mode of OSIRIS. Four masks (slits of each mask are indicated with a particular color) were designed to perform the observations in the MOS mode. The slits (colored rectangles) were oriented along the north-south direction, each slit is 1.2~ arcsec width. Due to astronomical and instrumental limitations only fifteen GC candidates were spectroscopically observed, with some clusters observed more than once. Each run included bias, flat-fields, calibration lamps, and a standard star (see Table~\ref{tab:sp}).




The data reduction was carried out using the 
GTCMOS\footnote{GTCMOS pipeline: \url{https://www.inaoep.mx/~ydm/gtcmos/gtcmos.html}} package, an IRAF-based pipeline developed by
one of us (Y.D. Mayya, see \citealt{Gomez2016} for details). OSIRIS uses two detectors with a small separation along the spatial axis, therefore the first step of the reduction process was to create a mosaic image by joining the images of the two detectors. Subsequently, all the bias frames for each night were combined using the median to obtain a master bias, which was subtracted from all the object frames. The spectra of the comparison lamps were employed to calibrate in wavelength each spectrum, whereas the standard star spectra were used to perform the flux calibration. The task {\it apall} was used to extract sky-subtracted 1D-spectrum of each object.

A total of 23 spectra of 15 GC candidates were extracted. All the spectra were corrected for Doppler effect using the {\sc iraf} task {\it dopcor}, at least three absorption features,  
mainly H and K of CaII, and iron lines,  
distributed along the observed wavelength interval were employed to measure the recessional velocity.
{ The spectra of these 15 candidates are shown in Figure~\ref{fig:Ss}, where we have identified the location of the absorption features typical of old stellar systems such as GCs with vertical gray bars. In two spectra (IDs 124 and 329), most of these features are hardly recognizable due to poor SNR, because of which we dropped these candidates from further spectroscopic analysis. Three of our spectroscopic targets turned out to be stars after our GAIA DR3 data analysis (see the next section), leaving us with 10 GC candidates with good quality spectra.}



\begin{figure*}
    \centering
\includegraphics[{trim= 0 0 0 0},clip,width=17.cm]{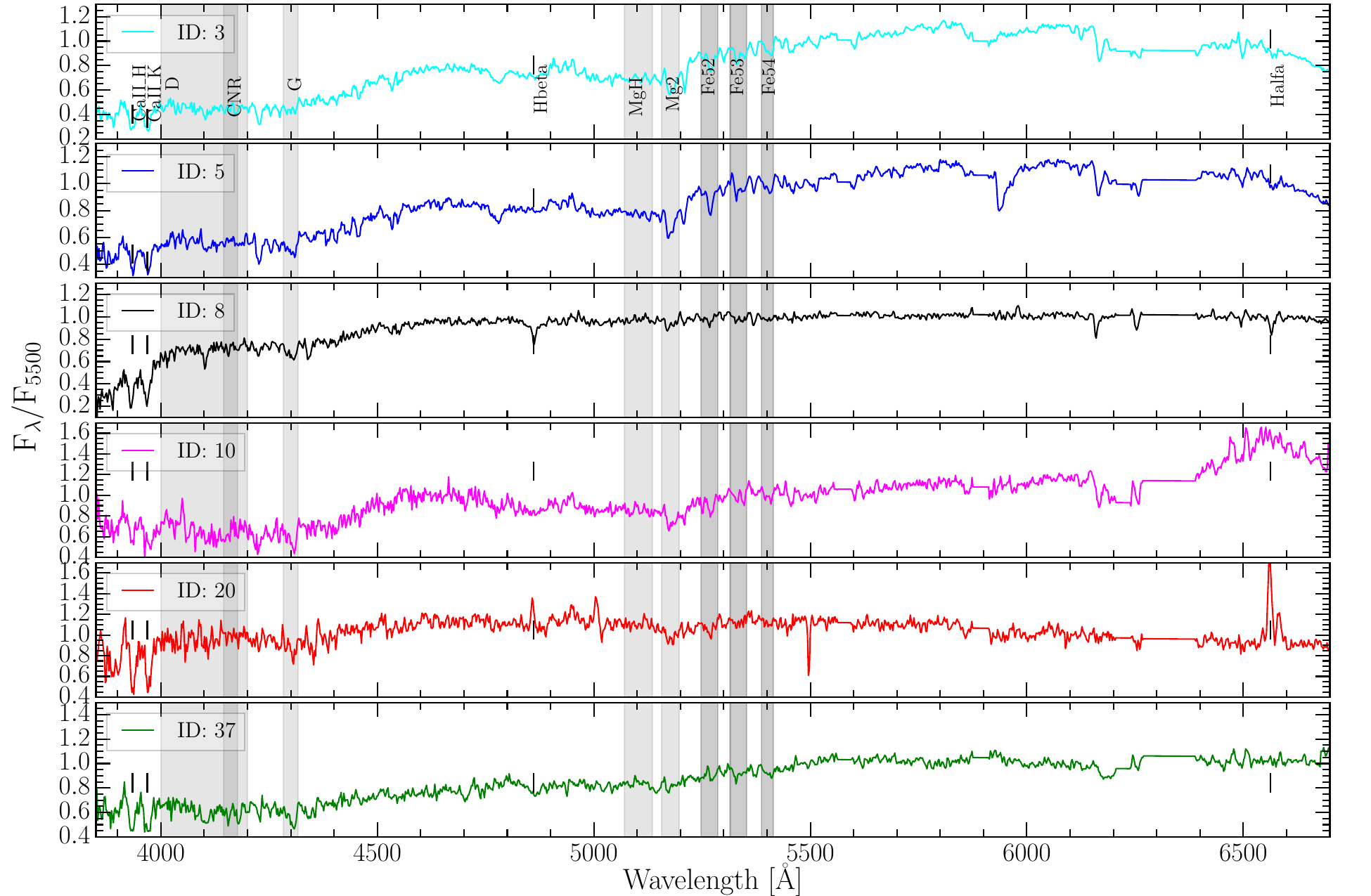}
    \caption{GTC/OSIRIS-MOS spectra of all observed GC candidates, the three foreground stars are included (IDs 3, 5, and 10, see Section~\ref{Pst}). The fluxes are corrected for the Galactic reddening and Doppler effect. To facilitate visualization the fluxes have been normalized to $F_{5500}$ value. The shaded areas indicate regions where spectral indices (labels in the top panel) were defined by
    \citet{Burstein84}, \citet{Brodie1990} and \citet{Trager1998}.}
    \label{fig:Ss}
\end{figure*}

\begin{figure*}
\ContinuedFloat
\centering
\includegraphics[{trim= 0 1.42cm 0 0},clip,width=17.cm]{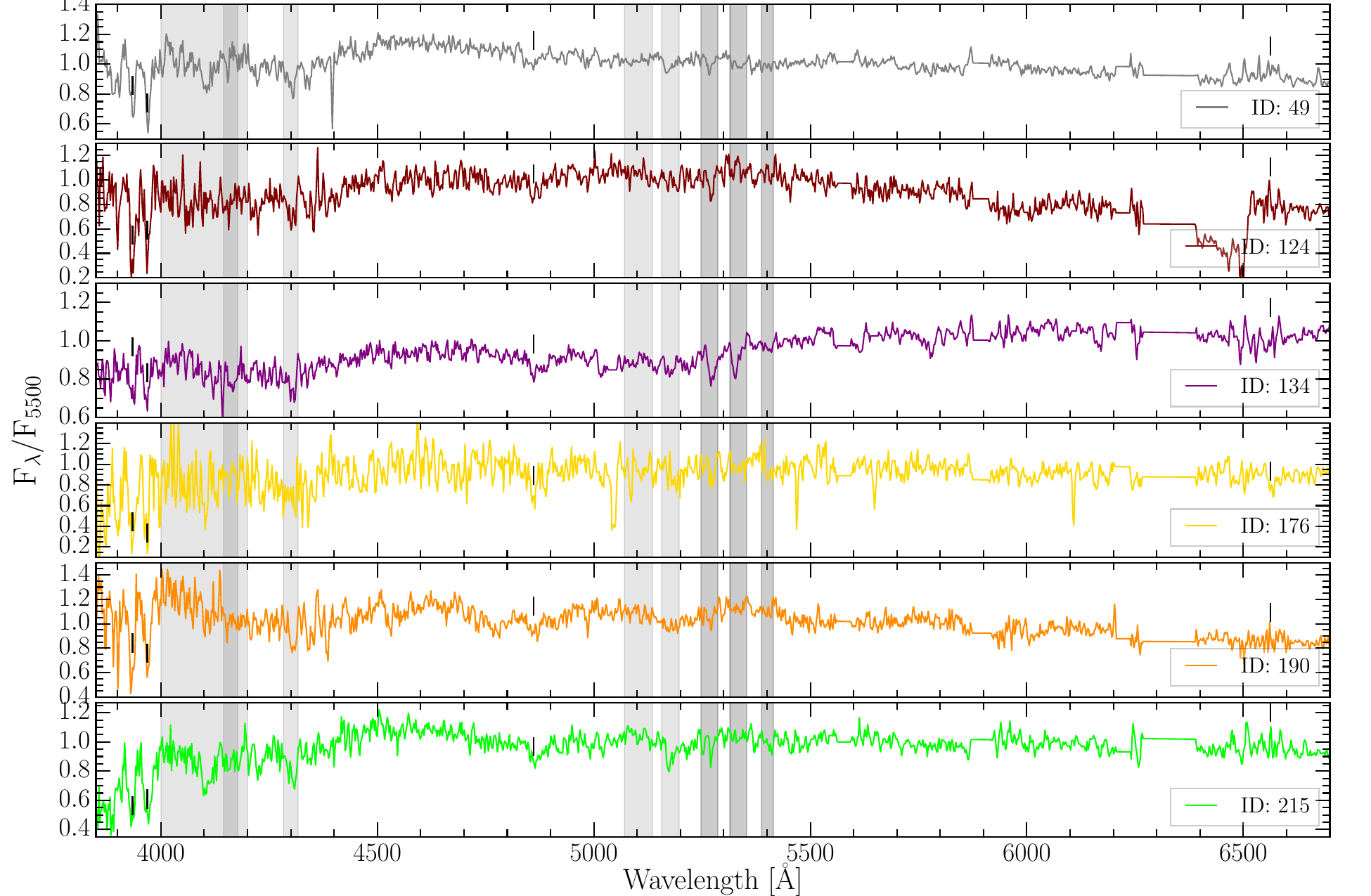}
\includegraphics[{trim= 0 10cm 0 0},clip,width=17.cm]{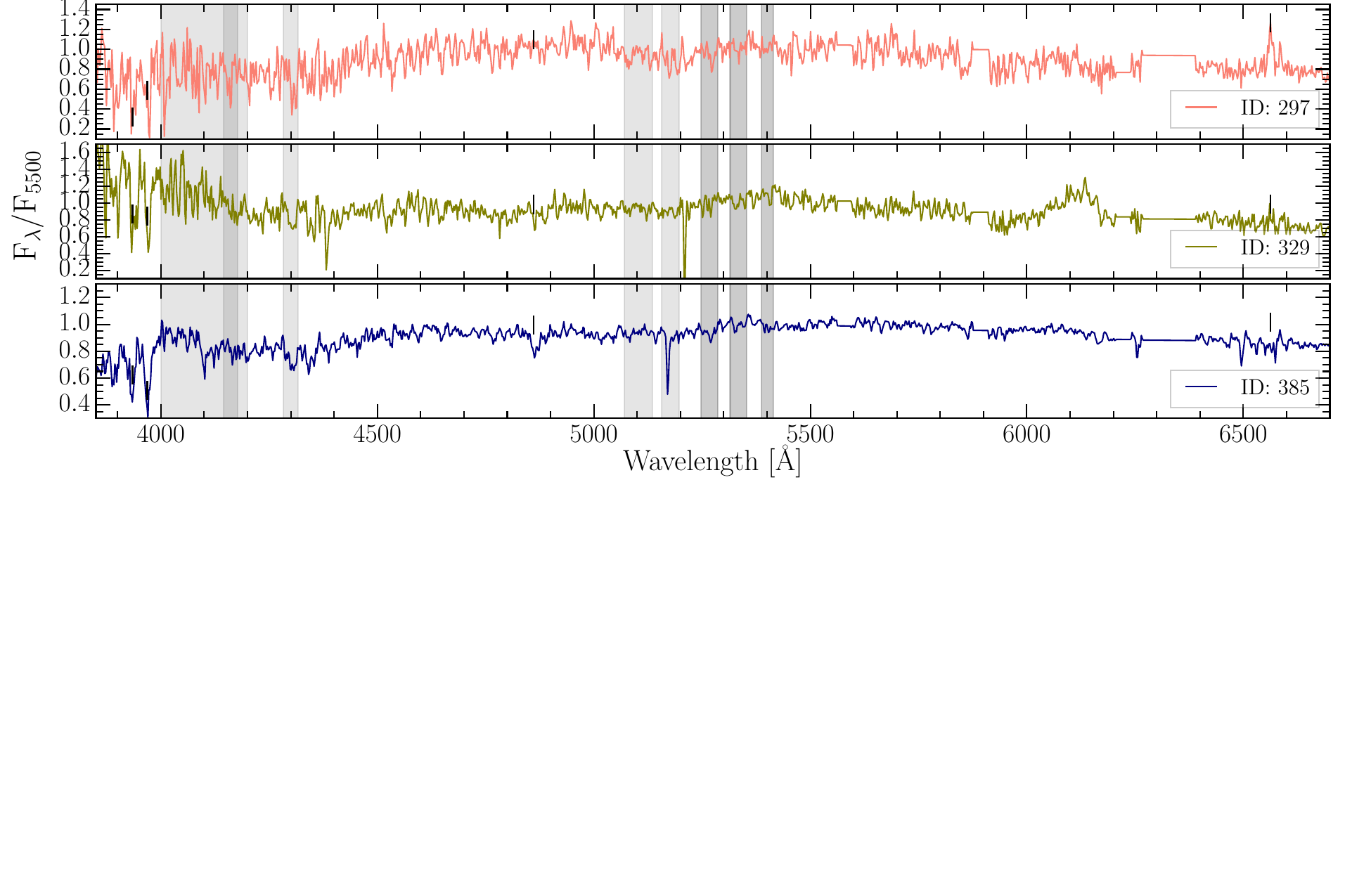}
\caption{Figure~\ref{fig:Ss} (continued).}
\end{figure*}

\section{The sample of confirmed GC candidates}
\label{Pst}

\subsection{Star-GC discrimination using GAIA parameters}

\begin{figure}
    \centering
\includegraphics[height=6.3cm]{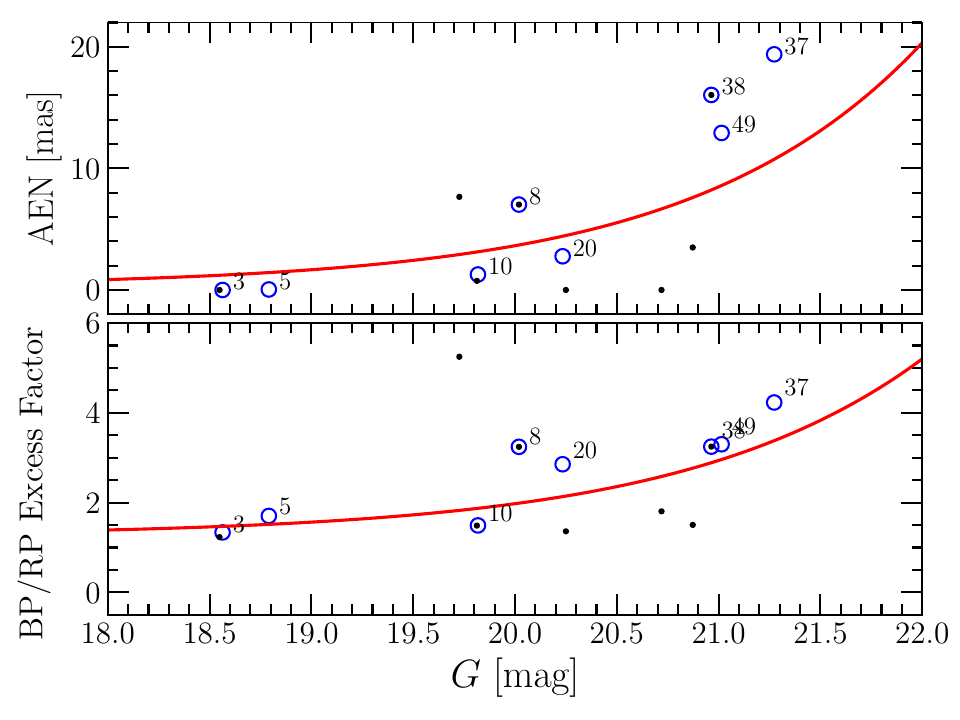}
\caption{Astrometric and photometric data provided by GAIA is employed to  confirm GC candidates (blue circles). Two diagrams, astrometric excess noise (AEN) versus $G$-band magnitude and blue photometry/red photometry (BP/RP) versus $G$-band magnitude, are used to distinguish between foreground stars from extragalatic extended sources.
GC candidates of \citetalias{Davidge2004} with entry in GAIA list 
are represented as black points (see text). 
Foreground stars lay under the red line \citep[see][]{Hughes2021}.}
    \label{fig:GAIA}
\end{figure}

\begin{figure}
    \centering
    \includegraphics[{trim= 0 0 0 0.2cm},clip,height=6.2cm]{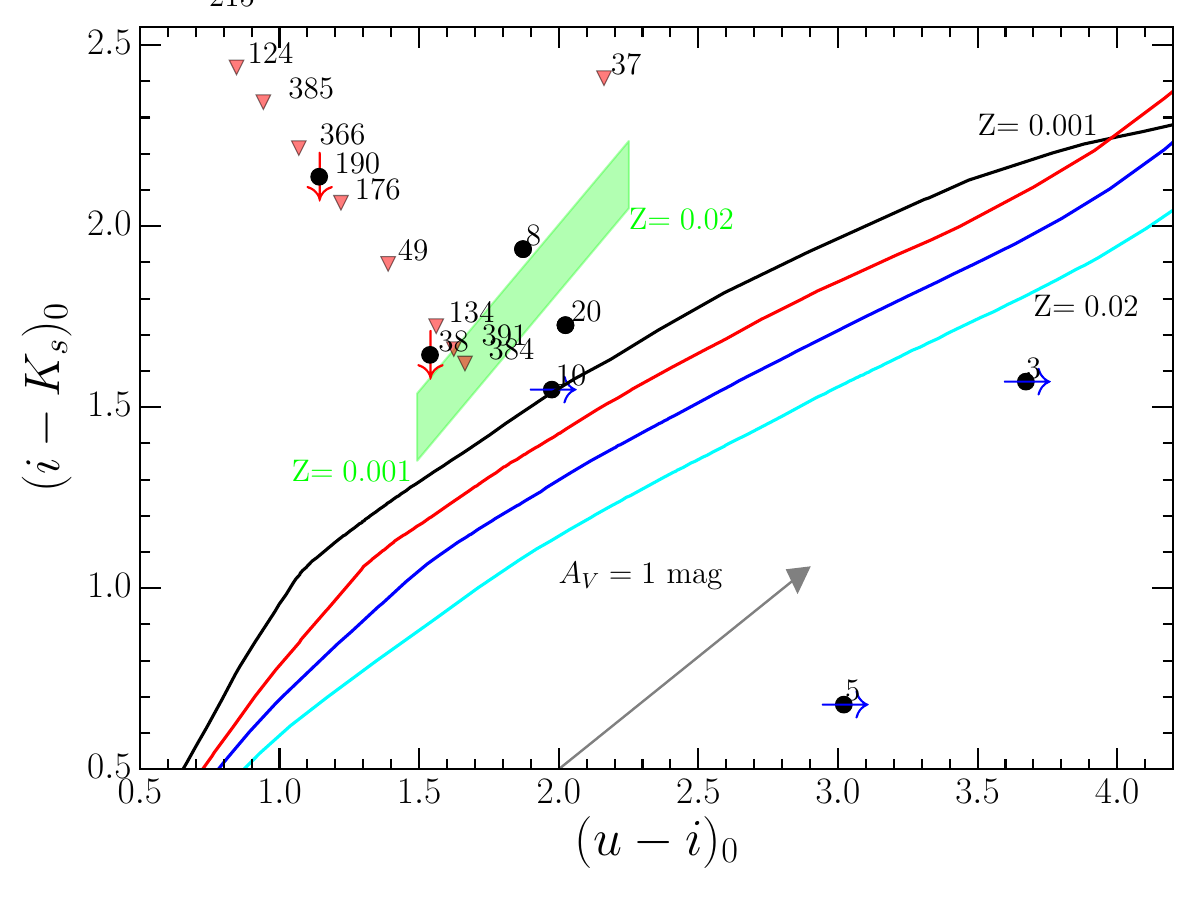}
\caption{{ Color-Color diagram $(i-K_s)_0$ vs $(u-i)_0$ comparing  the colors of the GC candidates (circles and triangles). 
Many candidate GCs are undetected in either the $u$-band (right arrows) or the $K_s$-band (down arrows) or in both these (inverted triangles).
Colors obtained with SSP models span over this diagram (green area),
they cover the age interval from 8 to 13~Gyr and four metallicities: Z= 0.001, 0.004, 0.008 and 0.02.
Four theoretical 
ZAMS models are included, they are indicated with colored continuous lines and 
have the same metallicities as SSPs. The extinction vector corresponds to $A_V = 1$ mag.
}}
    \label{fig:CCD}
\end{figure}

\begin{figure}
    \centering
    \includegraphics[{trim= 0 0 0 0.2cm},clip,height=6.2cm]{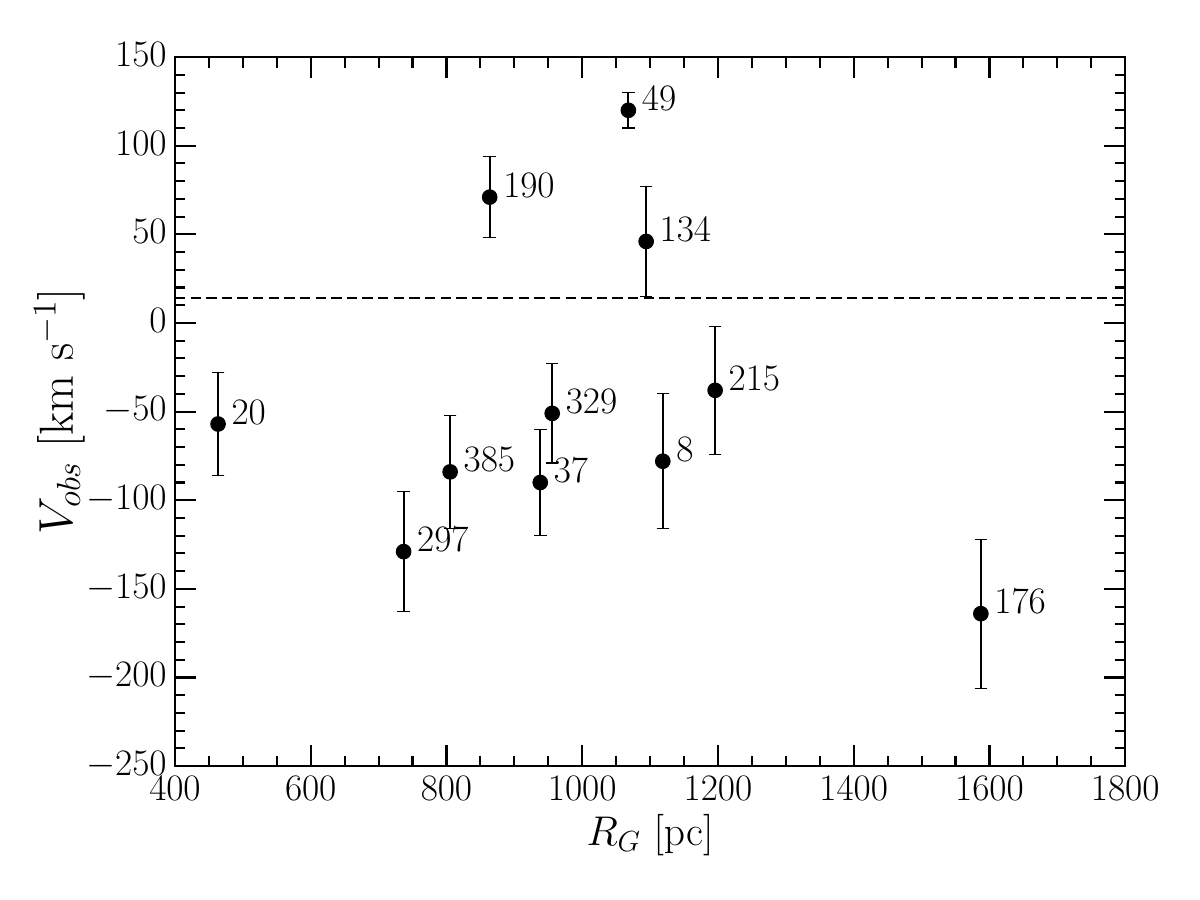}
\caption{Radial velocities of the confirmed GC candidates
against galactatocentric radius ($R_G$). The velocity
of galaxy is represented with  
dashed-line (see Table~\ref{tab:0}).}
    \label{fig:Vel3}
\end{figure}



We find that three of our GC candidates (IDs 3, 5, and 10) were classified as stars by \citetalias{Davidge2004} using 
ground-based NIR photometry.
SExtractor and IRAF give  us FWHMs for these objects significantly greater than the expected value for a point source 
(FWHM < 2.1 pix) on the HST images. 
We found  that their peak fluxes on the HST images are close to saturation limit, which makes the measured FWHM larger than that expected for a point source.
The availability of GAIA { DR3} data offers a modern method to verify the nature of these three suspected stars, as well as to identify any other Galactic interlopers in our sample. Eight of our GC candidates are bright enough to have a GAIA entry. 
In Table~\ref{tab:GAIA3}, we give the relevant GAIA parameters for these eight GC candidates. The three suspected stars (IDs 3, 5, and 10) have parallaxes that correspond to distances below $\sim$2~kpc confirming them as the  Galactic stars. 


\cite{Hughes2021} showed that there are two GAIA parameters  that help to identify the Galactic stars in extragalactic GC samples even when parallax measurements are unreliable. These two additional parameters are the astrometric excess noise (AEN) and the BP/RP (blue photometry/red photometry) excess factor. Extended and marginally extended sources have higher values of these parameters with respect to point sources, with a tendency for fainter objects to have larger values of these parameters. \cite{Hughes2021} defined the dependence of these parameters with magnitude by curves that discriminate point sources from extragalactic GCs in nearby galaxies. 
In Figure~\ref{fig:GAIA}, we show our GC candidates (empty circles) in plots involving these GAIA parameters. The three objects previously noted as stars (IDs 3, 5, and 10) all lie below or very close to the curve in both plots.
ID 20 lies marginally below the curve in the AEN plot, but lies well above the curve in the BR/RP excess factor plot. This source does not have measurable parallax and hence we consider it as a GC candidate rather than a Galactic star. Other four objects lie above the curve in both plots confirming them as marginally resolved extragalactic GCs.

We make use of these GAIA star cluster discriminators to check the possibility that some of the GC candidates of \citetalias{Davidge2004} are foreground Galactic stars. We show in Figure~\ref{fig:GAIA}, all the GC candidates of \citetalias{Davidge2004} by black points. 
The only objects that pass the criteria for extragalactic GCs (marginally-resolved objects) are 
three,  
the same two objects that we 
classified as GC candidates  
(IDs 8 and 38). 
The third object is classified as young SSC  (color F475W$-$F814W < 1.5 mag, see Section \ref{Sample}) by us.  
The rest of the objects have FWHM on the HST images consistent with them being stars. Implications of this will be discussed in Section \ref{GCis}.

{ 
\subsection{Star-GC discrimination using Color-Color Diagram}
\label{C-CD}

The Color-Color diagram $(i-K_s)_0$ vs $(u-i)_0$ has been established to be a useful diagram to discriminate the Galactic 
 interlopers from genuine GCs (\citealt{Munoz2014}, \citealt{Gonzalez2017}). 
The resulting plot is shown in Figure~\ref{fig:CCD}. Some of our candidates are not detected either at the shortest ($u$) or/and the longest ($K_s$) bands, in which case the observed colors are either bluer or redder limits, respectively.
A short arrow is shown when it is undetected in only one of the $u$ or $K_s$ bands, whereas inverted red triangles are put when it is not detected in both filters. 
The reddening vector for $A_V=$ 1~mag using the
\citetalias{Cardelli1989} 
extinction curve 
is shown by the arrow. 
The plotted colors have been corrected for the Galactic reddening of 
$E(B-V)=0.07$ mag \citep{Schlegel1998}, but not any internal reddening in the host galaxy. 
We also show as the green area the expected location of GCs, modelled as old SSPs between Z= 0.001 and 0.02 (\citetalias{Girardi2002}) and the locus of the zero age main sequence (ZAMS) where the majority of the Galactic field stars lie.
The plotted ZAMS locus is taken from the theoretical stellar evolutionary code PARSEC\footnote{\url{http://stev.oapd.inaf.it/cgi-bin/cmd}} at the same four metallicities 
\citep{Bressan2012}.

 Objects in the diagram with colors that are consistent with the colors of the green area will be identified as GCs, whereas those close to the ZAMS locus are likely to be the Galactic stars. 
 Except the three candidates identified as stars (IDs 3, 5 and 10), all the rest are consistent as GCs after taking into account the detection-limit arrows. 
 Objects that are undetected in the $u$ and $K$ bands fall along a diagonal with their non-detections consistent with the colors expected for GCs.}


\subsection{Final list of GC candidates}

Finally, a careful visual inspection of each GC candidate allowed us to reject a fourth object, ID 124, as it shows morphological signatures of being composed by two point sources. In summary, we removed four objects from the initial list to get a total sample of 17 objects that have colors and sizes suggesting they are genuine GC candidates.

{ We could obtain  radial velocity measurements from our spectroscopic observations for 11 of these objects, which are  plotted against the galactocentric radius 
$R_G$ in Figure~\ref{fig:Vel3}.  The recessional velocity of galaxy
is shown by the dashed-line (see Table~\ref{tab:0}).
All the GC candidates have values within 200~km\,s$^{-1}$ of the recessional velocities.
}

\begin{table}
\centering
\setlength{\tabcolsep}{1.2pt}
\caption{Parameters of the GC candidates from GAIA DR3.
}
\label{tab:GAIA3}
\resizebox{\columnwidth}{!}
{
\begin{tabular}{lccrrcrc} 
\hline
ID &ID$_{\rm G}$ & Parallax &pmRA & pmDec & $G$ & AEN & BP/RP \\
(1) & (2)& (3) &(4) & (5)& (6) & (7) & (8) \\
\hline \hline                                             
 3*  &0528 & 1.397$\pm$0.149  &10.56$\pm$0.15  &$-$7.34$\pm$0.13 &18.56 & 0.00  & 1.34   \\
 5*  &5248 & 0.877$\pm$0.174  &2.15$\pm$0.17   &4.27$\pm$0.15    &18.79 & 0.04  & 1.70   \\
 8  &6912 & \null            &$-$2.10$\pm$0.99&$-$0.21$\pm$0.87  &20.02 & 7.03  & 3.24   \\
 10* &7408 & 0.495$\pm$0.381  &0.30$\pm$0.38   &$-$4.50$\pm$0.33 &19.82 & 1.27  & 1.49   \\
 20 &5008 & \null            &0.92$\pm$0.60   &$-$8.95$\pm$0.62  &20.23 & 2.77  & 2.85   \\
 37 &4528 &  \null           & \null          &  \null           &21.27 & 19.38 & 4.23   \\  
 38 &1968 &  \null           & \null          &  \null           &20.96 & 16.04 & 3.25   \\ 
 49 &3824 &   \null          & \null          &  \null           &21.01 & 12.91 & 3.30   \\

			\hline
\end{tabular}
}
\footnotesize{\\Notes: (2) Last four numbers
of GAIA ID, (3) parallax ($\pi$) in [mas], (4-5) proper motion of right ascension and declination  in [mas yr$^{-1}$], (6) $G$-band magnitude in [mag], (7) astrometric excess noise in [mas], (8)BP/RP excess factor.}
\end{table}

\section{Metallicity and age from spectro-photometric analysis}
\label{Analy}


\defcitealias{Mayya2013}{M13}
\defcitealias{Luis2024}{LN24}
The observed spectrum of each cluster was used to determine two of the most important physical parameters of a stellar population: 
age and metallicity. 
{ Two methods are in common use to obtain these two quantities from spectroscopic data --- (1) simultaneous measurement of age and metallicity by fitting the entire spectrum with a model spectrum, (2) use of age-insensitive spectral features among the Lick indices to obtain metallicity and then use the entire spectrum to get the best determination of the age. 
The former technique is greatly in use to analyse large spectral databases such as SDSS spectra 
\citep[e.g.][]{Perez2013}, whereas the latter technique gives more reliable  results for individual objects especially when the SNR is $\sim$20 or less (e.g.
\citealt{Larsen2001}; \citealt[hereafter \citetalias{Mayya2013}]{Mayya2013}). We used the latter method in this study for analysing the spectroscopic data of our GC candidates.}

For the  GC candidates with no observed spectrum, we obtained ages and extinctions by SED-fitting to the multi-band photometry with SSP models (section \ref{SED}).

\subsection{Metallicity determination}
\label{SpecM}

Absorption lines of cool stars allow the determination of stellar metallic abundances and ages.
Among all the absorption lines, the iron lines Fe5270, Fe5335 and Fe5406 (Fe52, Fe53 and Fe54, respectively) are 
sensitive primarily to metallicity for stellar systems older 
than $\sim$3~Gyr (\citetalias{Mayya2013}; \citealt{Luis2024}) 
independent of their ages. 
We used this property of Fe indices to 
obtain metallicity of our GC candidates.
These three spectroscopic indices were defined by 
\citet{Burstein84}, \citet{Brodie1990} and \citet{Trager1998} and 
empirically calibrated 
as function of the metallicity, [Fe/H],
using the spectra of Galactic GCs provided by \cite{Schiavon2005}. 
For each index, a second order equation was fitted to the relation of the index against the metallicity (see Figure 4 in \citetalias{Mayya2013} and also \citealt{Luis2024}). 

\begin{figure}
    \centering
    \includegraphics[height=6.cm]{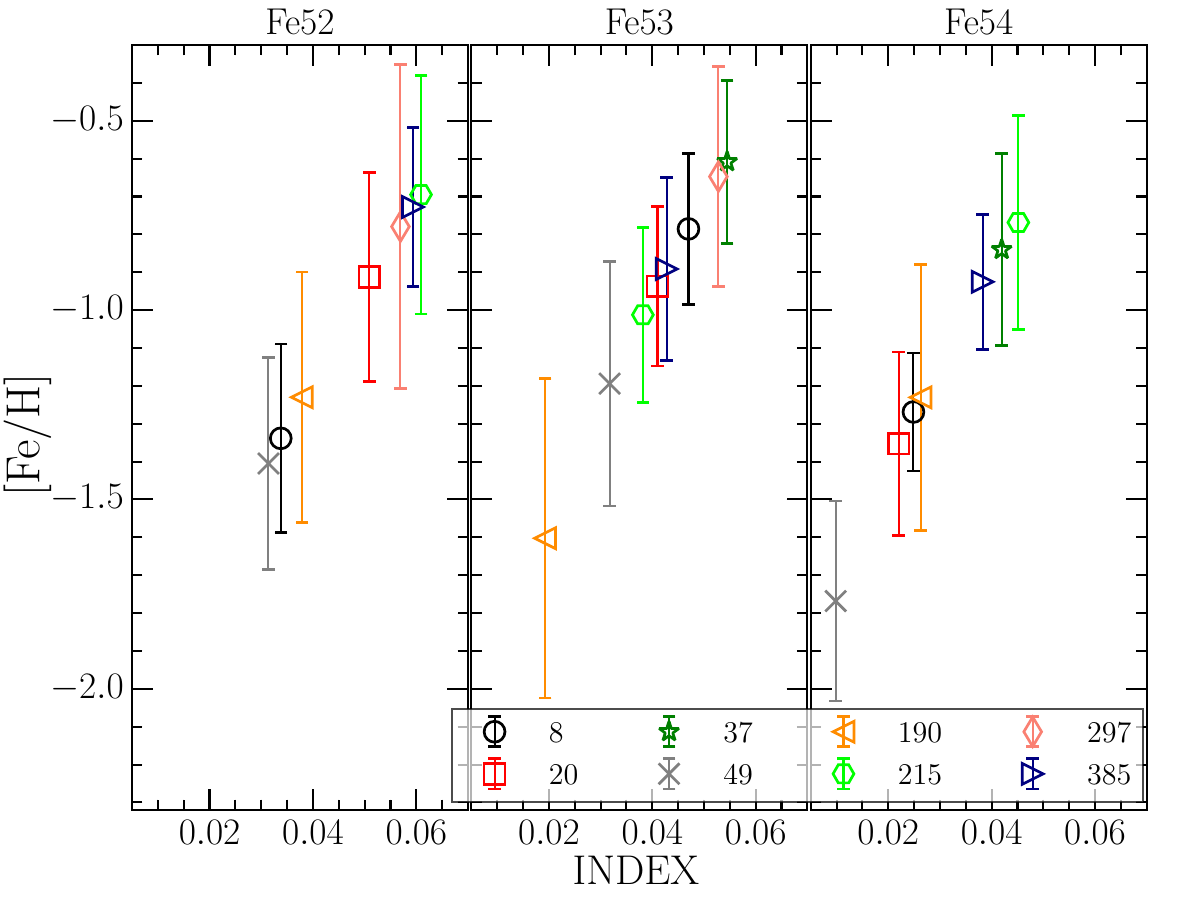}
    \caption{{ Three spectral indices: Fe52, Fe53 and Fe54, were employed to measure the  metallicity (see text) of the GC candidates (see box). The mean metallicity is shown in Table~\ref{tab:R}.}}
    \label{fig:Fe}
\end{figure}

Each spectral index gives us an independent determination of the iron abundance. 
In some cases the extracted spectrum, with poor SNR, does not behave very well near the spectral features showing small jumps or is dominated by noise, making the index unusable. Hence, the metallicity of each  cluster was obtained as the mean of the three values, { weighted by the inverse of their variance.}
Figure~\ref{fig:Fe} shows the metallicity of the GC candidates determined as a function of the measured indices Fe52, Fe53 and Fe54.
Each panel displays the results obtained with one index. 
%
%
%
We conducted a Monte Carlo simulation to estimate the errors in the metallicity. 
For each observed spectrum, we generated one thousand of simulated observations by randomly 
adding noise with an RMS value calculated for each index interval. 
Equations 1 and 2 of \citet{Brodie1990} were employed to measure the spectral indices of each simulated spectrum. 
The metallicity estimates [Fe/H] along with their errors are given in Table~\ref{tab:R}.

Additionally we determined the [$\alpha/$Fe] ratios (alpha element enrichment) of the GC candidates using the relation between Mg${\it b}/\langle$Fe$\rangle$ and [$\alpha/$Fe] (where $\langle$Fe$\rangle= 0.5 \times$(Fe5270 + Fe5335) defined by \cite{Thomas2003}. 
For this purpose we have degraded the observed spectra to mimic the resolution of the Lick/IDS system \citep[see][]{Worthey1994}, and we employed the definitions according to the Lick/IDS system to measure the Mg{\it b},  Fe5270, and Fe5335 indices. 
A new Monte Carlo simulation was applied to calculate the uncertainties of these indices. 

We have parameterized the Mg{\it b}/$\langle$Fe$\rangle$ -- [$\alpha$/Fe]
(dark area of Figure 4 in \citealt{Thomas2003}) 
relation with the second { order} 
equation,

\begin{equation}
    [\alpha/{\rm Fe}]= \frac{-1+\sqrt{{\rm Mg{\it b}/\langle Fe\rangle}-0.2}}{0.75},
\end{equation}
{ which is applicable for }
 Mg{\it b}/$\langle$Fe$\rangle > 1$ and [$\alpha/$Fe] $> -0.05$ . 
 { We could employ this equation to determine the alpha element enhancement in six clusters for which the measured
Mg{\it b}/$\langle$Fe$\rangle > 1$}. 
Table~\ref{tab:Lick} shows the values of F52, Fe53, Mg${\it b}$ indices and Mg${\it b}/\langle$Fe$\rangle$ and [$\alpha/$Fe] ratios.





\begin{table*}
	\centering
	\caption{Physical parameters of the clusters obtained with the analysis of the spectroscopic data.}
	\label{tab:R}
\resizebox{2\columnwidth}{!}{ 
	\begin{tabular}{rcccrcrccc} 
		\hline
ID & [Fe/H]$_{\text{\normalfont Fe}52}$ & [Fe/H]$_{\text{\normalfont Fe}53}$ & [Fe/H]$_{\text{\normalfont Fe}54}$ & [Fe/H]$_{\text{\normalfont mean}}$ & [Fe/H]$_{\text{\normalfont SSP}}$ &  $\log$(Age)& $E(B-V)$ &
$A_V$ & $\chi^2_{min}$\\
	
 (1) & (2) &(3) &(4) & (5) &(6)  & (7) & (8) & (9) & (10) \\
		\hline \hline

8   & $-1.34\pm$0.25   & $-0.79\pm$0.20   & $-1.27\pm$0.15 & $-1.13\pm$0.11 &$-1.31$& 10.18$\pm$0.05 &0.18$\pm$0.01 &0.56$\pm$0.03 &2.93 \\
20  & $-0.91\pm$0.28   & $-0.94\pm$0.21   & $-1.35\pm$0.24 & $-1.06\pm$0.14 &$-1.31$& 10.12$\pm$0.10 &0.04$\pm$0.04 &0.12$\pm$0.12 &3.80 \\
37  & \null            & $-0.61\pm$0.21   & $-0.84\pm$0.25 & $-0.71\pm$0.16 &$-0.71$& \null          &\null         &\null         &\null \\
49  & $-1.40\pm$0.28   & $-1.19\pm$0.32   & $-1.77\pm$0.26 & $-1.49\pm$0.16 &$-1.31$& 9.62 $\pm$0.10 &0.00$\pm$0.01 &0.00$\pm$0.03 &4.65 \\
134 & \null            & \null            & \null          & $-0.71$        &$-0.71$& 9.39 $\pm$0.20 &0.12$\pm$0.05 &0.37$\pm$0.15 &10.85 \\
176 & \null            & \null            & \null          & $-0.71$        &$-0.71$& 9.62 $\pm$0.20 &0.00$\pm$0.01 &0.00$\pm$0.03 &23.48 \\
190 & $-1.23\pm$0.33   & $-1.60\pm$0.42   & $-1.23\pm$0.35 & $-1.32\pm$0.21 &$-1.31$& 9.48 $\pm$0.20 &0.00$\pm$0.01 &0.00$\pm$0.03 &10.56 \\
215 & $-0.69\pm$0.31   & $-1.01\pm$0.23   & $-0.77\pm$0.28 & $-0.86\pm$0.16 &$-0.71$& 9.47 $\pm$0.10 &0.00$\pm$0.01 &0.00$\pm$0.03 &6.53 \\
297 & $-0.78\pm$0.43   & $-0.65\pm$0.29   & \null          & $-0.69\pm$0.24 &$-0.71$&                &              &              & \\
385 & $-0.73\pm$0.21   & $-0.89\pm$0.24   & $-0.92\pm$0.18 & $-0.85\pm$0.12 &$-0.71$& 9.64 $\pm$0.10 &0.02$\pm$0.01 &0.62$\pm$0.03 &7.88 \\

        \hline
	\end{tabular}    
}
\footnotesize{
\\Notes: (2) Metallicity calculated using F52 index, (3) metallicity calculated using F53 index, (4) metallicity calculated using F54 index, (5) weighted mean metallicity (metallicity of IDs 134 and 174 was fixed), (6) metallicity of the model employed to determine $\log$(Age) and reddening ($E(B-V)$), (7) $\log$(Age) of the best fitted model, (8) $E(B-V)$ of best fitted model, (9) extinction of the best fitted model ($R_V=$ 3.1), (10) minimal $\chi^2$ obtained with the best fitted model. IDs 37 and 297 did not provide reliable estimates.
}
\end{table*}

\begin{table}
\centering
\caption{$\alpha$-enhancement of the clusters}
\label{tab:Lick}
 \setlength{\tabcolsep}{1.5pt}
 \resizebox{\columnwidth}{!}
 {
\begin{tabular}{rrcccr}
\hline
ID &Fe52 & Fe53 & Mg$b$ &  Mg$b/\langle$Fe$\rangle$ & [$\alpha/$Fe] \\
(1)&(2) & (3)  & (4) &  (5) & (6)\\
  \hline
\hline
 
8   &1.16$\pm$0.19 &1.40$\pm$0.15 &1.58$\pm$0.19 &1.24$\pm$0.19 &$0.03\pm0.12$ \\
20  &1.49$\pm$0.28 &1.04$\pm$0.17 &2.02$\pm$0.25 &1.60$\pm$0.28 &$0.24\pm0.16$ \\
37  &\null         &1.56$\pm$0.24 &0.37$\pm$0.30 &\null         &\null \\
49  &1.17$\pm$0.23 &1.29$\pm$0.13 &1.54$\pm$0.19 &1.25$\pm$0.21 &$0.03\pm0.14$ \\
134 &2.00$\pm$0.58 &2.79$\pm$0.47 &0.94$\pm$0.26 &0.39$\pm$0.13 &\null \\
176 &1.05$\pm$0.62 &1.65$\pm$0.35 &0.61$\pm$0.40 &0.45$\pm$0.32 &\null \\
190 &0.97$\pm$0.26 &0.99$\pm$0.16 &1.12$\pm$0.10 &1.14$\pm$0.21 &$-0.04\pm0.14$\\
215 &1.78$\pm$0.31 &0.75$\pm$0.12 &2.52$\pm$0.40 &1.99$\pm$0.41 &$0.45\pm0.21$ \\
297 &1.83$\pm$0.52 &1.54$\pm$0.31 &1.40$\pm$0.43 &0.83$\pm$0.29 &\null \\
385 &1.72$\pm$0.25 &1.19$\pm$0.21 &2.88$\pm$0.79 &1.99$\pm$0.59 &$0.45\pm0.29$ \\

 \hline  
  \end{tabular}
  }
  \footnotesize{\\Notes: (2-4) Indices Fe52, Fe53, and Mg$b$,
  (5) Ratio Mg$b/<$Fe$>$, (6) $\alpha$-enhancement from \cite{Thomas2003} (see text).}
  \end{table}

\subsection{Spectroscopic age and reddening determination}
\label{SpecA}

Once the chemical abundances of the clusters were determined, we proceeded to calculate the age and reddening of the GC candidates. 
The first step to deredden the observed spectra was to correct them for the Galactic extinction using $A_V$= 0.21~mag \citep{Schlegel1998}, and the extinction curve established by 
\citetalias{Cardelli1989}
 with $R_V=$ 3.1. The best-fitting method was employed to determine the age and the local reddening, $E(B-V)$, of the GC candidates. The software tool known as Analyzer of Spectra for Age Determination ({\tiny ASAD}; \citealt{Asad2014}) was used to determine the $\log$(Age) and $E(B-V)$ values for the observed spectra.

The {\tiny ASAD} code is designed to determine the age and reddening of an observed spectrum by seeking the best-fit match through a comparison of the observed spectrum with a set of SSP models (spectra). The process of identifying the best fit is accomplished by minimizing the $\chi^2$, which is defined as follows (\citealt{Asad2014}):
\begin{equation}
\label{eq:chi}
\chi^2= \sum^{\lambda_{max( \text{\normalfont \AA} ) }
}_{\lambda=\lambda_{min( \text{\normalfont \AA} )}} \left( 
\frac{( \text{\normalfont OF})_{\lambda}-
(\text{\normalfont MF})_{\lambda}}{(\text
{\normalfont OF})_{\lambda}}  \right)^2,
\end{equation}
where OF and MF are the observed and the model flux  
(normalized at 5500 \AA), respectively.

The synthesis of the SSP models is based on the work  by 
\cite{Vazdekis2010}. 
These models were developed using the Kroupa initial mass function (IMF, \citealt{Kroupa2001}), the evolutionary tracks known as \textquoteleft\textit{Padova+00}\textquoteright\ tracks by \cite{Girardi2000}, 
and the Medium-resolution INT Library of Empirical Spectra (MILES) library of observed spectra provided by \cite{Sanchez2006}. 
The SSP models were acquired from the MILES web page\footnote{\url{http://research.iac.es/proyecto/miles/pages/webtools/tune-ssp-models.php}}
and subsequently custom-built to the specific requirements of our spectroscopic data analysis.

Based on the results previously calculated, we conducted the analysis of each spectrum by utilizing reddened models with chemical enrichment that closely approximates the already determined metallicity. Precisely, we employed SSP models with metallicities of [Fe/H] = $-$1.31 and $-$0.71 dex, and ages ranging from $\log$(Age) = 7.8 to 10.25 yr in increments of 0.05 dex, the reddening of spectra spans over a range of $E(B-V)$= 0 to 1 mag, in steps of 0.01 mag.

To ensure the robustness and accuracy of the fitting process, we narrowed the analysis to a specific wavelength range, which extended from 3900 to 6700 \AA~ (up to 5600 \AA~ for ID 20).  The uncertainties associated to the ages and reddening values calculated were determined with a series of Monte Carlo simulations. Similarly to previous Monte Carlo simulations, each observed spectrum produce one thousand of simulated \textquoteleft observed\textquoteright ~spectra. We determined the age and reddening for each mock spectrum. 

\begin{figure*}
    \centering
   \includegraphics[{trim= 0 1.4cm 0 1.cm},clip,width=\columnwidth]{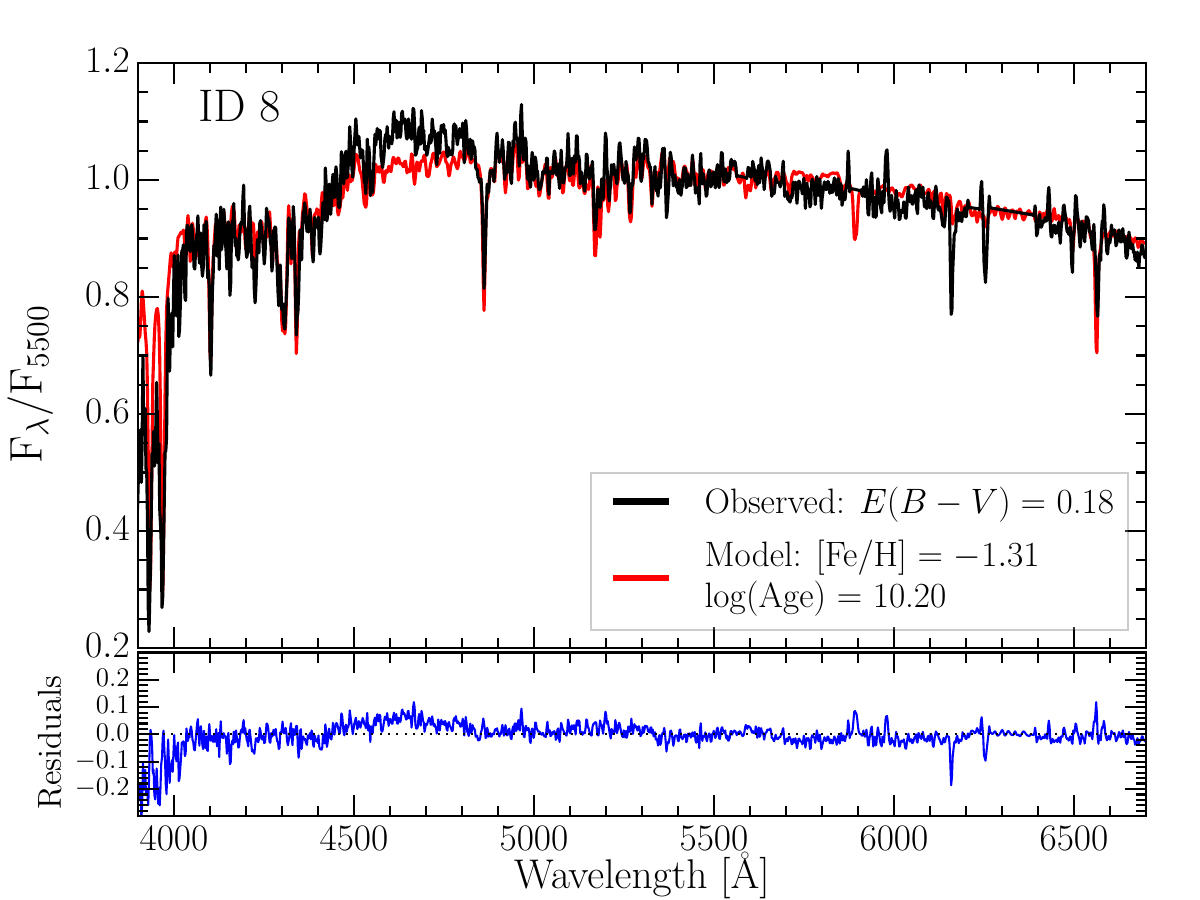}
   \includegraphics[{trim= 0 1.4cm 0 1.cm},clip,width=\columnwidth]{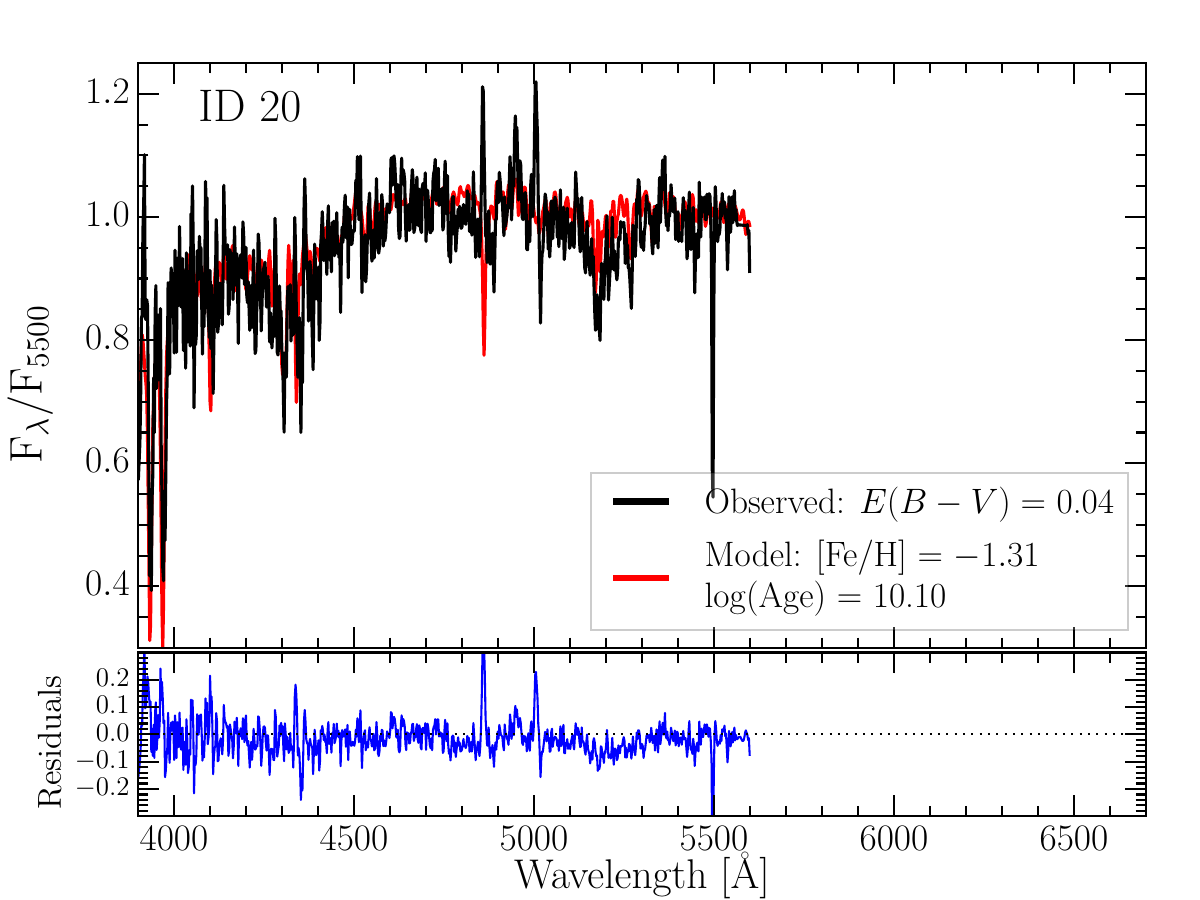}
   \includegraphics[{trim= 0 1.4cm 0 1.cm},clip,width=\columnwidth]{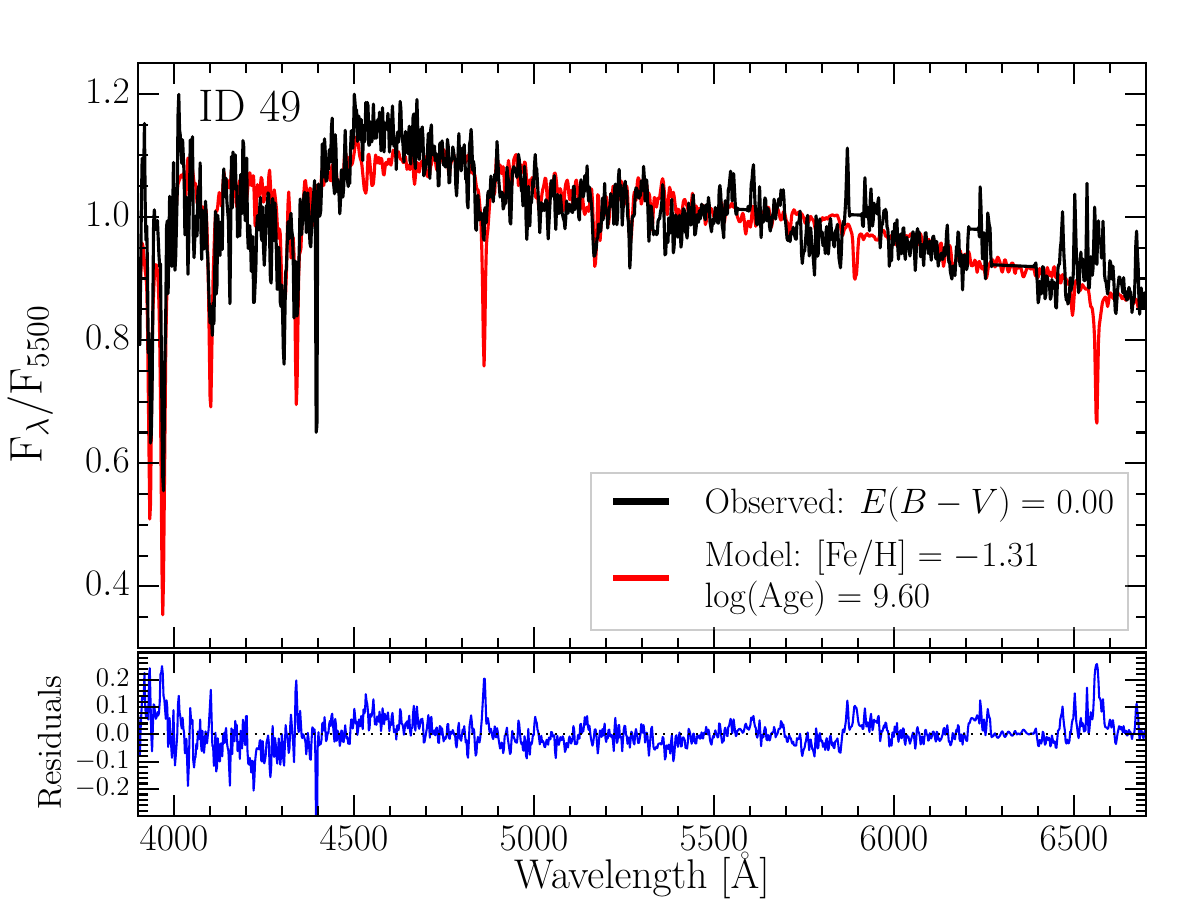}
   \includegraphics[{trim= 0 1.4cm 0 1.cm},clip,width=\columnwidth]{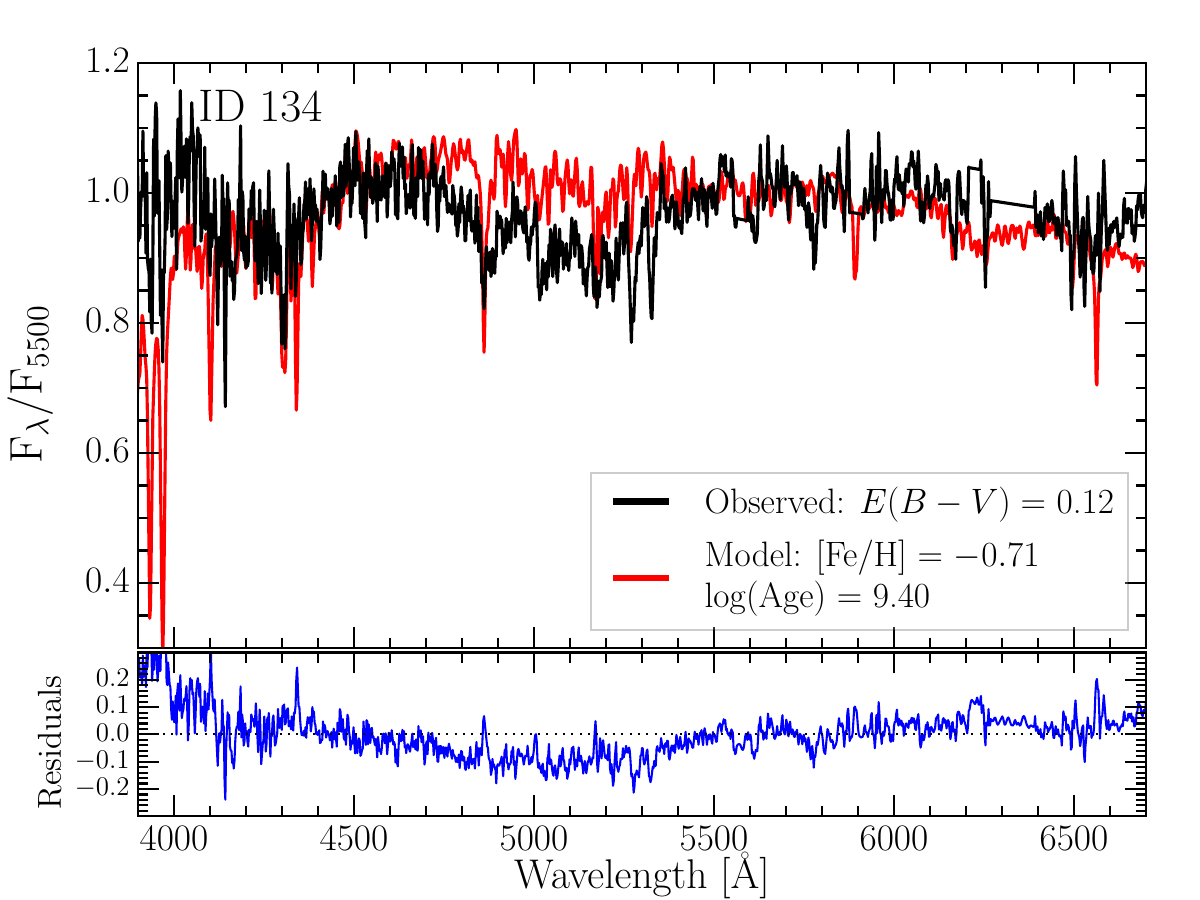}
   \includegraphics[{trim= 0 1.3cm 0 1.cm},clip,width=\columnwidth]{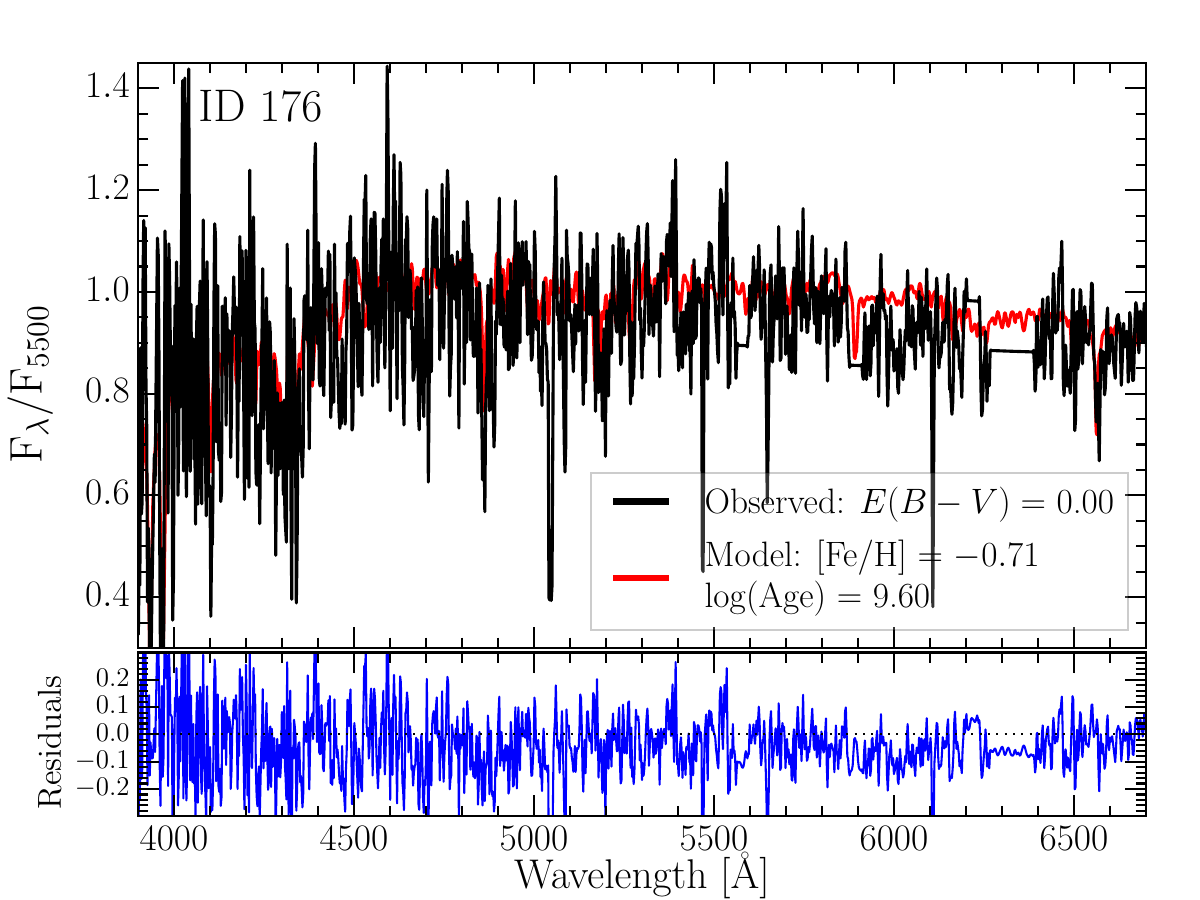}
    \includegraphics[{trim= 0 1.3cm 0 1.cm},clip,width=\columnwidth]{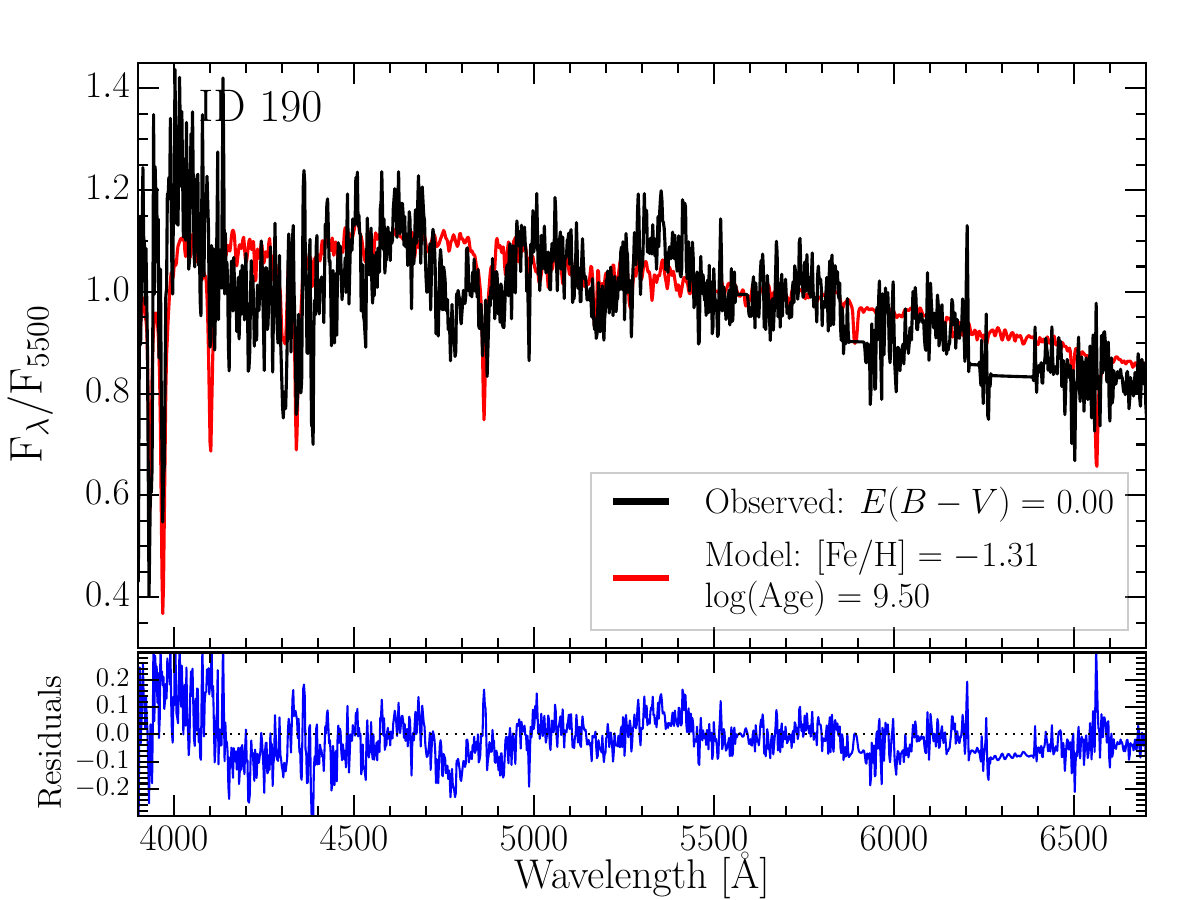}
   \includegraphics[{trim= 0 0.cm 0 1.cm},clip,width=\columnwidth]{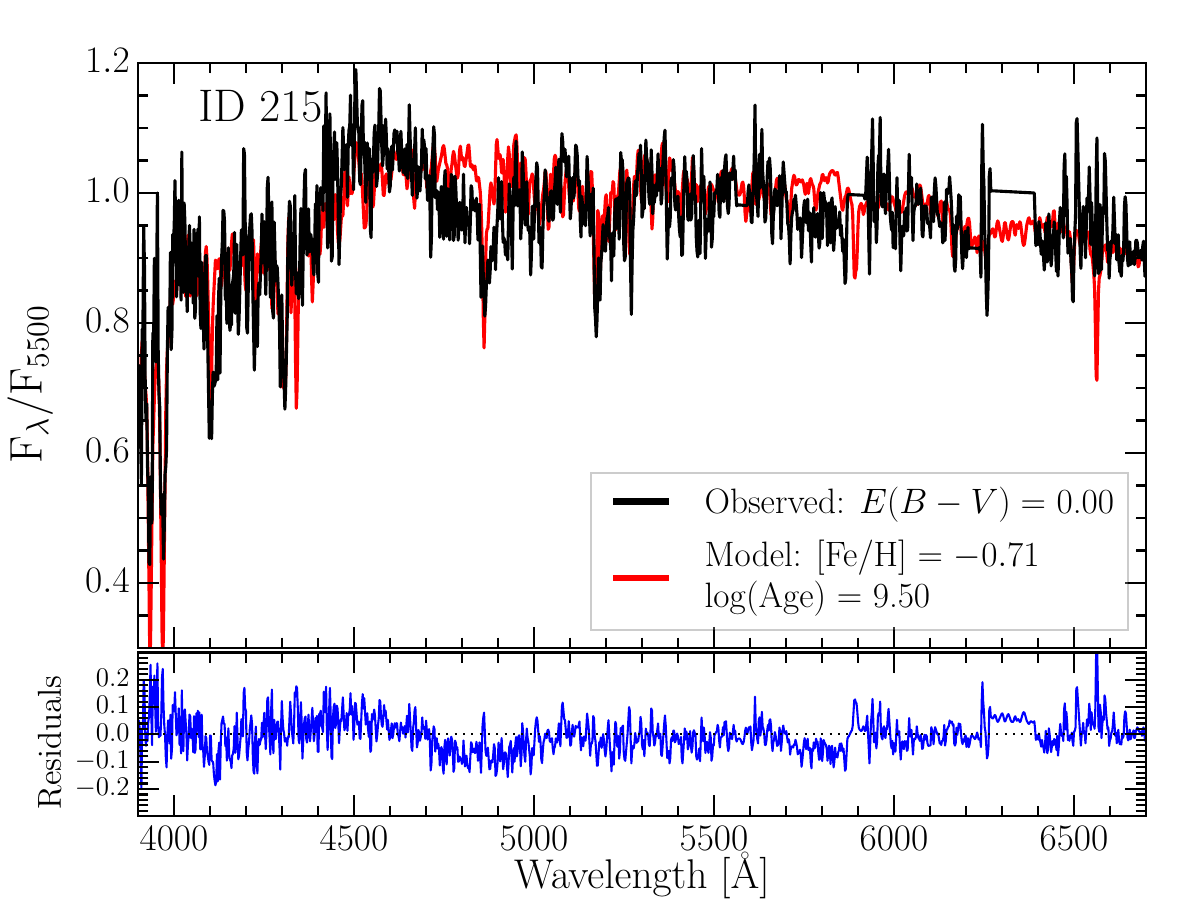}
   \includegraphics[{trim= 0 0.cm 0 1.cm},clip,width=\columnwidth]{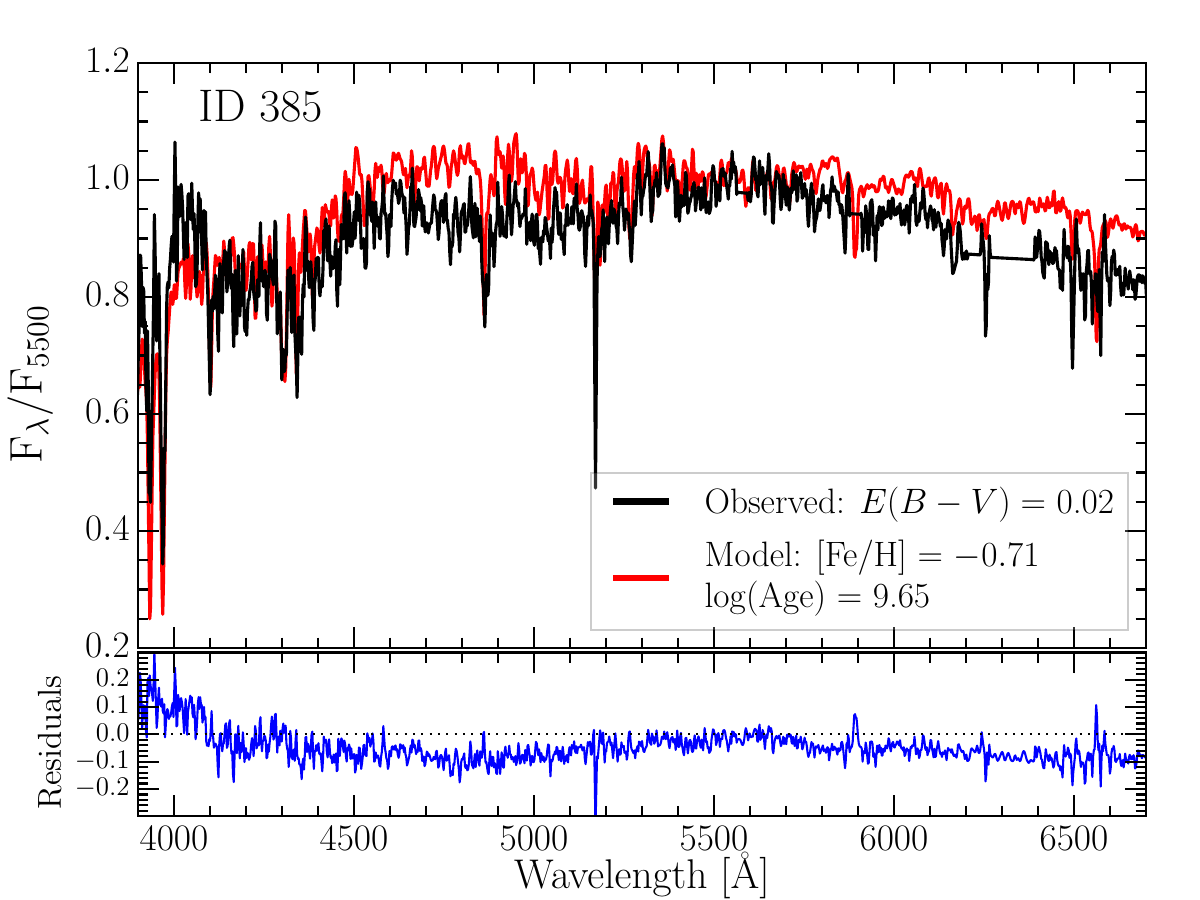}
    \caption{The pannels display the results obtained with the best-fitting method (see box) using {\tiny ASAD} code. Each result is described with two panels, the upper panel shows the observed spectrum and best-fitting model (fluxes are normalized to F$_{5500}$), while the lower panel shows the residual spectrum, given by F${\text{\normalfont{obs}}}- $F${\text{\normalfont{mdl}}}$.
Clusters IDs 8 and 20 are identified as classical GCs (see text), 
while the rest 
are estimated to be younger, with ages $\lesssim$ 4 Gyr. 
    }
    \label{fig:Sp}
\end{figure*}

Values of
$\log$(Age) and extinction obtained for the  clusters are presented in Table~\ref{tab:R} and illustrated in Figure~\ref{fig:Sp}. This figure illustrates both the observed spectra (black line) and their corresponding best-fitted models (red line). 
The boxes in each panel display the values of the best-fit $\log(\text{Age})$ and reddening. The best fitted model reproduces pretty well several features of the observed spectrum (see upper part of each panel). Residuals are shown in the lower part of each panel, which are less than 10\% in majority of cases. 

Based on this spectral analysis, objects IDs 8 and 20 are identified as classical GCs with ages exceeding 13~Gyr and poor in metals, while the rest are determined to be considerably younger, with ages between 3 and 5~Gyr and metal-rich ([Fe/H]$\sim -0.71$~dex). We refer the latter group of clusters as the Intermediate-Age Clusters (IACs). The detailed implications and discussions regarding these classical GCs and IACs will be explored and presented in Section \ref{Discu}.


\subsection{Physical parameters from SED-fitting}
\label{SED}


We could obtain age, metallicity and extinction for only ten of the 17 GC candidates. The rest of the objects are not observed spectroscopically, or the observed spectra are of bad quality. We used the popular technique of SED-fitting approach using the multi-band photometry ranging from 3000~\AA\ to 2.2~$\mu$m in filters listed in Table~\ref{tab:MB}. The use of this range of colors, especially the $g-K_s$ color, is particularly useful to break the age–metallicity degeneracy for old populations as illustrated in Figure 7 of \citetalias{Mayya2013}. Thus, we obtain the age, metallicity, and extinction for all  globular cluster candidates. The availability of independent determination 
 of these quantities using spectroscopy for 10 of the objects allows us to authenticate the detrmined values.


We constructed a theoretical grid of colors employing SSP models calculated
by \citetalias{Girardi2002} using the Kroupa IMF. 
{ The ages are varied from  1~Myr to 12.6~Gyr at logarithmic steps of 0.05~dex at the Z= 0.02 and 0.008. We increased the starting point of the age grid to 4 and 5.6~Gyr at Z=0.004 and 0.001, respectively, given that we do not expect young metal-poor clusters in our sample.}
The grid incorporates these four metallicities.
To simulate the effects of internal extinction, the synthetic colors were reddened using the \citetalias{Cardelli1989} extinction curve. We varied the $A_V$ value from 0.0 to 3.0 mag in steps of 0.01 mag. The number of these synthetic colors is $\sim$3000.
%
%
%

The age, metallicity, and extinction of each star cluster are determined using a $\chi^2$ minimization approach. The $\chi^2$ value is defined as:

\begin{equation}
\label{chi}
\chi^2=\sum_i \omega_i \frac{({\rm color}_i^{\rm obs}-{\rm color}_i^{\rm SSP})^2}{(\sigma_i^{\rm obs})^2},
\end{equation}
where ${\rm color}_i^{\rm obs}$ and ${\rm color}_i^{\rm SSP}$ are the observed and synthetic colors, respectively, $\sigma_i^{\rm obs}$ is the uncertainty of ${\rm color}_i^{\rm obs}$, and the index $i$ runs over the number of colors, such as $F300W-r$ (when the GC candidate is inside the F300W image), $u-r$, $g-r$, $r-i$, $r-z$, $r-J$, $r-H$, $r-K_s$, $u-g$, and $g-K_s$. 
As commented above, 
this last color is particularly useful to break the age–metallicity degeneracy.
%
The $\omega_i$ factor represents the weight assigned to each color. After conducting various tests, it was determined that the best fit (minimum value of $\chi^2$) is achieved by assigning equal weight to all colors. Thus, $\omega_i$ is set to 1/10 (or 1/9 when F300W is not available), indicating an equal contribution from each color.

We performed a Monte Carlo simulation to derive the physical parameters ($\log$(Age), Z and $A_V$) along with their associated uncertainties. This involved introducing Gaussian noise into each observed set of colors, resulting in the generation of 10$^4$ sets of \textquoteleft observed\textquoteright ~colors. 
The $\chi^2$-minimization method was applied to each set of \textquoteleft observed\textquoteright ~colors. This resulted in a Gaussian distribution of ages for the best-fitting models, where the mean ($\mu_{Age}$) represents the $\log$(Age) of the cluster, and its uncertainty is given by the standard deviation ($\sigma_{Age}$) of the age distribution. Similarly, the $\chi^2$-minimization process generated 10$^4$ extinction values, which were used to calculate $A_V$ and its associated uncertainty for each cluster. The metallicity of each cluster corresponds to the metallicity of the model that provides the best fit. The obtained results are presented in Table~\ref{tab:RSED}.

The Figure~\ref{fig:SED} displays the results obtained with this SED-fitting analysis for 
all GC candidates. The observed photometric data are represented by solid circles, while theoretical reddened colors are displayed with open circles. Arrows indicate upper limits in the data. 
 
For the sake of illustration, in each panel we plot a model spectrum\footnote{SSP models provided by A. Bressan, private communication} of best-fit age, metallicity and extinction values. 
Photometric and spectroscopic data are normalized to the flux at the $g$-band (flux at 4770 \AA). 
The residuals, calculated as ${\rm F_{obs}}-{\rm F_{SSP}}$, are displayed at the bottom of each panel. The obtained parameters for the remaining clusters are provided in Table~\ref{tab:RSED}.

In Figure~\ref{fig:Comp}, we compare the ages determined from the SED-fitting procedure with the spectroscopic ages for the eight clusters in common. 
Two of the clusters inferred as classical GCs (IDs 8 and 20) using the spectroscopic ages have ages $\log$(Age) > 9.90  
with the SED method. The derived metallicities of these two clusters are also among the lowest, which authenticates the values derived from the SED-fitting method.
For the other 6 clusters classified as IACs spectroscopically, the SED-fitting derived ages are also consistent with them being younger than classical GCs, with the SED-fitting derived ages systematically higher by up to 0.3~dex. Among the remaining nine clusters without spectroscopic data, one (ID 38) is likely a classical GC ($\log$(Age) $> 10.05\pm0.10$; Z= 0.001), with the rest being IACs with a mean age of 4~Gyr and Z $\lesssim$ 0.004. The derived $A_V \lesssim$ 1~mag. 
%
%

The Figure~\ref{fig:HistAgeAvF} shows the age and extinction values of all star clusters, {  either  
through spectral analysis when available (eight objects) or  through SED-fitting in cases where spectral analysis was not feasible (nine objects). } 
 of $\log$(Age) and $A_V$ for both the samples are very similar, which ensures that the conclusions that we derive do not depend on the method of analysis. 
 The bulk of the red clusters are IACs with ages spread around $\log$(Age) $\sim$9.6, i.e. 4~Gyr. The $A_V$ distribution is dominated by values of $<0.5$~mag.

\begin{table}
	\centering
	\caption{Physical parameters determined with SED-fitting using multi-band photometric data.}
	\label{tab:RSED}
  \resizebox{\columnwidth}{!}
{
	\begin{tabular}{rrrcc} 
		\hline
ID & Z & $\log$(Age)& $A_V$ & $\chi^2_{min}$ \\
(1)& (2)& (3)       & (4)   & (5)       \\ 
		\hline \hline
  
8   & 0.001 & 9.93$\pm$0.10 & 0.27$\pm$0.08 & 0.88 \\
20  & 0.001 & 9.91$\pm$0.05 & 1.17$\pm$0.14 & 8.61 \\
37  & 0.004 & 9.63$\pm$0.10 & 1.82$\pm$0.19 & 0.61 \\
38  & 0.001 & 10.05$\pm$0.10 & 0.08$\pm$0.10 & 0.04 \\
49  & 0.001 & 9.90$\pm$0.05 & 0.61$\pm$0.15 & 1.00 \\
134 & 0.004 & 9.69$\pm$0.15 & 1.38$\pm$0.17 & 0.65 \\
176 & 0.004 & 9.60$\pm$0.05 & 0.74$\pm$0.17 & 0.84 \\
190 & 0.001 & 9.79$\pm$0.10 & 0.22$\pm$0.20 & 0.06 \\
215 & 0.004 & 9.60$\pm$0.05 & 0.43$\pm$0.24 & 0.32 \\
297 & 0.004 & 9.60$\pm$0.05 & 1.36$\pm$0.35 & 1.42 \\
329 & 0.004 & 9.60$\pm$0.05 & 1.98$\pm$0.37 & 1.10 \\
366 & 0.02  & 8.36$\pm$0.25 & 0.72$\pm$0.25 & 0.19 \\
376 & 0.001 & 9.80$\pm$0.05 & 0.78$\pm$0.27 & 0.25 \\
384 & 0.008 & 8.70$\pm$0.40 & 0.22$\pm$0.39 & 0.35 \\
385 & 0.004 & 9.60$\pm$0.05 & 0.79$\pm$0.24 & 0.37 \\
390 & 0.004 & 9.60$\pm$0.05 & 1.55$\pm$0.36 & 1.56 \\
391 & 0.004 & 9.70$\pm$0.15 & 0.09$\pm$0.14 & 0.05 \\

        \hline
	\end{tabular}  
 }
 \footnotesize{\\Notes: (2) Metallicity, (3) $\log$(Age) [yr], (4) extinction [mag] (5) minimal $\chi^2$ of the best fitting.}
	\end{table}

\begin{figure*}
    \centering
    \includegraphics[{trim= 0 1.51cm 0 0.9cm},clip,width=0.8\columnwidth]{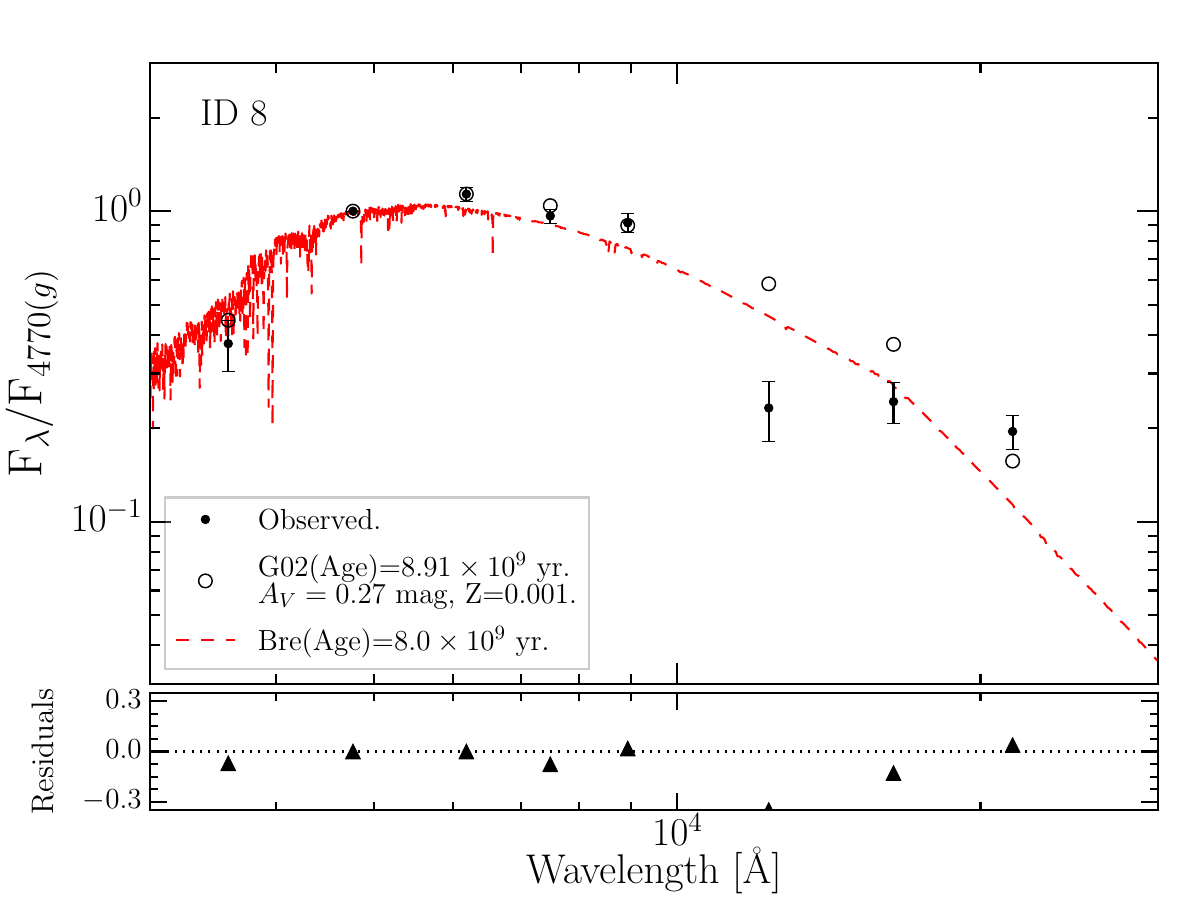}
    \includegraphics[{trim= 0. 1.51cm 0 0.9cm},clip,width=0.8\columnwidth]{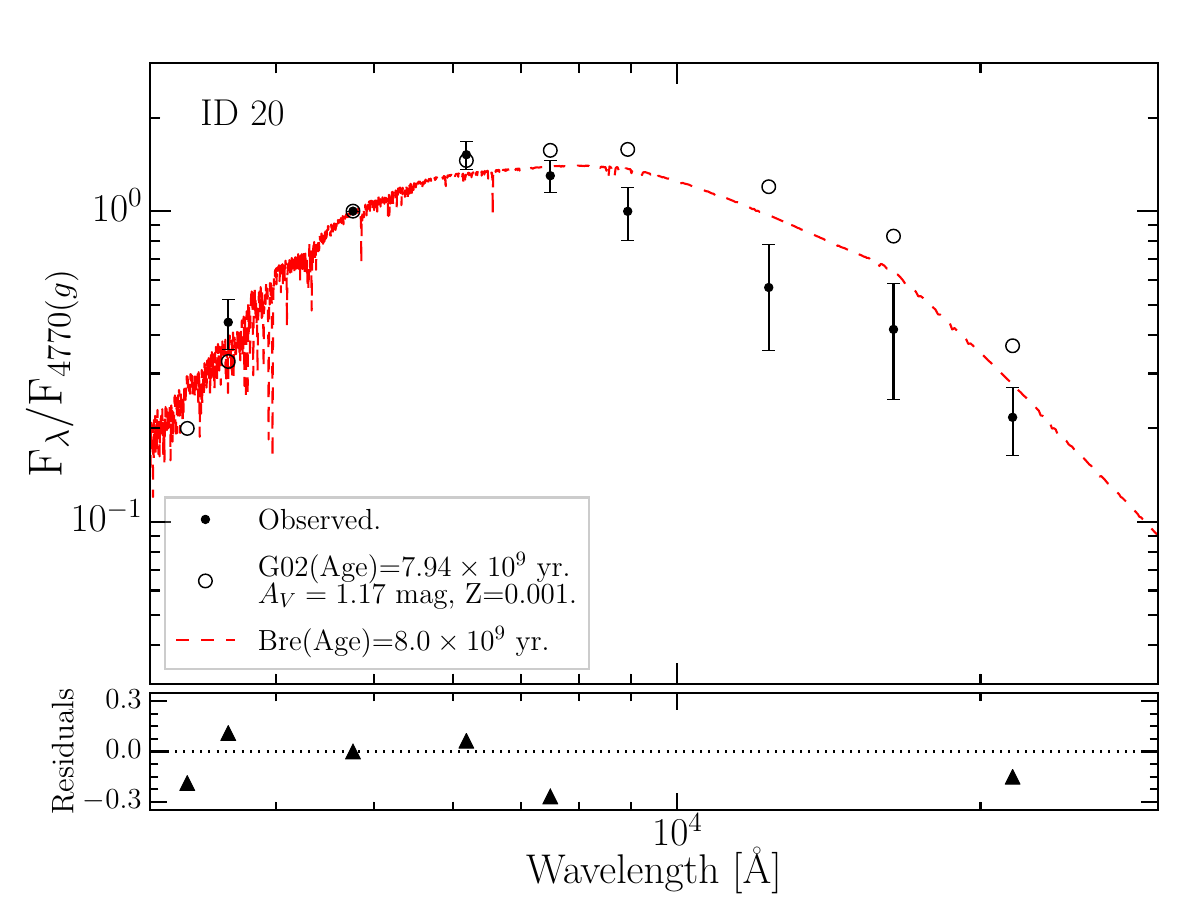}
    \includegraphics[{trim= 0 1.51cm 0 0.9cm},clip,width=0.8\columnwidth]{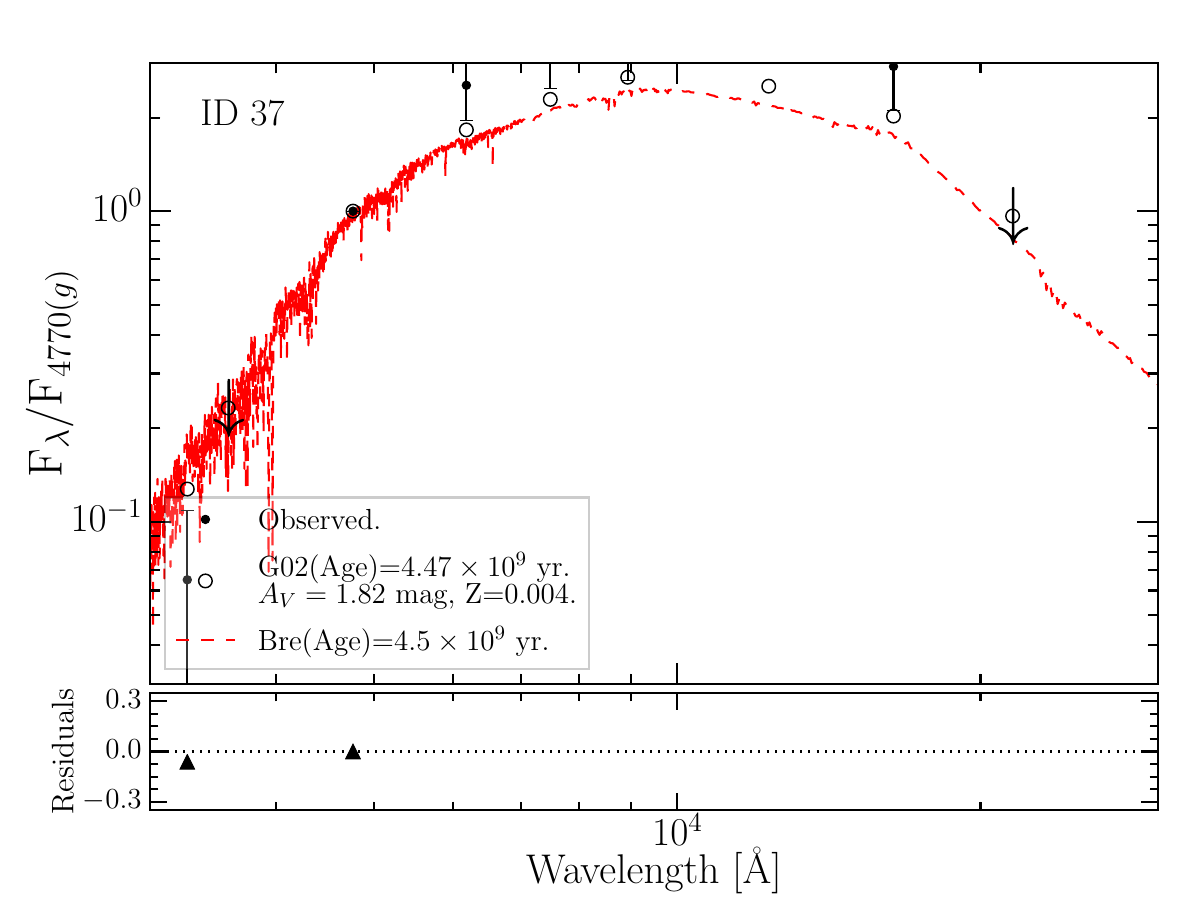}
   \includegraphics[{trim= 0 1.51cm 0 0.9cm},clip,width=0.8\columnwidth]{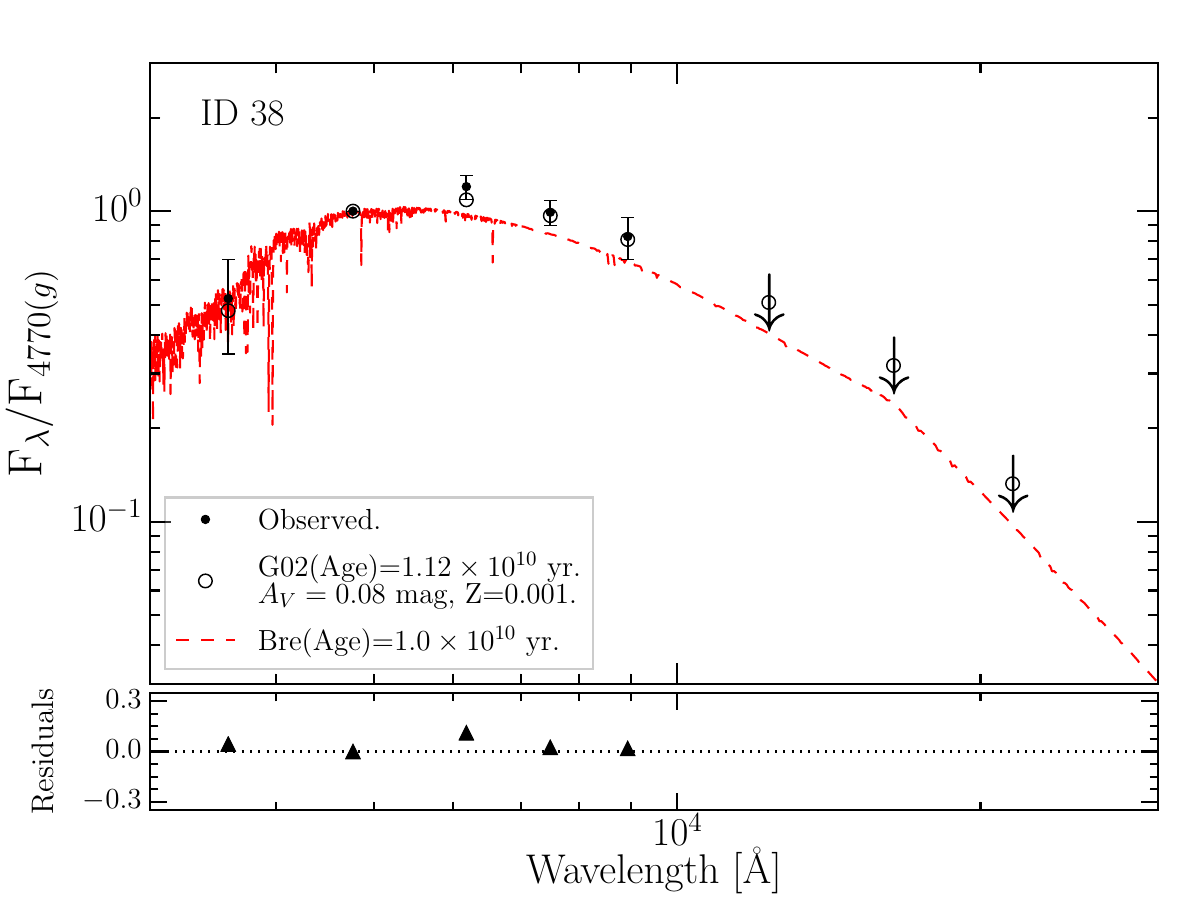}
   \includegraphics[{trim= 0 1.51cm 0 0.9cm},clip,width=0.8\columnwidth]{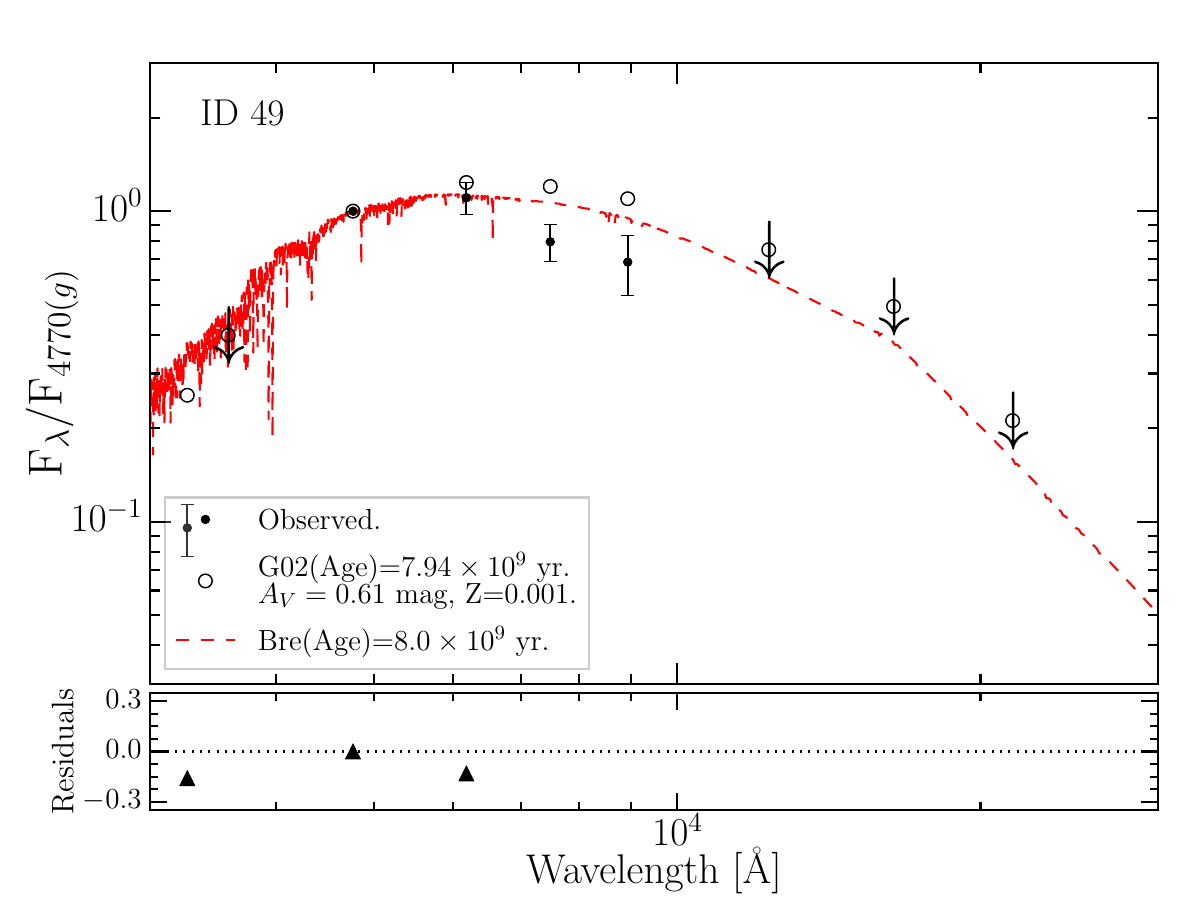}
     \includegraphics[{trim= 0 1.51cm 0 0.9cm},clip,width=0.8\columnwidth]{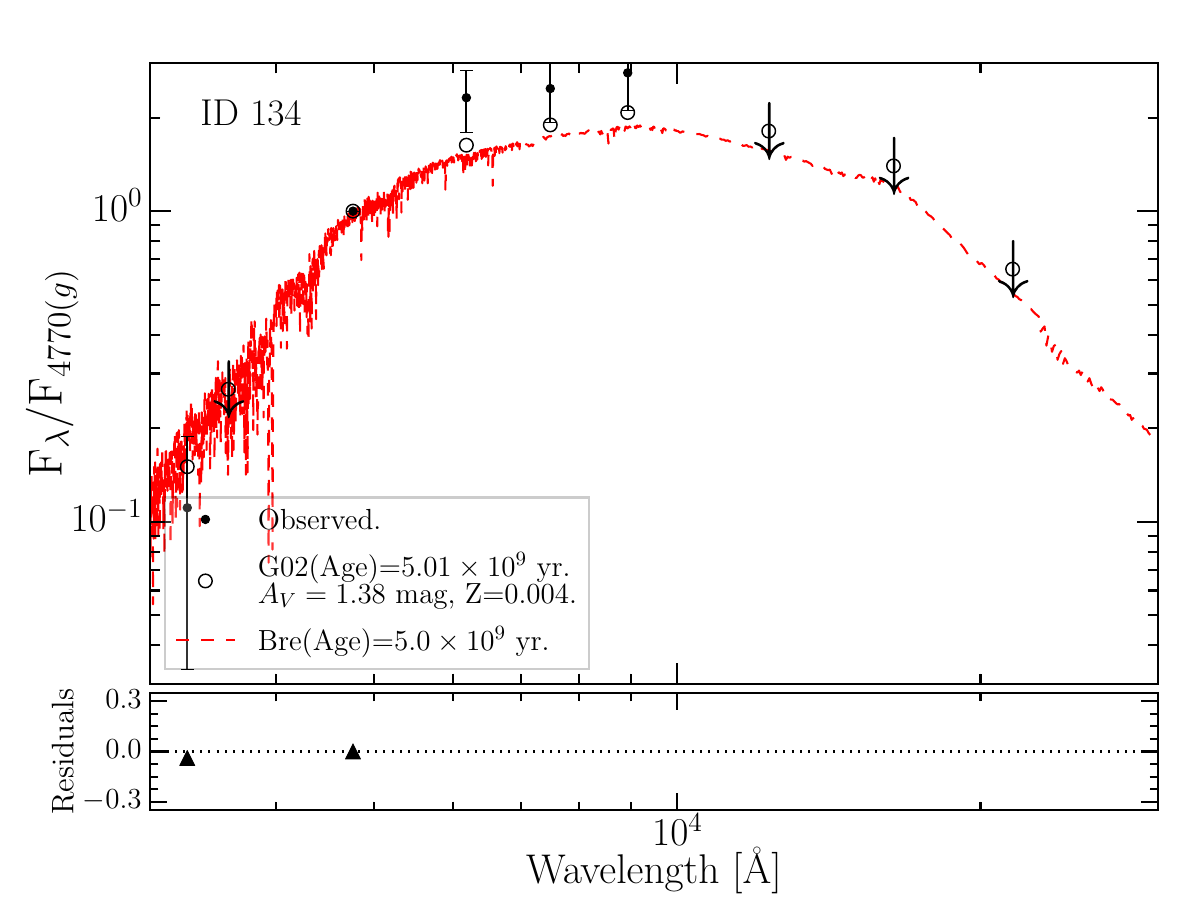}
   \includegraphics[{trim= 0 1.51cm 0 0.9cm},clip,width=0.8\columnwidth]{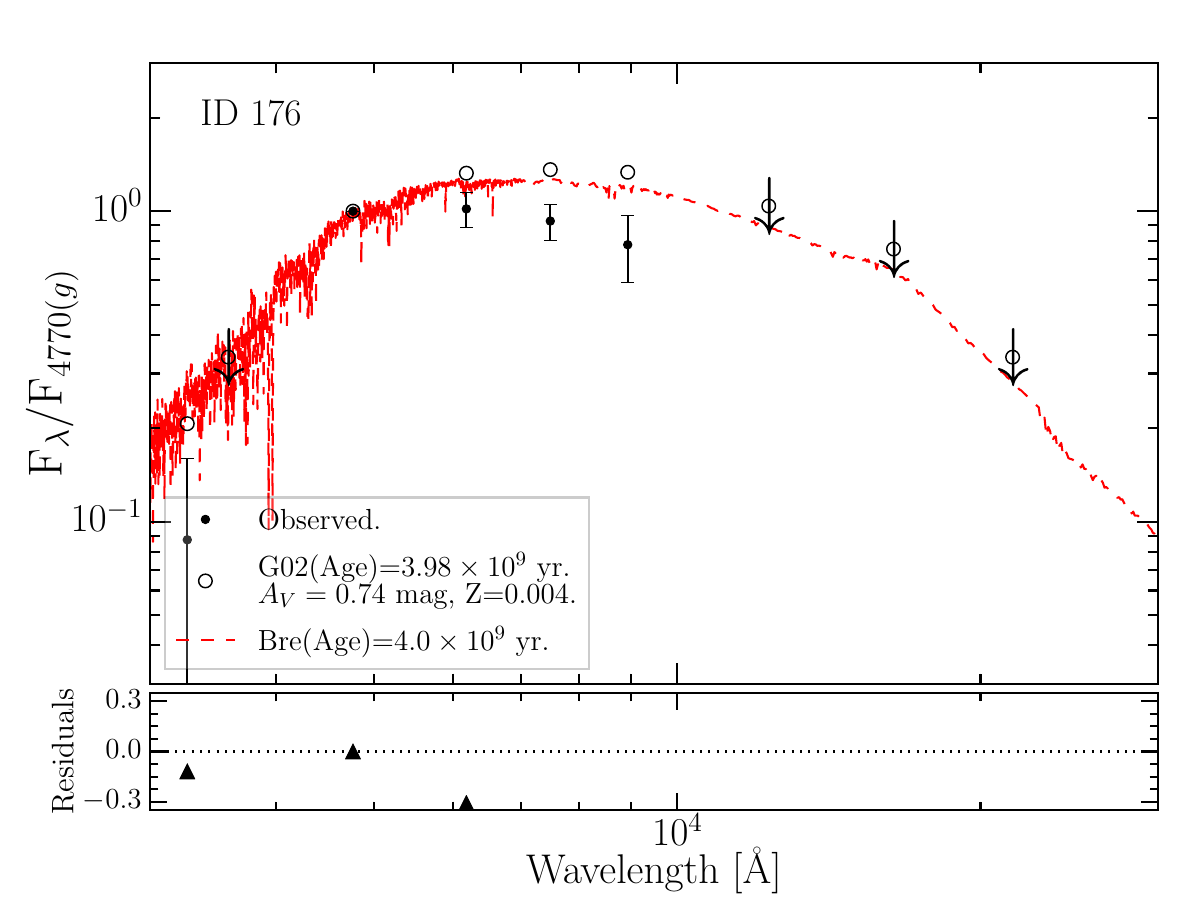}
   \includegraphics[{trim= 0 1.51cm 0 0.9cm},clip,width=0.8\columnwidth]{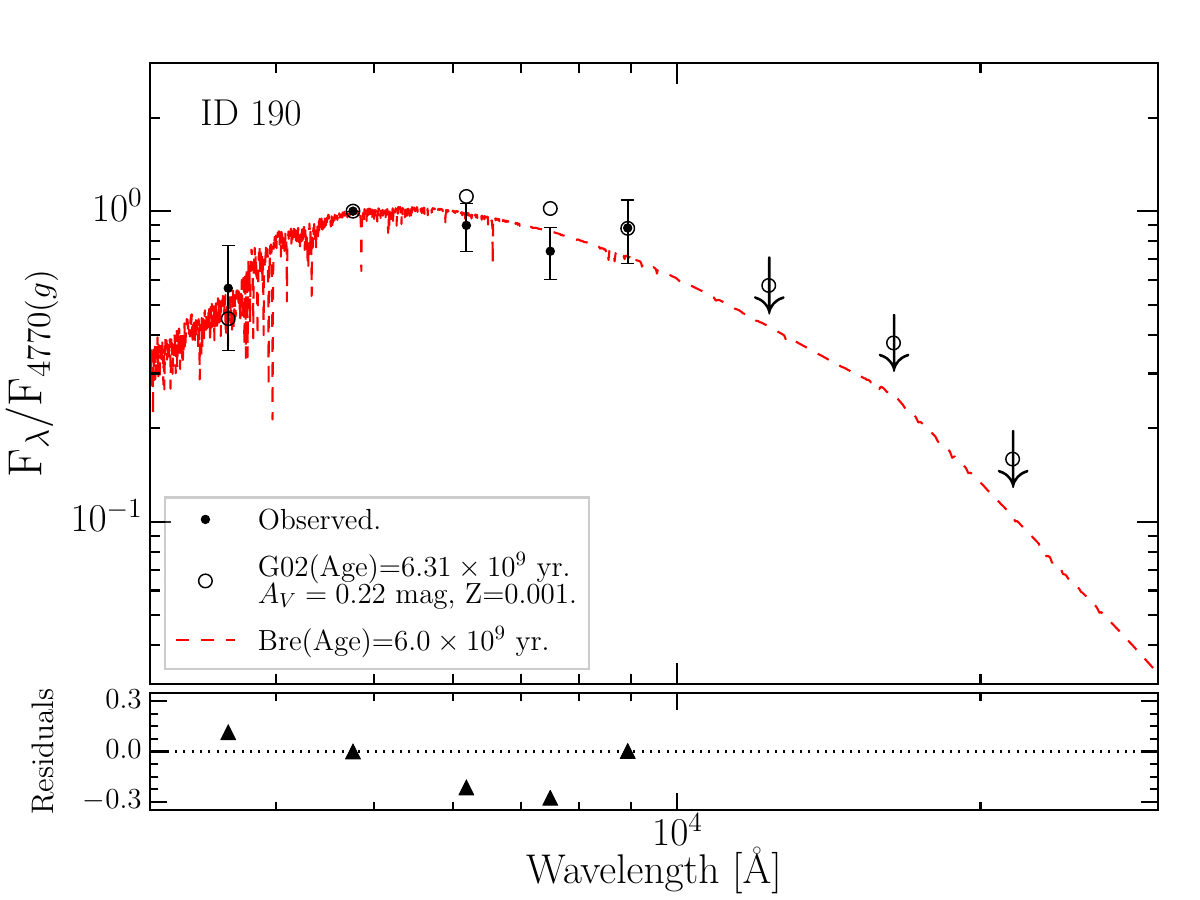}
   \includegraphics[{trim= 0 0 0 0.9cm},clip,width=0.8\columnwidth]{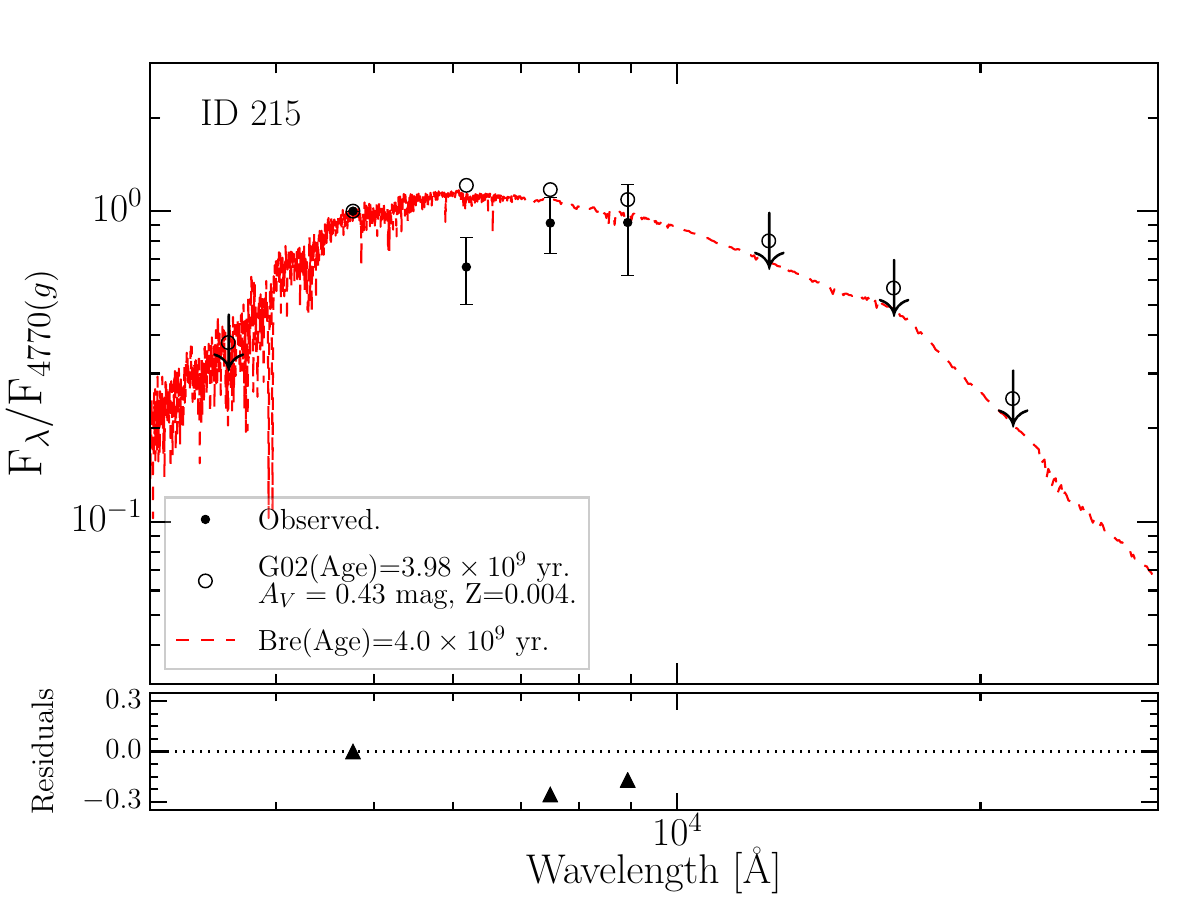}
   \includegraphics[{trim= 0 0 0 0.9cm},clip,width=0.8\columnwidth]{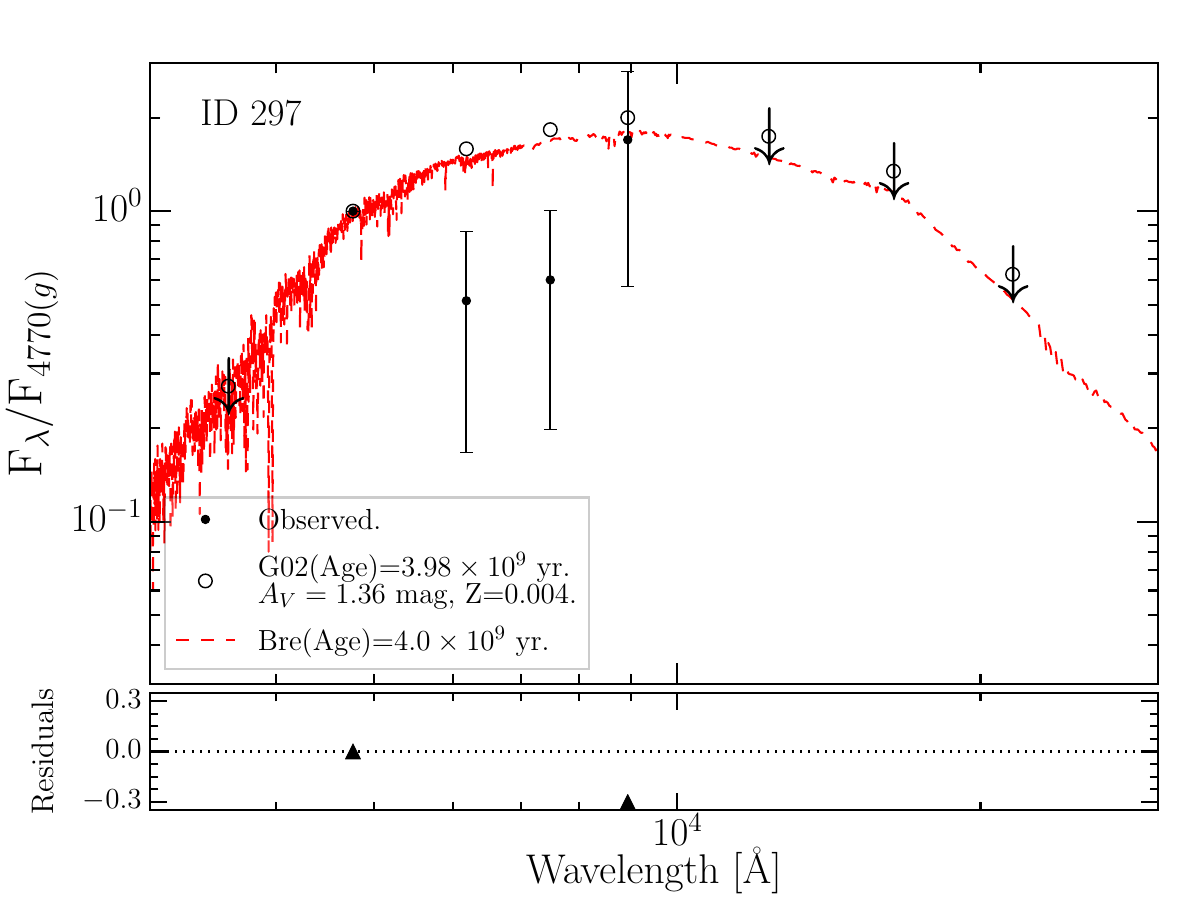}

    \caption{The plots show the results obtained with SED-fitting method using multi-band photometry. The photometric data of the best-fitted model (open circles) is compared with the observational data of the cluster (solid circles), arrows represent upper limit in the respective band. The spectrum of a SSP with physical parameters similar to the best-fitted model is shown (red line) to illustrate the possible form of the spectrum of each cluster. Residuals, ${\rm F_{obs}} - {\rm F_{SSP}}$,  are shown in the bottom of each panel. 
    In 
    Table~\ref{tab:RSED} lists the values the physical parameters obtained for all clusters are shown.}
    \label{fig:SED}
\end{figure*}

\begin{figure*}
\ContinuedFloat
\centering

\includegraphics[{trim= 0 1.51cm 0 0.9cm},clip,width=0.8\columnwidth]{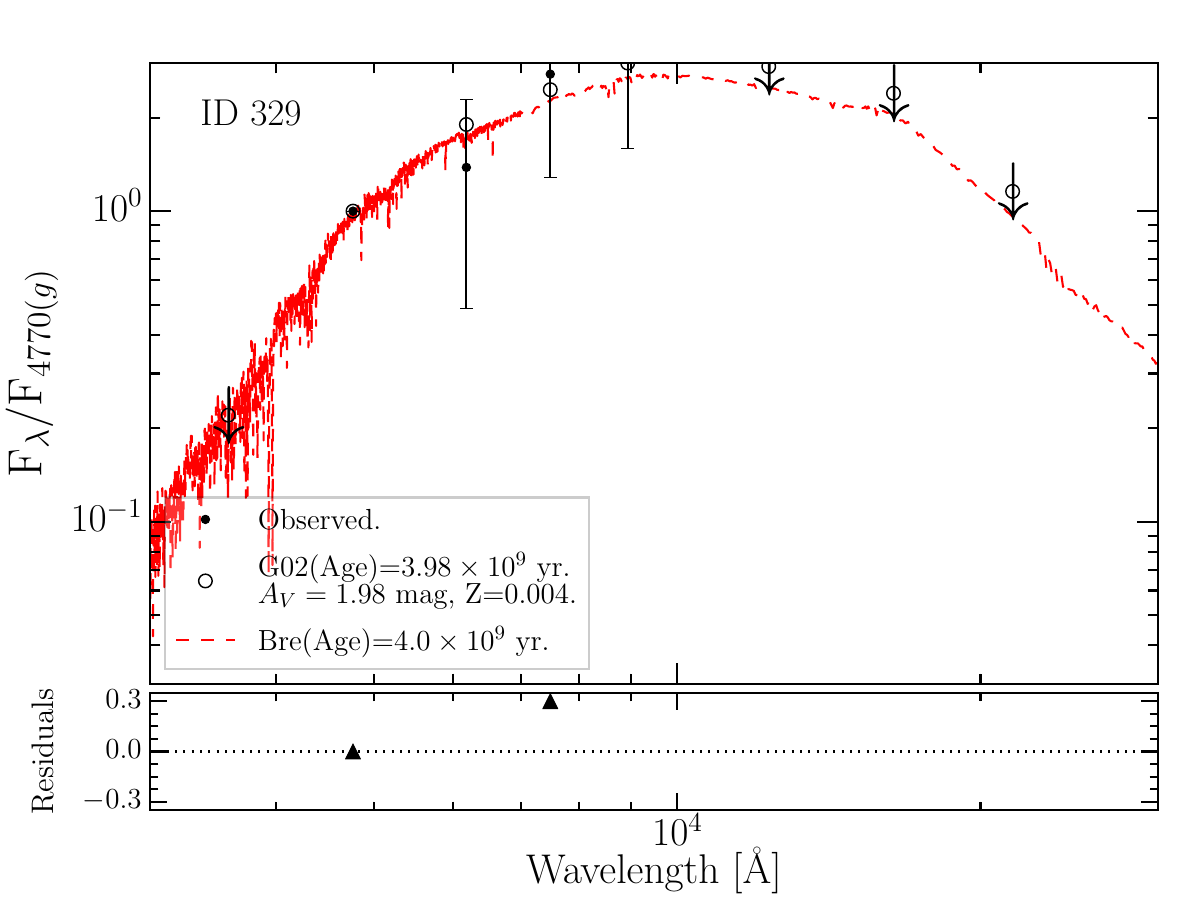}
\includegraphics[{trim= 0 1.51cm 0 0.9cm},clip,width=0.8\columnwidth]{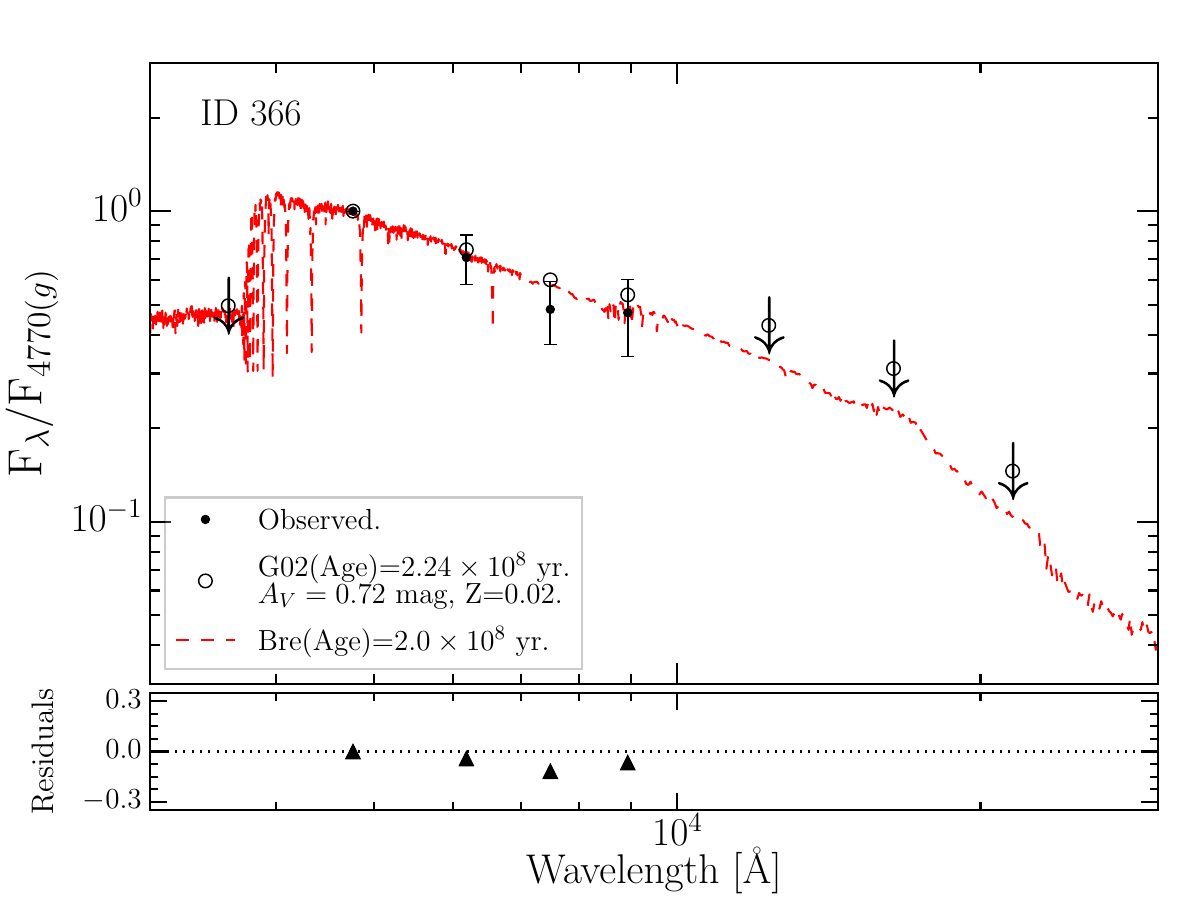}
\includegraphics[{trim= 0 1.51cm 0 0.9cm},clip,width=0.8\columnwidth]{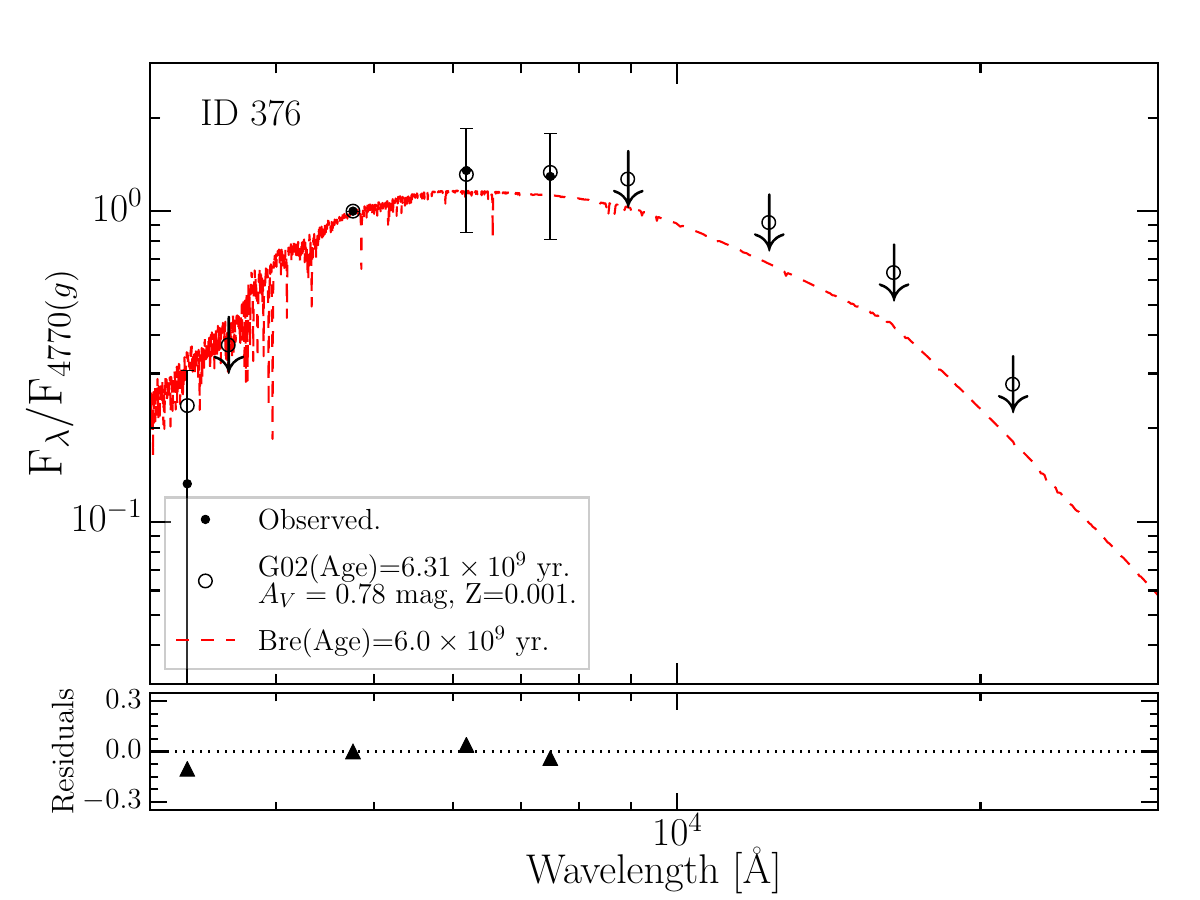}
\includegraphics[{trim= 0 1.51cm 0 0.9cm},clip,width=0.8\columnwidth]{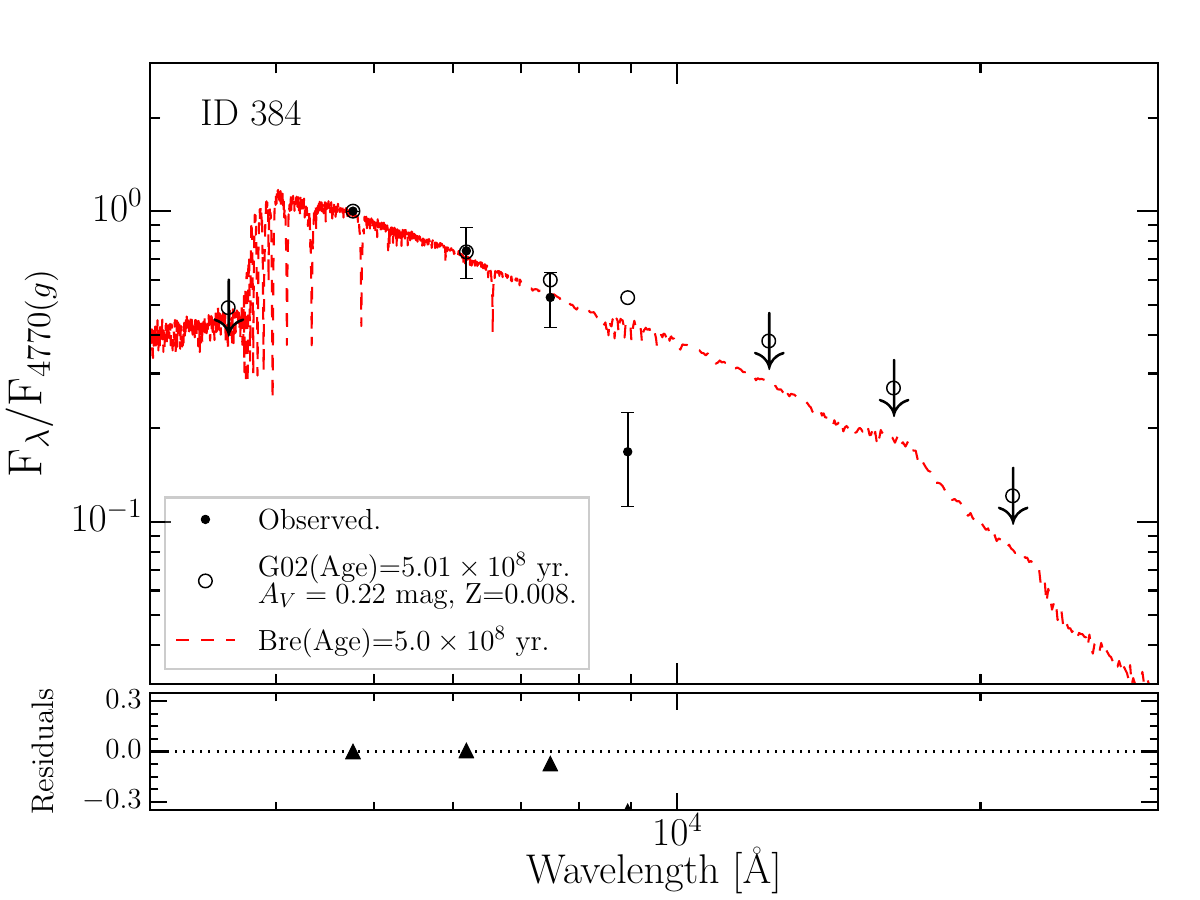}
\includegraphics[{trim= 0 1.51cm 0 0.9cm},clip,width=0.8\columnwidth]{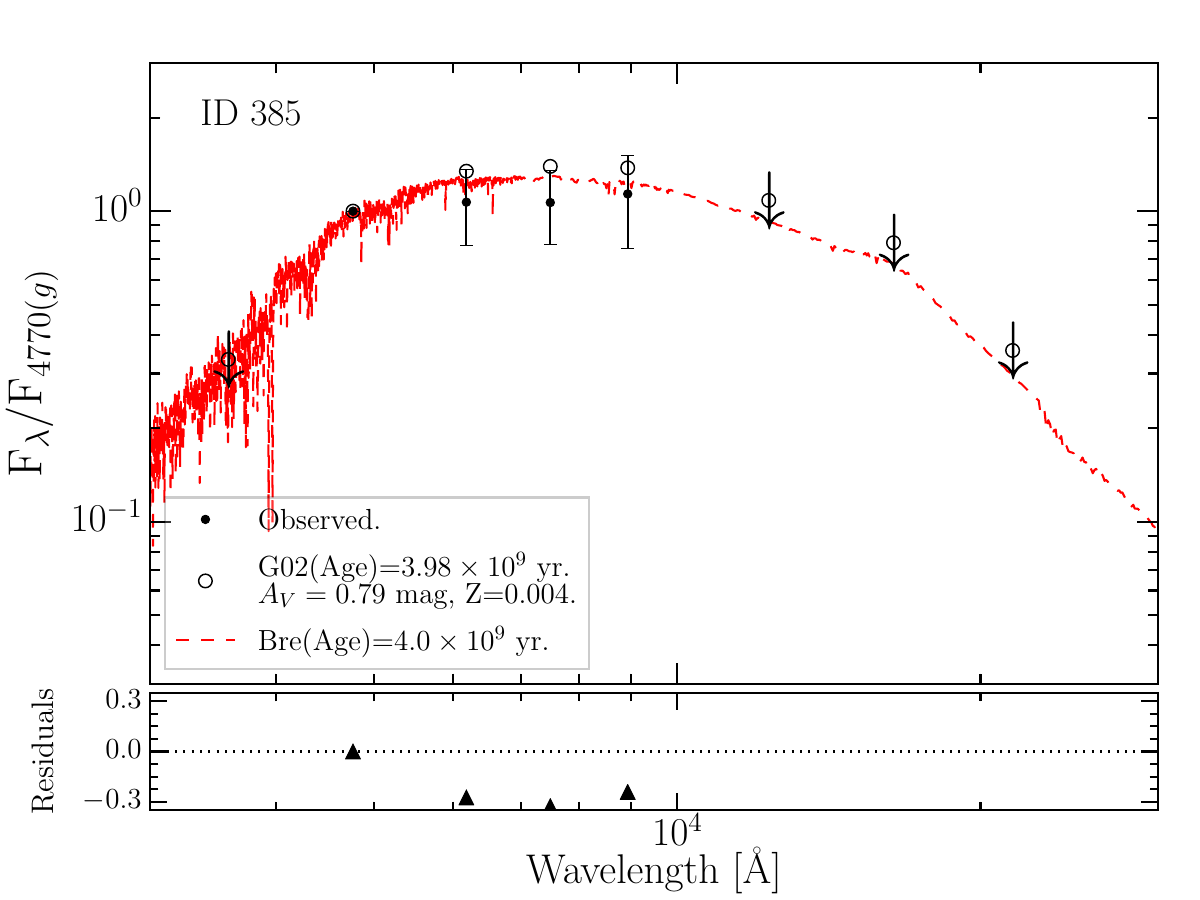}
\includegraphics[{trim= 0 1.51cm 0 0.9cm},clip,width=0.8\columnwidth]{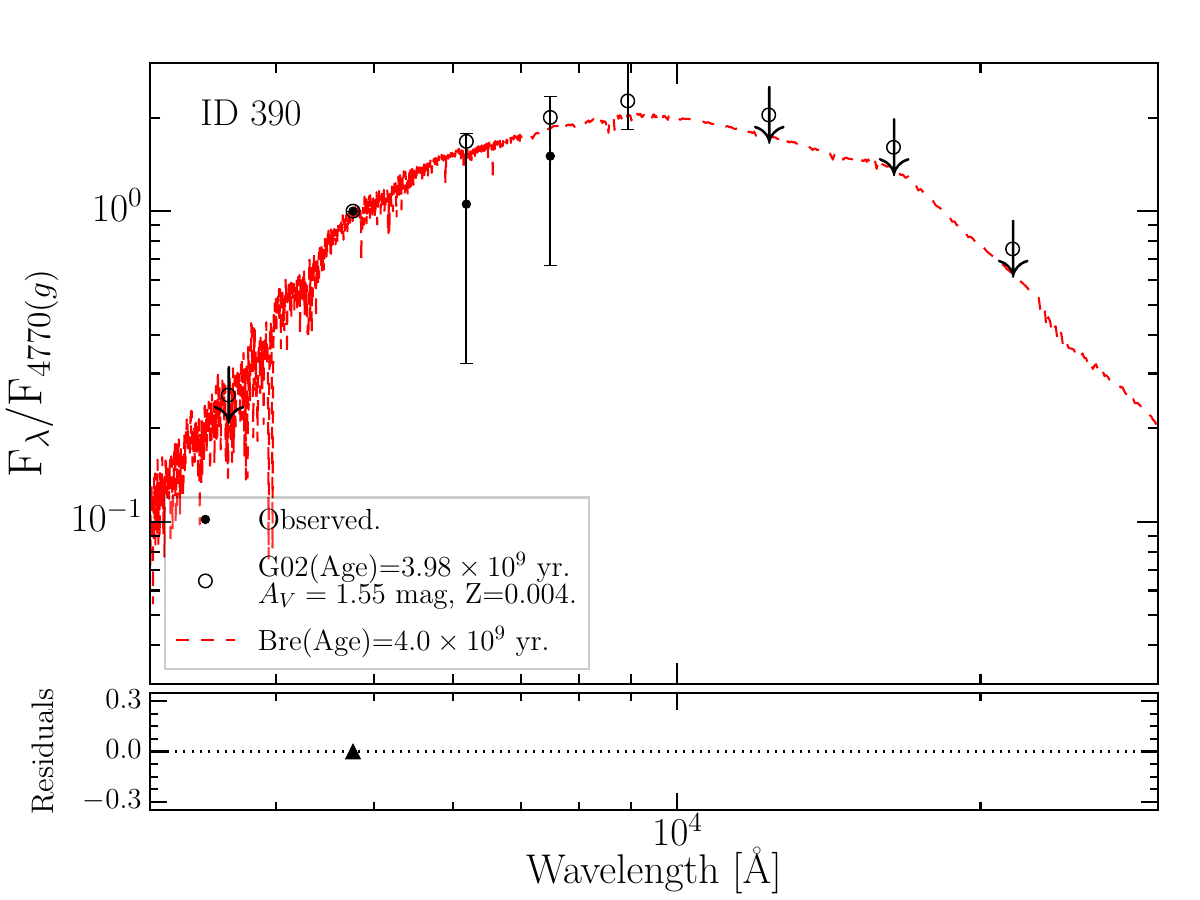}
\includegraphics[{trim= 0 .1cm 0 0.9cm},clip,width=0.8\columnwidth]{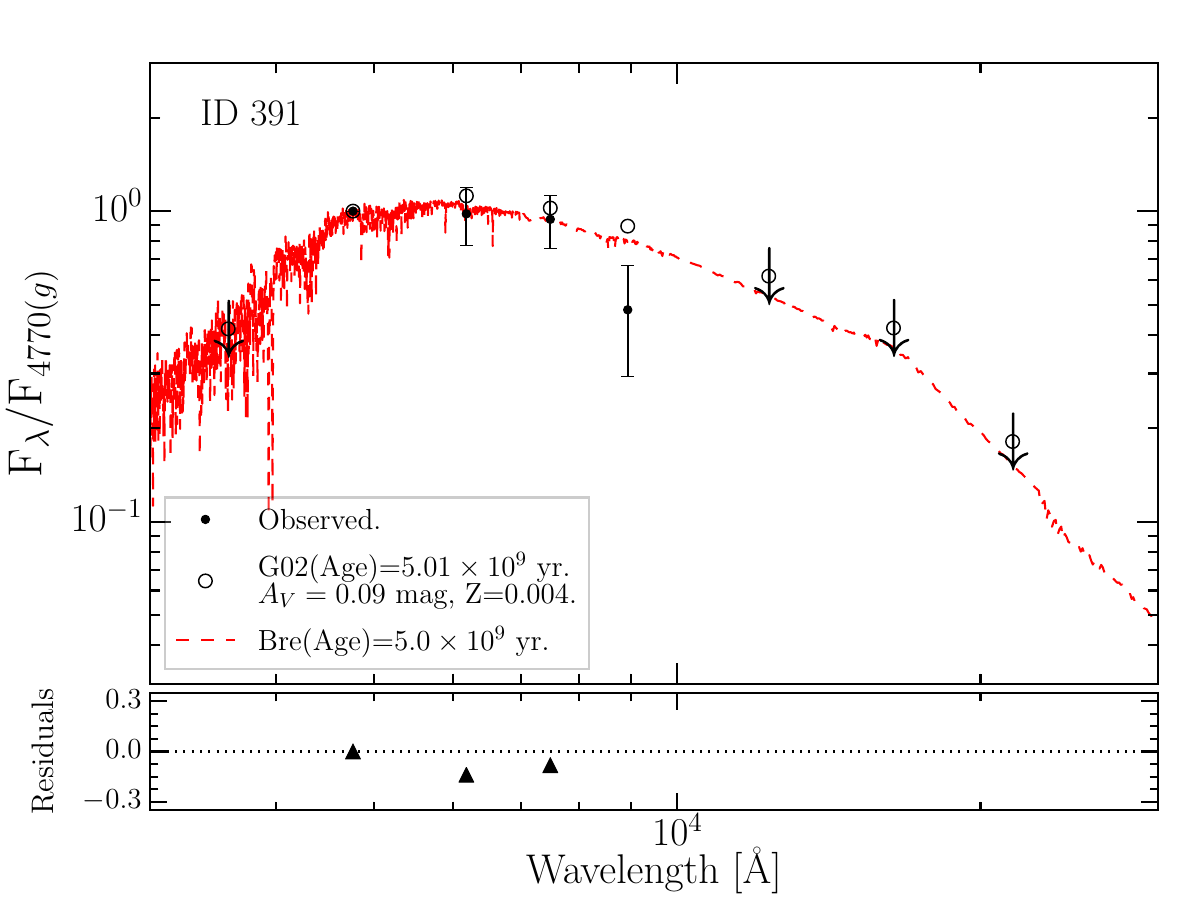}

\caption{Figure~\ref{fig:SED} (continued).}
\end{figure*}

\subsection{Photometric mass}
\label{pMa}


The determination of age, metallicity and $A_V$ for each cluster allows us to determine the photometric mass of the star clusters. We used the relation:

\begin{equation}
\label{eq:mass}
M_{\text{F814W},0} = {\rm F814W}_{\text{SSP}}(t) - 2.5\log\left(\frac{Mass}{M_\odot}\right),
\end{equation}
where $M_{\text{F814W,0}}$ is the reddening-corrected absolute magnitude in the F814W filter (see subsection \ref{Sample}),
${\rm F814W}_{\text{SSP}}(t)$ is the F814W magnitude for a cluster of 1~M$_\odot$ total mass predicted by the SSP model \citetalias{Girardi2002} of age $t= \log$(Age). 
The values of $\log(Mass)$, along with all the derived parameters for all clusters, are summarized in Table~\ref{tab:final}.
The implications of these findings 
will be  discussed in the next section.

\begin{figure}
\centering
\includegraphics[width=7cm]{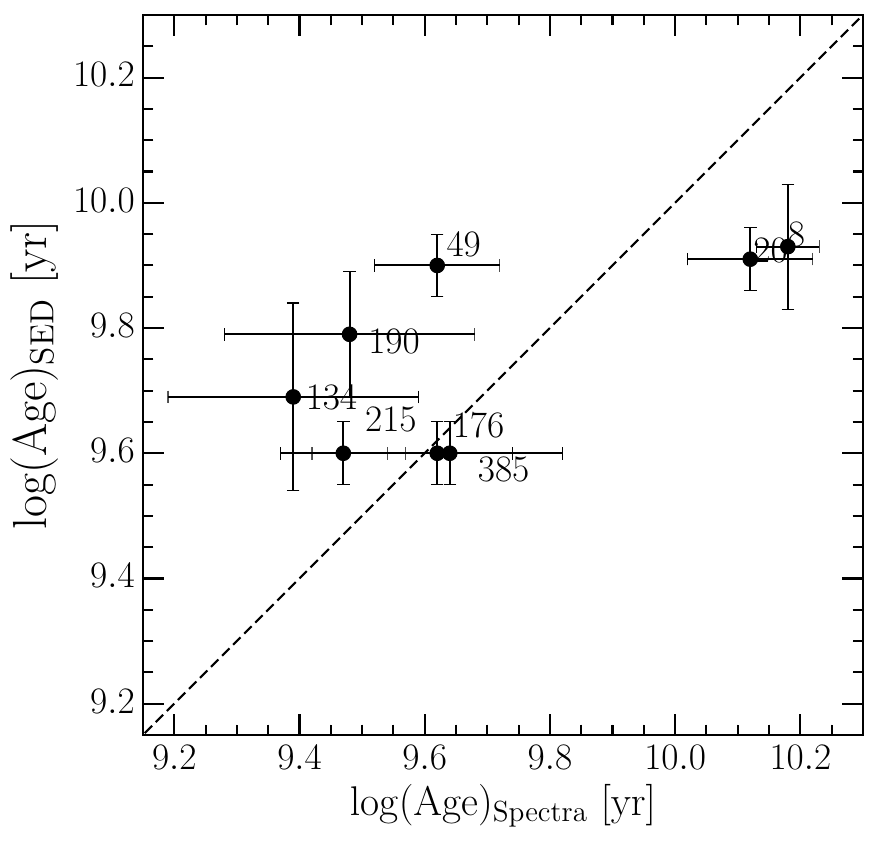}
    \caption{Comparison of $\log$(Age) obtained through two methods: spectra analysis (x-axis) and SED-fitting (y-axis). The dashed line represents a one-to-one correspondence.}
    \label{fig:Comp}
\end{figure}

\begin{figure}
    \centering
    \includegraphics[height=6.cm]{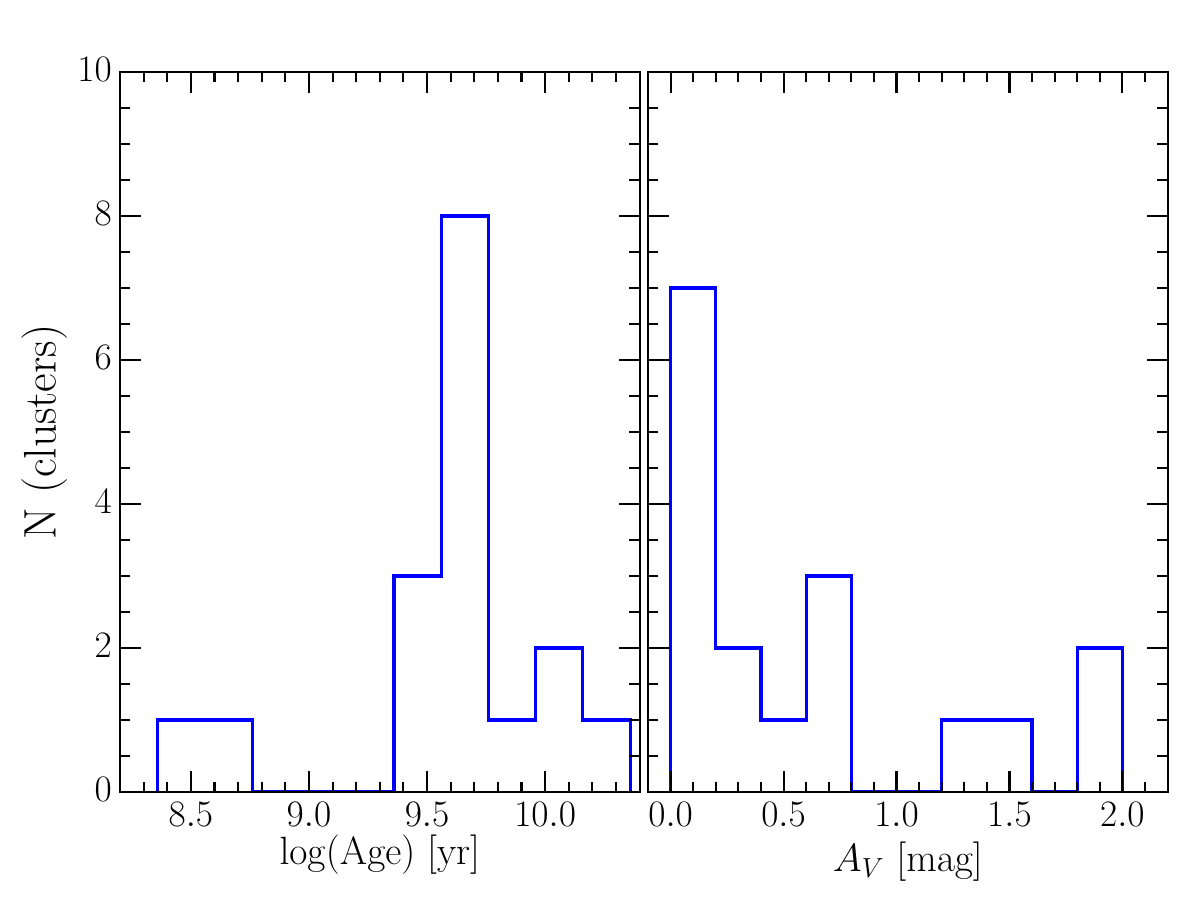}
    \caption{{ Histograms 
    for $\log$(Age) and $A_V$,  primarily derived through spectral analysis when available and through SED-fitting in cases where spectral analysis was not feasible 
    Distribution of $\log$(Age) exhibits an important population of IACs (age $\lesssim5$ Gyr) together with the classical GCs (Age $>$10 Gyr).}}
    \label{fig:HistAgeAvF}
\end{figure}

\section{Discussion}
\label{Discu}

\begin{figure}
    \centering
    \includegraphics[{trim= 1.1cm .1cm 0.5cm 0},clip,height=6.5cm]
    {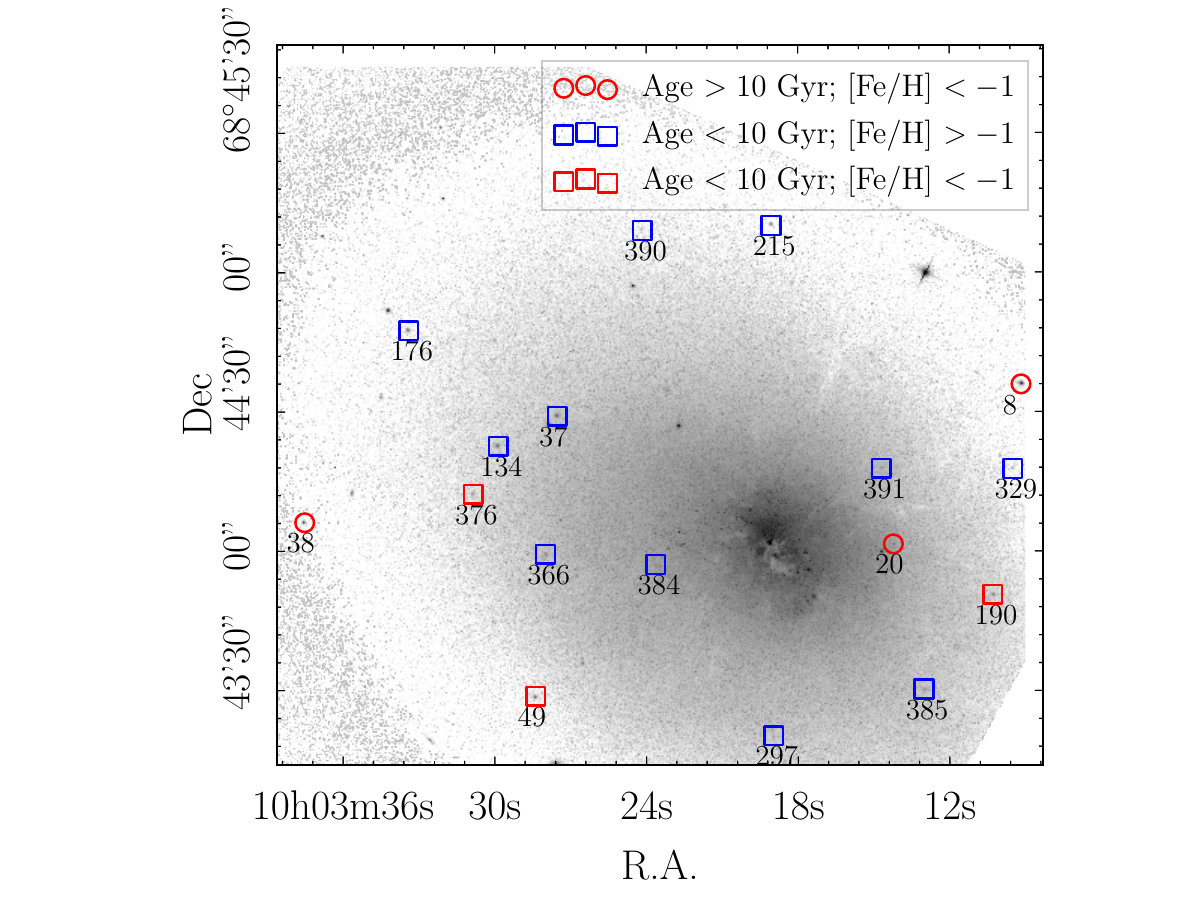}
    \caption{{ Spatial distribution of
    all clusters analyzed in this work, with the shapes and color of the symbols indicating their age and metallicities, respectively: circles 
    for classical GCs (age > 10 Gyr), 
    squares for IACs (age < 10 Gyr); red for 
    metal-poor ([Fe/H] < $-1$) and
    blue for metal-rich ([Fe/H] > $-1$). The three classical GCs are metal-poor, whereas the majority of the IACs are metal-rich.
    North is at the top and East is to the left.}}
    \label{fig:RADec}
\end{figure}

\begin{table}
\centering
    \caption{Derived parameters for clusters in NGC 3077.}
     \setlength{\tabcolsep}{1.5pt}
     \resizebox{\columnwidth}{!}
 {
    \begin{tabular}{rcccrrc}
    \hline
    
ID &$A_V$ &$M_{{\text F814W},0}$ &$M_{V,0}$ & [Fe/H] &$\log$(Age) & $\log(Mass)$ \\
(1)&   (2)&       (3) &      (4) &    (5) &        (6) &        (7)  \\
\hline
\hline

8	 & 	0.56$\pm$0.03 & $-9.68\pm0.02$& $-8.72$  &$-1.13\pm0.11$ &10.18$\pm$0.05  &$6.05^{+0.04}_{-0.04}$\\
20	 & 	0.12$\pm$0.12 & $-8.59\pm0.07$& $-7.63$  &$-1.06\pm0.14$ &10.12$\pm$0.10  &$5.62^{+0.04}_{-0.04}$\\
37	 & 	1.82$\pm$0.19 & $-9.49\pm0.11$& $-8.49$  &$-0.71\pm0.16$ & 9.63$\pm$0.10  &$5.64^{+0.03}_{-0.05}$\\
38	 & 	0.08$\pm$0.10 & $-8.47\pm0.06$& $-7.51$  &$-1.31 \null$  &10.05$\pm$0.10  &$5.53^{+0.04}_{-0.07}$\\
49	 & 	0.00$\pm$0.03 & $-8.19\pm0.02$& $-7.19$  &$-1.49\pm0.16$ & 9.62$\pm$0.10  &$5.12^{+0.04}_{-0.01}$\\
134	 & 	0.37$\pm$0.15 & $-8.25\pm0.09$& $-7.23$  &$-0.71 \null$  & 9.39$\pm$0.20  &$5.05^{+0.16}_{-0.15}$\\
176	 & 	0.00$\pm$0.03 & $-8.14\pm0.02$& $-7.15$  &$-0.71 \null$  & 9.62$\pm$0.20  &$5.10^{+0.17}_{-0.09}$\\
190	 & 	0.00$\pm$0.03 & $-7.96\pm0.03$& $-6.92$  &$-1.32\pm0.21$ & 9.48$\pm$0.20  &$5.03^{+0.06}_{-0.16}$\\
215	 & 	0.00$\pm$0.03 & $-8.02\pm0.02$& $-6.99$  &$-0.86\pm0.16$ & 9.47$\pm$0.10  &$5.05^{+0.01}_{-0.09}$\\
297	 & 	1.36$\pm$0.35 & $-7.86\pm0.21$& $-6.86$  &$-0.69\pm0.24$ & 9.60$\pm$0.05  &$4.99^{+0.05}_{-0.01}$\\
329	 & 	1.98$\pm$0.37 & $-8.05\pm0.22$& $-7.06$  &$-0.71 \null$  & 9.60$\pm$0.05  &$5.07^{+0.05}_{-0.01}$\\
366	 & 	0.72$\pm$0.25 & $-7.44\pm0.16$& $-7.01$  &$0.0   \null$  & 8.36$\pm$0.25  &$4.19^{+0.15}_{-0.21}$\\
376	 & 	0.78$\pm$0.27 & $-7.63\pm0.16$& $-6.67$  &$-1.31 \null$  & 9.80$\pm$0.05  &$4.98^{+0.07}_{-0.03}$\\
384	 & 	0.22$\pm$0.39 & $-6.75\pm0.24$& $-6.11$  &$-0.40 \null$  & 8.70$\pm$0.40  &$4.02^{+0.16}_{-0.27}$\\
385	 & 	0.62$\pm$0.03 & $-7.54\pm0.05$& $-6.55$  &$-0.85\pm0.12$ & 9.64$\pm$0.10  &$4.92^{+0.08}_{-0.06}$\\
390	 & 	1.55$\pm$0.36 & $-7.31\pm0.21$& $-6.31$  &$-0.71 \null$  & 9.60$\pm$0.05  &$4.77^{+0.05}_{-0.01}$\\
391	 & 	0.09$\pm$0.14 & $-6.59\pm0.09$& $-5.58$  &$-0.71 \null$  & 9.70$\pm$0.15  &$4.57^{+0.12}_{-0.08}$\\

\hline   
\end{tabular}
}
\footnotesize{\\Notes: (2) extinction of  cluster in [mag], (3) absolute magnitude in F814W band corrected by extinction using \citetalias{Cardelli1989} law, (4) magnitude in V band also extinction corrected, $M_{V,0}= M_{F814W,0} + (V-I)(t)_{\rm SSP}$, where $(V-I)(t)_{\rm SSP}$ is the color of \citetalias{Girardi2002} models, (5) metallicity in terms of [Fe/H] (see e.g. \href{http://miles.iac.es/pages/ssp-models.php} {\nolinkurl{miles.iac.es/pages/ssp-models}} for equivalence), (6) $\log$(Age) in [yr], (7) photometric $\log(Mass)$ in [$M_\odot$].}
\label{tab:final}
\end{table}

{ Study of the age and metallicity distribution of the surviving population of GCs in galaxies}
 gives useful clues to understand the build-up of galaxies. As outlined in the introduction, in a hierarchical scenario of galaxy
formation, metal-poor GCs are formed in the halos of low-mass galaxies \citep{Bekki2008}. NGC~3077 is a low-mass local galaxy that has avoided being  accreted, and hence it offers an opportunity to test the 
{ two-phase formation scenario that has been successful in explaining the observed properties of galaxies \citep{Oser2010}.}
In this section, we discuss 
{ the classical GCs and IAC populations in NGC~3077 in the context of hierarchical scenario of galaxy formation and evolution.}

{Though GCs have been reported previously in NGC~3077 (e.g. \citetalias{Davidge2004}), 
this is the first time that their ages and metallicities have been confirmed through spectroscopy.}
Additionally, { we discover an important  population}
of IACs with an age around 5 Gyr. 
{ In Figure~\ref{fig:RADec} all clusters
analyzed in this work are spatially 
displayed in the galaxy,
symbol shape is for age: circles 
    for classical GCs (age > 10 Gyr), 
    squares for IACs (age < 10 Gyr); while 
    color is for metallicity: red for 
    metal-poor ([Fe/H] < $-1$) and
    blue for metal-rich ([Fe/H] > $-1$).
Further details on these clusters will be discussed in below.}


\subsection{Properties of Classical GCs and IACs}
 \label{DGC}

 \begin{figure*}
\centering
\includegraphics[height=6.7cm]{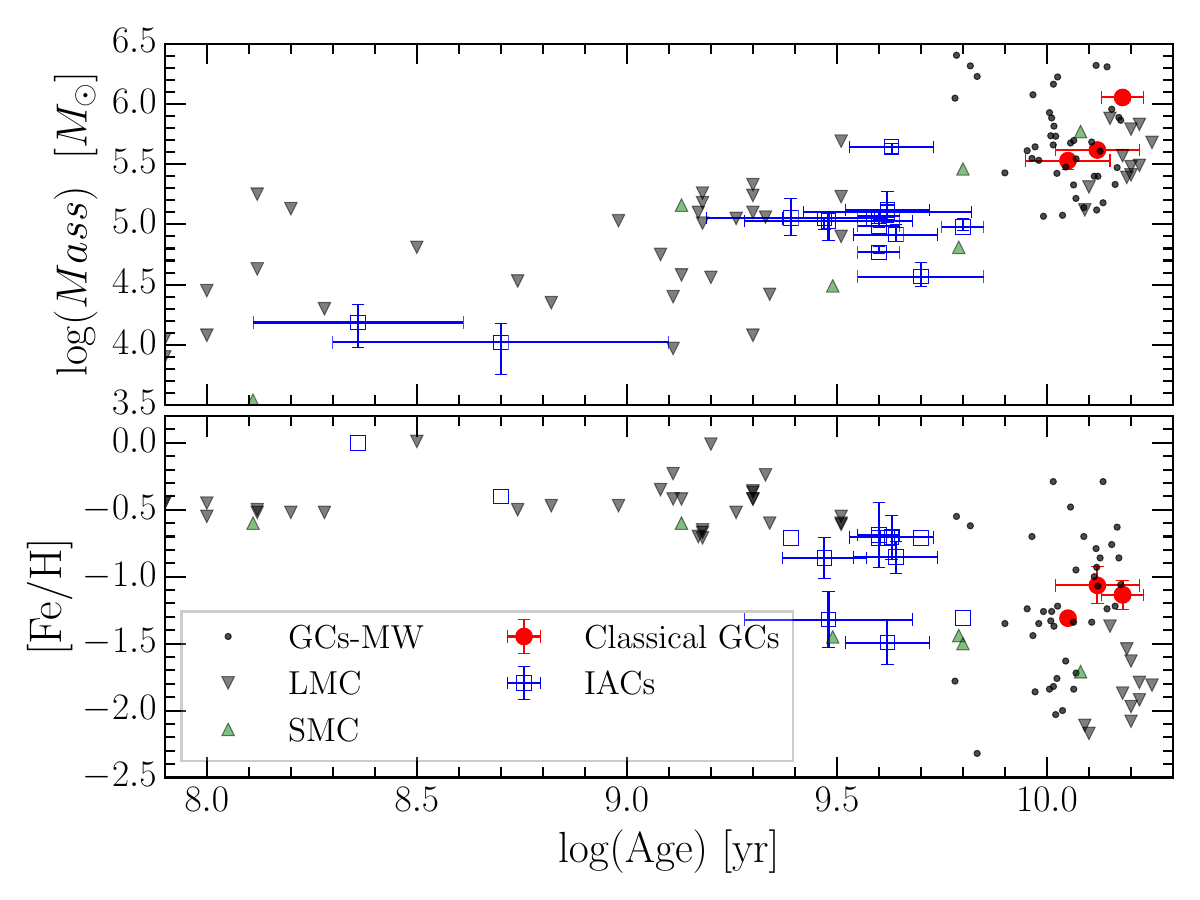}
\includegraphics[height=6.7cm]{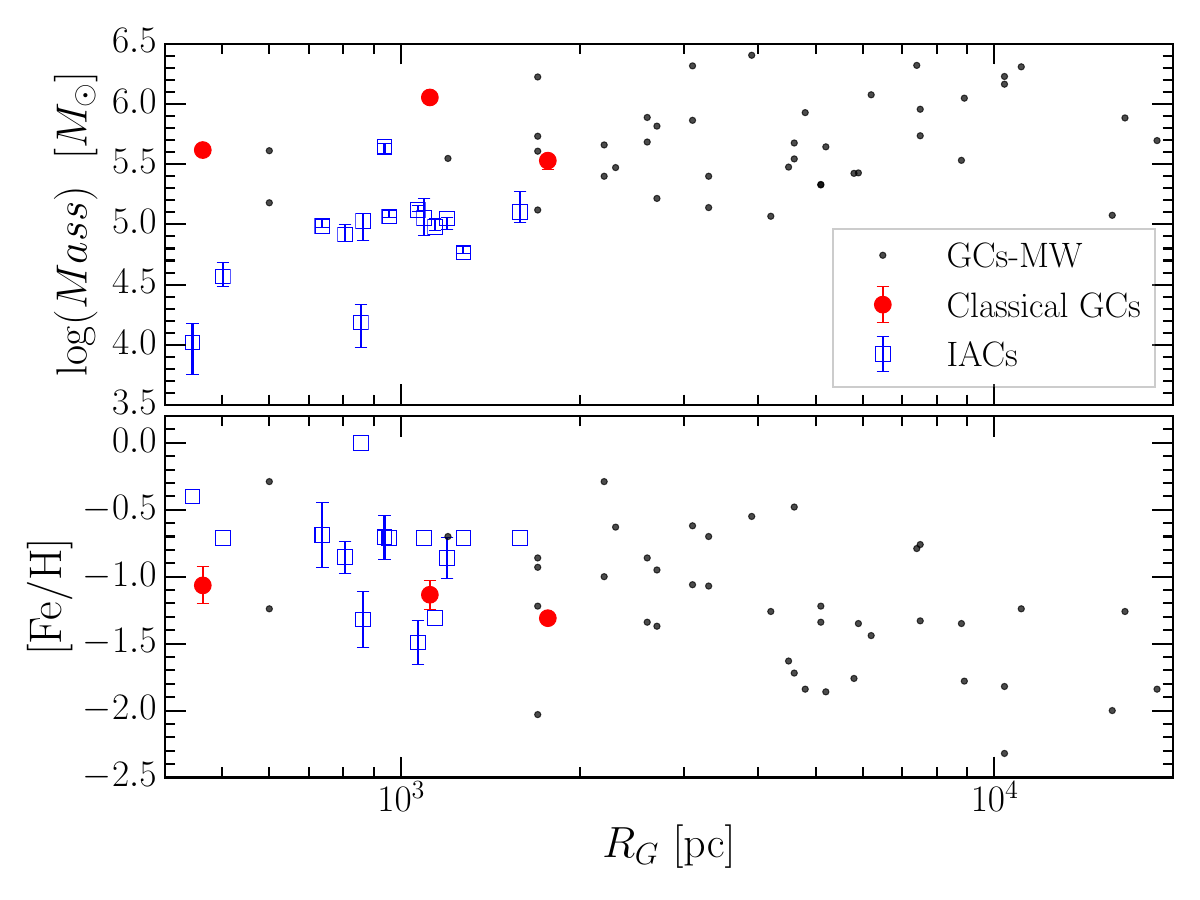}
    \caption{{ Diagrams for all clusters studied in this work, classical GCs 
    (red circles) and IACs (blue squares). 
({\it left}) $\log(Mass)$ and [Fe/H] are plotted against $\log$(Age), together with
Galactic GCs (black points; \citealt{Harris1996},
\citealt{Cezario2013}); LMC clusters
(inverted triangles; 
\citealt{Santos2004}; 
\citealt{McLaughlin2005};
\citealt{Pessev2006}; 
\citealt{Mucciarelli2010}; 
\citealt[and references therein]{Baumgardt2013}; 
\citealt{Ahumada2019})
and SMC (green triangles; 
\citealt{Pessev2006}; 
\citealt{Glatt2011}; 
\citealt{Gatto2021}). 
({\it right}) $\log(Mass)$ and [Fe/H] plotted against $R_G$, together with Galactic GCs (\citealt{Harris1996},
\citealt{Cezario2013}).
%
}}
    \label{fig:Mass_}
\end{figure*}

%

The three classical GCs in NGC~3077, namely IDs 8, 20 and 38, are old (age $>$ 10~Gyr), metal-poor ([Fe/H] $< -1.0$), and  massive (mass $> 10^5 \rm M_\odot$), characteristics similar to those found in GCs of the the Milky Way (MW, \citealt{Harris1996}). 
Their absolute magnitudes are on the brighter side of the turnover magnitude of $M_{V,0}=-7.4$~mag for GCs in the MW and spiral galaxies \citep{Lomeli2022}.
The remaining fourteen clusters exhibit intermediate ages, with the majority having ages between $\sim$2 and 6~Gyr, with a couple of them as young as $\sim$500 Myr. 
NGC~3077 is not alone in having more IACs than GCs. 
For example the red clusters of LMC are $\sim$3 Gyr old \citep{vandenBergh1991}. We here compare the properties of classical GCs and IACs in NGC~3077 to the properties of the population of old clusters in the MW and irregular galaxies.

In Figure~\ref{fig:Mass_}, we show the distribution of mass and metallicity against the age (left) and the 
galactocentric distance ($R_G$) of both the GC (red circles) and IAC (blue squares) populations studied in this work { together with the Galactic GCs 
(black points)
and clusters of LMC/SMC 
(inverted and green triangles, respectively).}

In the left panel, a clear correlation is observed between   mass and age. However, we should keep in mind that the plotted mass is the  photometric mass. For a fixed detection limit, the minimum mass that can be detected increases with age, which explains the absence of old low-mass clusters. So, the observed { trend} 
indicates that the most massive cluster formed in recent times is of lower mass as compared to those formed in the earliest epochs of galaxy formation. The right panel reveals that there is no galactocentric distance dependence of the mass and metallicity of the old clusters in NGC~3077.

The three classical GCs are situated at different { projected distances from the galaxy center} 
as depicted in the right panels of Figure~\ref{fig:Mass_}.
The most massive GC in this analysis, ID 8, is located in the northwest direction from the center of NGC~3077 toward M81. 
This GC has a photometric mass of $\sim 1.1\times 10^6 M_\odot$ which is comparable to the reported mass of $\omega$~Cen 
(M54, \citealt{Harris1996}),  
and to the most massive GC found in  irregular galaxies such as NGC 4449 \citep{Annibali2018}.


\begin{figure}
    \centering
    \includegraphics[height=6.cm]{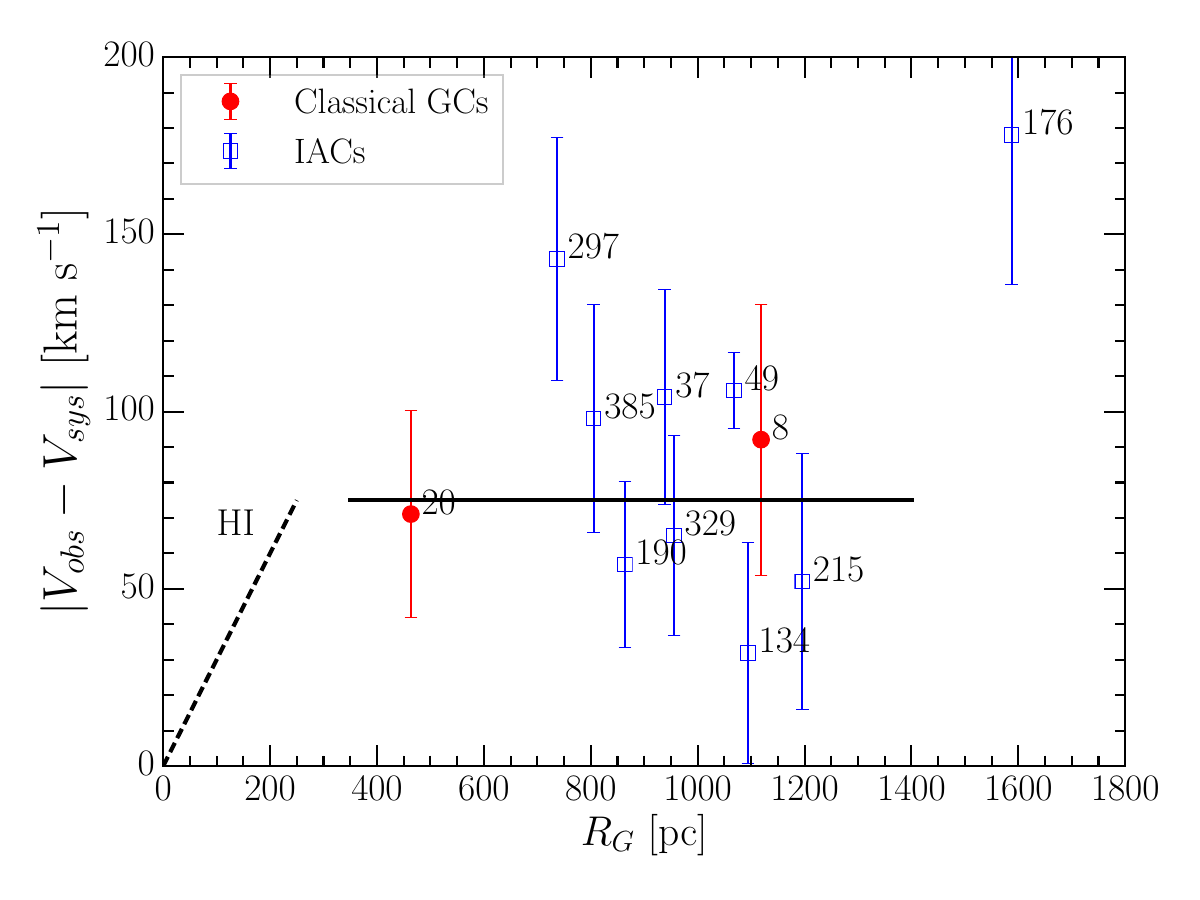}
    \caption{Rotation velocities of IACs (squares) and GCs 
    (red solid circles) against  
    $R_G$ with the average velocity shown by the solid line.
    An approximate HI rotation curve from 
    \citet{Walter2002} in the inner part is shown by a 
    dashed line. Except for IDs 297 and 176, the velocity 
    of the rest of IACs are consistent with a flat 
    rotation curve.}
    \label{fig:Vel_R}
\end{figure}

The availability of radial velocities of IACs allows us to address the question of whether they are in the disk or halo of NGC~3077. \citet{Walter2002} analyzed the HI, CO and H$\alpha$ kinematics of NGC~3077. They found the HI and CO gas within 70~km\,s$^{-1}$ of the systemic velocity. The H$\alpha$ velocity on the other hand extends to 100~km\,s$^{-1}$. It is likely that the H$\alpha$ velocities have a component from outflowing gas and hence serve as a maximum upper limit for the rotationally supported objects in the disk of NGC~3077. 
The measured radial velocities of IACs are well within the range of velocities expected for rotating systems in the disk of NGC~3077. 
In Figure~\ref{fig:Vel_R}, we plot the absolute value of the rotation velocities of IACs (squares) and GCs (red solid circles) after subtracting the systemic velocity as a function of galactocentric distance, both quantities without any corrections for the inclination. An approximate HI rotation curve from \cite{Walter2002} 
in the inner part is shown by a straight line.  Except two IACs (IDs 297 and 176), the rest have rotation velocities of 
70$\pm10$ km\,s$^{-1}$ similar to the HI velocities. This suggests that IACs are located in the disk of the galaxy at the flat part of the rotation curve. Two possible outliers are IDs 297 and 176. These two objects might not be formed {\it in situ}, instead could be accreting into NGC~3077. Interestingly, the two GCs also have radial velocities consistent with a flat rotation curve.

The IACs span a wide range of metallicities and show an age-metallicity relation with the two youngest clusters being the most metal-rich and three of the oldest IACs 
 (IDs 49, 190, and 376) with [Fe/H] $\sim-1.31$ are as metal-poor as the three GCs in NGC~3077. 
{ NGC\,3077 is not the first case of a metal-poor IAC; 
the metallicity of NGC 339, an IAC ($\sim$6~Gyr old) 
in the SMC is as metal-poor as its oldest cluster, NGC~121.} 
 The observed range of metallicities is in the range of 
 metallicities of metal-rich MW GCs. The median metallicity of [Fe/H]= $-0.71$ for our sample of IACs is similar to the mean metallicity of IACs of the central regions of the SMC \citep{Narloch2021} and of the disk arms and periphery of the LMC \citep{Narloch2022}.  
 Measured [$\alpha$/Fe] for our sample of GCs and IACs cover a wide range of values between $-${ 0.04} and 0.45, with no tendency for the oldest metal-poor clusters to have the highest $\alpha$-enhancement. In fact the two clusters IDs 385 and 215 with highest $\alpha$-enhancements are metal-rich and are of 3 to 4~Gyr age. Three LMC clusters (NGC 1786, NGC 2210, and NGC 2257)  have similarly high $\alpha$-enhancements, but they are old and metal-poor ([Fe/H]$\sim-$1.8; \citealt{Mucciarelli2010}). These metal-rich, $\alpha$-enhanced IACs are not expected in the standard chemical evolution models of galaxies (e.g. \citealt{Minchev2013}). Observations of more clusters would be required in order to postulate alternative scenarios of chemical evolution. 
 
 { In summary, all clusters older than 10~Gyr in NGC\,3077 are metal-poor suggesting that they are most-likely formed in metal-poor satellite galaxies and accreted later into NGC\,3077 in the second phase of galaxy formation \citep{Oser2010}. At the same time NGC\,3077 had a major cluster formation event as recently as 4~Gyr ago, that contains both metal-poor and metal-rich clusters. The cluster formation events are related to accretion of satellite galaxies, suggesting that the accretion phase lasted well into z $<$ 1 epochs. Such an extended accretion phase is consistent with the results of simulations by \citet{Oser2010}, who found accretion-driven {\it in situ} star formation extending all the way to z $<$ 1  in low-mass galaxies. The same models also predict a larger fraction of older clusters to be metal-rich in low-mass galaxies as compared to those in massive galaxies. However, NGC\,3077 lacks such metal-rich old clusters. Absence of metal-rich old clusters is a typical property of irregular galaxies such as the LMC and SMC (\citealt{Beasley2020}), which suggests that NGC\,3077 has a formation history similar to these dwarf galaxies rather than the low-mass galaxies modelled by \citet{Oser2010}.
 }

\subsection{On the nature of NGC~3077}
 \label{GCis}

NGC~3077 is a low-mass dusty star-forming galaxy morphologically classified as Irr II galaxy \citep{Sandage1961}, much like M82, the well-known starburst galaxy in the M81-group. Galaxies classified as Irr II do not follow the normal relations obeyed by the galaxies along the Hubble Sequence 
(e.g. \citealt{Sandage1975}). 
It is understood that the current appearance is as a result of recent interaction events. The galaxies 
M81, M82 and NGC~3077 had gone 
through a close interaction around 300~Myr ago, which has left behind streams of HI connecting these galaxies (\citealt{yun1999}). 
The interaction brought fresh gas into NGC~3077 leading to its current nuclear star formation activity 
(\citealt{Benacchio1981}).

The nature of NGC~3077 before its most recent interaction with members of M81 group is still unclear. \cite{Price1989} 
 suggested that it was most likely a dwarf elliptical (dE) galaxy, similar to M31's companion NGC~185. 
 \citetalias{Davidge2004} used the K-band radial surface brightness profiles, which are not greatly affected by the post-interaction star formation and hence trace the pre-interaction morphology, to test the hypothesis of dE origin. 
 He found the K-band radial surface brightness profiles is similar to that of NGC~205, another dE companion of M31.

The old stellar populations hold valuable insights into early nature of galaxies. \citet{Okamoto2023} found that the metallicity distribution function of resolved old stellar populations of NGC~3077 is similar to that of the dE 
NGC\,185 (\citealt{Crnojevic2014}) to suggest a dE origin for NGC~3077. The GC population offers an alternative way to address the nature of the pre-interaction galaxy. If NGC~3077 was a dE galaxy, it is expected to have a specific frequency of GCs, $S_{\rm N}$, of greater than 2 (\citealt{Harris1991}). This amounts to more than 10 GCs in NGC~3077. 

Our findings of only three classical GCs gives us a $S_{\rm N}$= 0.7, well below the values expected for dEs, but consistent with values reported by \citet{Harris1991} and \citet{Georgiev2008} for irregular galaxies with similar $M_V$ magnitude. 
\citetalias{Davidge2004} reported 12 GCs in NGC~3077 using ground-based NIR images, corresponding to $S_{\rm N}$= 2.5, 
{ a number close to that expected if it was a dE galaxy. However, our analysis of these 12 objects reported as GCs using the recently-available GAIA parameters (see Figure~\ref{fig:GAIA}) suggests only two objects, both  classified by us as classical GCs, are genuine GCs, with the rest 
}
foreground Galactic stars. The FWHM of these stars 
that fall inside the HST/ACS field of view (FoV) also satisfy our criteria as foreground stars at the HST resolution. Hence, the use of GAIA parameters would lower the $S_{\rm N}$ value from \citetalias{Davidge2004} study close to 0.7.

{ The $S_{\rm N}$-based arguments to infer intrinsic morphology of galaxies suffers from a caveat: each object}
in the study of \citet{Harris1991} was all not confirmed as a classical metal-poor old GC by detailed studies such as carried out here. { Hence, } 
if we relax the definition, most if not all of our 17 GC candidates would be classified as GCs, in which case the $S_{\rm N}$ for NGC~3077 could be in the range of values expected for dE systems. 
{ These numbers could be lower limits if some GCs were stripped during the interaction event that the members of M81 group had around 300~Myr ago.}
Massive SSCs that survive for gigayears are usually formed in intense star formation bursts, associated with past interaction and merging events. Hence, the presence of a rich population of 3–4 Gyr IACs in NGC~3077 suggests major events of star formation at these look back times. Study of star formation histories (SFH) of dE galaxies suggests that these galaxies formed bulk of their stars early in their lives, later on forming stars at low rates 
(\citealt{Weisz2014}; 
\citealt{Koleva2009}). 
{ The presence of a rich population of IACs suggests a SFH akin to irregular galaxies. }
Hence, if the pre-interaction  NGC~3077 was { indeed} a dE, the IACs wouldn't have been formed {\rm in situ} in NGC~3077, and instead acquired through accretion.

\section{Summary and Conclusions}
\label{Conc}


{ 
We here present the results of an elaborate search for GC candidates in the HST/ACS images of the nearby Irr II galaxy NGC 3077. New spectroscopic observations using the OSIRIS spectrograph at the GTC were carried out for a sub-sample of the selected candidates to obtain radial velocity, age, metallicity, extinction and photometric mass. Complementary data from SDSS and 2MASS missions, as well as proper motion and other parameters  from GAIA mission were compiled to reject stars and other non-cluster candidates to define a sample of 17 genuine GC-like objects. We find three of these objects have characteristics of classical metal-poor ([Fe/H] $<-$1.0) old (age $>$10~Gyr) GCs. Most of the remaining objects are of intermediate age, metal-rich ([Fe/H]$>-$1.0) clusters. The measured radial velocities of IACs are well within the range of velocities expected for rotating systems in the disk of NGC\,3077. The metallicities, ages and masses of GCs and IACs in NGC\,3077 compare well with those properties in the nearby dwarf galaxies, the LMC and SMC. 
}

The calculated value of $S_{\rm N}$ is  0.7, { which is similar to the values obtained in previous studies for irregular galaxies and suggest that the pre-interaction galaxy was actively forming stars and SSCs, and is unlikely to be dE as suggested in some of the previous studies.}

\section{Acknowledgments}

{ PAO thanks the Mexican CONAHCYT for the scholarship grant which enabled the realization of this work (doctoral fellowship 345527), and LLN thanks Funda\c{c}\~ao de Amparo \`a Pesquisa do Estado do Rio de Janeiro (FAPERJ) for granting the postdoctoral research fellowship E-40/2021(280692). We are grateful for an anonymous referee whose thoughtful suggestions helped to improve the presentation of the results obtained in this work.}

This study is based on observations carried out with the NASA/ESA Hubble Space
Telescope, which was accessed  
{ using the MAST interface available on Zenodo under an 
open-source 
Creative Commons Attribution license: 
\dataset[doi:10.17909/D7VY-5145]{https://doi.org/10.17909/d7vy-5145})}.
We also used Sloan Digital Sky Survey V data (SDSS), which was funded 
by the Alfred P. Sloan Foundation, the Heising-Simons Foundation, the National Science Foundation, and the Participating Institutions. SDSS acknowledges support and resources from the Center for High-Performance Computing at the University of Utah. SDSS telescopes are located at Apache Point Observatory, funded by the Astrophysical Research Consortium and operated by New Mexico State University, and at Las Campanas Observatory, operated by the Carnegie Institution for Science. The SDSS web site is \url{www.sdss.org}.
This publication makes use of data products from the Two Micron All Sky Survey, which is a joint project of the University of Massachusetts and the Infrared Processing and Analysis Center/California Institute of Technology, funded by the National Aeronautics and Space Administration and the National Science Foundation.
This work has also made use of data from the European Space Agency (ESA) mission
{\it Gaia} (\url{https://www.cosmos.esa.int/gaia}), processed by the {\it Gaia}
Data Processing and Analysis Consortium (DPAC,
\url{https://www.cosmos.esa.int/web/gaia/dpac/consortium}). Funding for the DPAC
has been provided by national institutions, in particular the institutions
participating in the {\it Gaia} Multilateral Agreement.

\bibliography{Ref}{}

\begin{thebibliography}{}
\expandafter\ifx\csname natexlab\endcsname\relax\def\natexlab#1{#1}\fi
\providecommand{\url}[1]{\href{#1}{#1}}
\providecommand{\dodoi}[1]{doi:~\href{http://doi.org/#1}{\nolinkurl{#1}}}
\providecommand{\doeprint}[1]{\href{http://ascl.net/#1}{\nolinkurl{http://ascl.net/#1}}}
\providecommand{\doarXiv}[1]{\href{https://arxiv.org/abs/#1}{\nolinkurl{https://arxiv.org/abs/#1}}}

\bibitem[{{Abazajian} {et~al.}(2003){Abazajian}, {Adelman-McCarthy}, {Ag{\"u}eros}, {Allam}, {Anderson}, {Annis}, {Bahcall}, {Baldry}, {Bastian}, {Berlind}, {Bernardi}, {Blanton}, {Blythe}, {Bochanski}, {Boroski}, {Brewington}, {Briggs}, {Brinkmann}, {Brunner}, {Budav{\'a}ri}, {Carey}, {Carr}, {Castander}, {Chiu}, {Collinge}, {Connolly}, {Covey}, {Csabai}, {Dalcanton}, {Dodelson}, {Doi}, {Dong}, {Eisenstein}, {Evans}, {Fan}, {Feldman}, {Finkbeiner}, {Friedman}, {Frieman}, {Fukugita}, {Gal}, {Gillespie}, {Glazebrook}, {Gonzalez}, {Gray}, {Grebel}, {Grodnicki}, {Gunn}, {Gurbani}, {Hall}, {Hao}, {Harbeck}, {Harris}, {Harris}, {Harvanek}, {Hawley}, {Heckman}, {Helmboldt}, {Hendry}, {Hennessy}, {Hindsley}, {Hogg}, {Holmgren}, {Holtzman}, {Homer}, {Hui}, {Ichikawa}, {Ichikawa}, {Inkmann}, {Ivezi{\'c}}, {Jester}, {Johnston}, {Jordan}, {Jordan}, {Jorgensen}, {Juri{\'c}}, {Kauffmann}, {Kent}, {Kleinman}, {Knapp}, {Kniazev}, {Kron}, {Krzesi{\'n}ski}, {Kunszt}, {Kuropatkin}, {Lamb}, {Lampeitl}, {Laubscher}, {Lee},
  {Leger}, {Li}, {Lidz}, {Lin}, {Loh}, {Long}, {Loveday}, {Lupton}, {Malik}, {Margon}, {McGehee}, {McKay}, {Meiksin}, {Miknaitis}, {Moorthy}, {Munn}, {Murphy}, {Nakajima}, {Narayanan}, {Nash}, {Neilsen}, {Newberg}, {Newman}, {Nichol}, {Nicinski}, {Nieto-Santisteban}, {Nitta}, {Odenkirchen}, {Okamura}, {Ostriker}, {Owen}, {Padmanabhan}, {Peoples}, {Pier}, {Pindor}, {Pope}, {Quinn}, {Rafikov}, {Raymond}, {Richards}, {Richmond}, {Rix}, {Rockosi}, {Schaye}, {Schlegel}, {Schneider}, {Schroeder}, {Scranton}, {Sekiguchi}, {Seljak}, {Sergey}, {Sesar}, {Sheldon}, {Shimasaku}, {Siegmund}, {Silvestri}, {Sinisgalli}, {Sirko}, {Smith}, {Smol{\v c}i{\'c}}, {Snedden}, {Stebbins}, {Steinhardt}, {Stinson}, {Stoughton}, {Strateva}, {Strauss}, {SubbaRao}, {Szalay}, {Szapudi}, {Szkody}, {Tasca}, {Tegmark}, {Thakar}, {Tremonti}, {Tucker}, {Uomoto}, {Vanden Berk}, {Vandenberg}, {Vogeley}, {Voges}, {Vogt}, {Walkowicz}, {Weinberg}, {West}, {White}, {Wilhite}, {Willman}, {Xu}, {Yanny}, {Yarger}, {Yasuda}, {Yip}, {Yocum}, {York},
  {Zakamska}, {Zehavi}, {Zheng}, {Zibetti}, \& {Zucker}}]{Abazajian2003}
{Abazajian}, K., {Adelman-McCarthy}, J.~K., {Ag{\"u}eros}, M.~A., {et~al.} 2003, \aj, 126, 2081, \dodoi{10.1086/378165}

\bibitem[{{Adamo} {et~al.}(2024){Adamo}, {Bradley}, {Vanzella}, {Claeyssens}, {Welch}, {Diego}, {Mahler}, {Oguri}, {Sharon}, {Abdurro'uf}, {Hsiao}, {Xu}, {Messa}, {Lassen}, {Zackrisson}, {Brammer}, {Coe}, {Kokorev}, {Ricotti}, {Zitrin}, {Fujimoto}, {Inoue}, {Resseguier}, {Rigby}, {Jim{\'e}nez-Teja}, {Windhorst}, {Hashimoto}, \& {Tamura}}]{Adamo2024}
{Adamo}, A., {Bradley}, L.~D., {Vanzella}, E., {et~al.} 2024, arXiv e-prints, arXiv:2401.03224, \dodoi{10.48550/arXiv.2401.03224}

\bibitem[{{Ahumada} {et~al.}(2019){Ahumada}, {Vega-Neme}, {Clari{\'a}}, \& {Minniti}}]{Ahumada2019}
{Ahumada}, A.~V., {Vega-Neme}, L.~R., {Clari{\'a}}, J.~J., \& {Minniti}, J.~H. 2019, \pasp, 131, 024101, \dodoi{10.1088/1538-3873/aae660}

\bibitem[{{Alamo-Mart{\'\i}nez} {et~al.}(2021){Alamo-Mart{\'\i}nez}, {Chies-Santos}, {Beasley}, {Flores-Freitas}, {Furlanetto}, {Trevisan}, {Schnorr-M{\"u}ller}, {Leaman}, \& {Bonatto}}]{Alamo2021}
{Alamo-Mart{\'\i}nez}, K.~A., {Chies-Santos}, A.~L., {Beasley}, M.~A., {et~al.} 2021, \mnras, 503, 2406, \dodoi{10.1093/mnras/stab538}

\bibitem[{{Annibali} {et~al.}(2018){Annibali}, {Morandi}, {Watkins}, {Tosi}, {Aloisi}, {Buzzoni}, {Cusano}, {Fumana}, {Marchetti}, {Mignoli}, {Mucciarelli}, {Romano}, \& {van der Marel}}]{Annibali2018}
{Annibali}, F., {Morandi}, E., {Watkins}, L.~L., {et~al.} 2018, \mnras, 476, 1942, \dodoi{10.1093/mnras/sty344}

\bibitem[{{Asa'd}(2014)}]{Asad2014}
{Asa'd}, R.~S. 2014, \mnras, 445, 1679, \dodoi{10.1093/mnras/stu1874}

\bibitem[{{Ashman} \& {Zepf}(1998)}]{Ashman1998}
{Ashman}, K.~M., \& {Zepf}, S.~E. 1998, Globular Cluster Systems (Cambridge University Press)

\bibitem[{{Babusiaux} {et~al.}(2023){Babusiaux}, {Fabricius}, {Khanna}, {Muraveva}, {Reyl{\'e}}, {Spoto}, {Vallenari}, {Luri}, {Arenou}, {{\'A}lvarez}, {Anders}, {Antoja}, {Balbinot}, {Barache}, {Bauchet}, {Bossini}, {Busonero}, {Cantat-Gaudin}, {Carrasco}, {Dafonte}, {Diakit{\'e}}, {Figueras}, {Garcia-Gutierrez}, {Garofalo}, {Helmi}, {Jim{\'e}nez-Arranz}, {Jordi}, {Kervella}, {Kostrzewa-Rutkowska}, {Leclerc}, {Licata}, {Manteiga}, {Masip}, {Mongui{\'o}}, {Ramos}, {Robichon}, {Robin}, {Romero-G{\'o}mez}, {S{\'a}ez}, {Santove{\~n}a}, {Spina}, {Torralba Elipe}, \& {Weiler}}]{GAIA2022}
{Babusiaux}, C., {Fabricius}, C., {Khanna}, S., {et~al.} 2023, \aap, 674, A32, \dodoi{10.1051/0004-6361/202243790}

\bibitem[{{Bastian} {et~al.}(2016){Bastian}, {Niederhofer}, {Kozhurina-Platais}, {Salaris}, {Larsen}, {Cabrera-Ziri}, {Cordero}, {Ekstr{\"o}m}, {Geisler}, {Georgy}, {Hilker}, {Kacharov}, {Li}, {Mackey}, {Mucciarelli}, \& {Platais}}]{Bastian2016}
{Bastian}, N., {Niederhofer}, F., {Kozhurina-Platais}, V., {et~al.} 2016, \mnras, 460, L20, \dodoi{10.1093/mnrasl/slw067}

\bibitem[{{Baumgardt} {et~al.}(2013){Baumgardt}, {Parmentier}, {Anders}, \& {Grebel}}]{Baumgardt2013}
{Baumgardt}, H., {Parmentier}, G., {Anders}, P., \& {Grebel}, E.~K. 2013, \mnras, 430, 676, \dodoi{10.1093/mnras/sts667}

\bibitem[{{Beasley}(2020)}]{Beasley2020}
{Beasley}, M.~A. 2020, in Reviews in Frontiers of Modern Astrophysics; From Space Debris to Cosmology, ed. P.~{Kab{\'a}th}, D.~{Jones}, \& M.~{Skarka}, 245--277, \dodoi{10.1007/978-3-030-38509-5_9}

\bibitem[{{Beasley} {et~al.}(2018){Beasley}, {Trujillo}, {Leaman}, \& {Montes}}]{Beasley2018}
{Beasley}, M.~A., {Trujillo}, I., {Leaman}, R., \& {Montes}, M. 2018, \nat, 555, 483, \dodoi{10.1038/nature25756}

\bibitem[{{Bekki} {et~al.}(2008){Bekki}, {Yahagi}, {Nagashima}, \& {Forbes}}]{Bekki2008}
{Bekki}, K., {Yahagi}, H., {Nagashima}, M., \& {Forbes}, D.~A. 2008, \mnras, 387, 1131, \dodoi{10.1111/j.1365-2966.2008.13318.x}

\bibitem[{{Benacchio} \& {Galletta}(1981)}]{Benacchio1981}
{Benacchio}, L., \& {Galletta}, G. 1981, \apjl, 243, L65, \dodoi{10.1086/183444}

\bibitem[{{Bertin} \& {Arnouts}(1996)}]{Bertin1996}
{Bertin}, E., \& {Arnouts}, S. 1996, \aaps, 117, 393

\bibitem[{{Blakeslee} {et~al.}(2012){Blakeslee}, {Cho}, {Peng}, {Ferrarese}, {Jord{\'a}n}, \& {Martel}}]{Blakeslee2012}
{Blakeslee}, J.~P., {Cho}, H., {Peng}, E.~W., {et~al.} 2012, \apj, 746, 88, \dodoi{10.1088/0004-637X/746/1/88}

\bibitem[{{Bressan} {et~al.}(2012){Bressan}, {Marigo}, {Girardi}, {Salasnich}, {Dal Cero}, {Rubele}, \& {Nanni}}]{Bressan2012}
{Bressan}, A., {Marigo}, P., {Girardi}, L., {et~al.} 2012, \mnras, 427, 127, \dodoi{10.1111/j.1365-2966.2012.21948.x}

\bibitem[{{Brodie} \& {Huchra}(1990)}]{Brodie1990}
{Brodie}, J.~P., \& {Huchra}, J.~P. 1990, \apj, 362, 503, \dodoi{10.1086/169288}

\bibitem[{{Brodie} \& {Strader}(2006)}]{Brodie2006}
{Brodie}, J.~P., \& {Strader}, J. 2006, \araa, 44, 193, \dodoi{10.1146/annurev.astro.44.051905.092441}

\bibitem[{{Burstein} {et~al.}(1984){Burstein}, {Faber}, {Gaskell}, \& {Krumm}}]{Burstein84}
{Burstein}, D., {Faber}, S.~M., {Gaskell}, C.~M., \& {Krumm}, N. 1984, \apj, 287, 586, \dodoi{10.1086/162718}

\bibitem[{{Caldwell} \& {Romanowsky}(2016)}]{Caldwell2016}
{Caldwell}, N., \& {Romanowsky}, A.~J. 2016, \apj, 824, 42, \dodoi{10.3847/0004-637X/824/1/42}

\bibitem[{{Calzetti} {et~al.}(2000){Calzetti}, {Armus}, {Bohlin}, {Kinney}, {Koornneef}, \& {Storchi-Bergmann}}]{Calzetti00}
{Calzetti}, D., {Armus}, L., {Bohlin}, R.~C., {et~al.} 2000, \apj, 533, 682, \dodoi{10.1086/308692}

\bibitem[{{Calzetti} {et~al.}(2004){Calzetti}, {Harris}, {Gallagher}, {Smith}, {Conselice}, {Homeier}, \& {Kewley}}]{Calzetti2004}
{Calzetti}, D., {Harris}, J., {Gallagher}, III, J.~S., {et~al.} 2004, \aj, 127, 1405, \dodoi{10.1086/382095}

\bibitem[{{Cardelli} {et~al.}(1989){Cardelli}, {Clayton}, \& {Mathis}}]{Cardelli1989}
{Cardelli}, J.~A., {Clayton}, G.~C., \& {Mathis}, J.~S. 1989, \apj, 345, 245, \dodoi{10.1086/167900}

\bibitem[{{Cezario} {et~al.}(2013){Cezario}, {Coelho}, {Alves-Brito}, {Forbes}, \& {Brodie}}]{Cezario2013}
{Cezario}, E., {Coelho}, P.~R.~T., {Alves-Brito}, A., {Forbes}, D.~A., \& {Brodie}, J.~P. 2013, \aap, 549, A60, \dodoi{10.1051/0004-6361/201220336}

\bibitem[{{Chies-Santos} {et~al.}(2022){Chies-Santos}, {de Souza}, {Caso}, {Ennis}, {de Souza}, {Barbosa}, {Chen}, {Javier Cenarro}, {Ederoclite}, {Crist{\'o}bal-Hornillos}, {Hern{\'a}ndez-Monteagudo}, {L{\'o}pez-Sanjuan}, {Mar{\'\i}n-Franch}, {Moles}, {Varela}, {V{\'a}zquez Rami{\'o}}, {Dupke}, {Sodr{\'e}}, \& {Angulo}}]{Chies2022}
{Chies-Santos}, A.~L., {de Souza}, R.~S., {Caso}, J.~P., {et~al.} 2022, \mnras, 516, 1320, \dodoi{10.1093/mnras/stac2002}

\bibitem[{{Crnojevi{\'c}} {et~al.}(2014){Crnojevi{\'c}}, {Ferguson}, {Irwin}, {McConnachie}, {Bernard}, {Fardal}, {Ibata}, {Lewis}, {Martin}, {Navarro}, {No{\"e}l}, \& {Pasetto}}]{Crnojevic2014}
{Crnojevi{\'c}}, D., {Ferguson}, A.~M.~N., {Irwin}, M.~J., {et~al.} 2014, \mnras, 445, 3862, \dodoi{10.1093/mnras/stu2003}

\bibitem[{{Davidge}(2004)}]{Davidge2004}
{Davidge}, T.~J. 2004, \aj, 127, 1460, \dodoi{10.1086/382096}

\bibitem[{{de Vaucouleurs}(1959)}]{Vaucouleurs1959}
{de Vaucouleurs}, G. 1959, Handbuch der Physik, 53, 275

\bibitem[{{de Vaucouleurs} {et~al.}(1991){de Vaucouleurs}, {de Vaucouleurs}, {Corwin}, {Buta}, {Paturel}, \& {Fouqu{\'e}}}]{deVaucouleurs1991}
{de Vaucouleurs}, G., {de Vaucouleurs}, A., {Corwin}, Jr., H.~G., {et~al.} 1991, {Third Reference Catalogue of Bright Galaxies. Volume I: Explanations and references. Volume II: Data for galaxies between 0$^{h}$ and 12$^{h}$. Volume III: Data for galaxies between 12$^{h}$ and 24$^{h}$.}

\bibitem[{{Forbes} {et~al.}(2011){Forbes}, {Spitler}, {Strader}, {Romanowsky}, {Brodie}, \& {Foster}}]{Forbes2011}
{Forbes}, D.~A., {Spitler}, L.~R., {Strader}, J., {et~al.} 2011, \mnras, 413, 2943, \dodoi{10.1111/j.1365-2966.2011.18373.x}

\bibitem[{{Freedman} {et~al.}(1994){Freedman}, {Hughes}, {Madore}, {Mould}, {Lee}, {Stetson}, {Kennicutt}, {Turner}, {Ferrarese}, {Ford}, {Graham}, {Hill}, {Hoessel}, {Huchra}, \& {Illingworth}}]{Freedman1994}
{Freedman}, W.~L., {Hughes}, S.~M., {Madore}, B.~F., {et~al.} 1994, \apj, 427, 628, \dodoi{10.1086/174172}

\bibitem[{{Gaia Collaboration} {et~al.}(2016){Gaia Collaboration}, {Prusti}, {de Bruijne}, {Brown}, {Vallenari}, {Babusiaux}, {Bailer-Jones}, {Bastian}, {Biermann}, {Evans}, \& et~al.}]{GAIA2016}
{Gaia Collaboration}, {Prusti}, T., {de Bruijne}, J.~H.~J., {et~al.} 2016, \aap, 595, A1, \dodoi{10.1051/0004-6361/201629272}

\bibitem[{{Gatto} {et~al.}(2021){Gatto}, {Ripepi}, {Bellazzini}, {Tosi}, {Cignoni}, {Tortora}, {Leccia}, {Clementini}, {Grebel}, {Longo}, {Marconi}, \& {Musella}}]{Gatto2021}
{Gatto}, M., {Ripepi}, V., {Bellazzini}, M., {et~al.} 2021, \mnras, 507, 3312, \dodoi{10.1093/mnras/stab2297}

\bibitem[{{Georgiev} {et~al.}(2008){Georgiev}, {Goudfrooij}, {Puzia}, \& {Hilker}}]{Georgiev2008}
{Georgiev}, I.~Y., {Goudfrooij}, P., {Puzia}, T.~H., \& {Hilker}, M. 2008, \aj, 135, 1858, \dodoi{10.1088/0004-6256/135/5/1858}

\bibitem[{{Girardi} {et~al.}(2002){Girardi}, {Bertelli}, {Bressan}, {Chiosi}, {Groenewegen}, {Marigo}, {Salasnich}, \& {Weiss}}]{Girardi2002}
{Girardi}, L., {Bertelli}, G., {Bressan}, A., {et~al.} 2002, \aap, 391, 195, \dodoi{10.1051/0004-6361:20020612}

\bibitem[{{Girardi} {et~al.}(2000){Girardi}, {Bressan}, {Bertelli}, \& {Chiosi}}]{Girardi2000}
{Girardi}, L., {Bressan}, A., {Bertelli}, G., \& {Chiosi}, C. 2000, \aaps, 141, 371, \dodoi{10.1051/aas:2000126}

\bibitem[{{Glatt} {et~al.}(2011){Glatt}, {Grebel}, {Jordi}, {Gallagher}, {Da Costa}, {Clementini}, {Tosi}, {Harbeck}, {Nota}, {Sabbi}, \& {Sirianni}}]{Glatt2011}
{Glatt}, K., {Grebel}, E.~K., {Jordi}, K., {et~al.} 2011, \aj, 142, 36, \dodoi{10.1088/0004-6256/142/2/36}

\bibitem[{{G{\'o}mez-Gonz{\'a}lez} {et~al.}(2016){G{\'o}mez-Gonz{\'a}lez}, {Mayya}, \& {Rosa-Gonz{\'a}lez}}]{Gomez2016}
{G{\'o}mez-Gonz{\'a}lez}, V.~M.~A., {Mayya}, Y.~D., \& {Rosa-Gonz{\'a}lez}, D. 2016, \mnras, 460, 1555, \dodoi{10.1093/mnras/stw1118}

\bibitem[{{Gonz{\'a}lez-L{\'o}pezlira} {et~al.}(2017){Gonz{\'a}lez-L{\'o}pezlira}, {Lomel{\'\i}-N{\'u}{\~n}ez}, {{\'A}lamo-Mart{\'\i}nez}, {{\'O}rdenes-Brice{\~n}o}, {Loinard}, {Georgiev}, {Mu{\~n}oz}, {Puzia}, {Bruzual A.}, \& {Gwyn}}]{Gonzalez2017}
{Gonz{\'a}lez-L{\'o}pezlira}, R.~A., {Lomel{\'\i}-N{\'u}{\~n}ez}, L., {{\'A}lamo-Mart{\'\i}nez}, K., {et~al.} 2017, \apj, 835, 184, \dodoi{10.3847/1538-4357/835/2/184}

\bibitem[{{Harris} {et~al.}(2004){Harris}, {Calzetti}, {Gallagher}, {Smith}, \& {Conselice}}]{Harris2004}
{Harris}, J., {Calzetti}, D., {Gallagher}, III, J.~S., {Smith}, D.~A., \& {Conselice}, C.~J. 2004, \apj, 603, 503, \dodoi{10.1086/381669}

\bibitem[{{Harris}(1991)}]{Harris1991}
{Harris}, W.~E. 1991, \araa, 29, 543, \dodoi{10.1146/annurev.aa.29.090191.002551}

\bibitem[{{Harris}(1996)}]{Harris1996}
---. 1996, \aj, 112, 1487, \dodoi{10.1086/118116}

\bibitem[{{Harris} \& {van den Bergh}(1981)}]{Harris1981}
{Harris}, W.~E., \& {van den Bergh}, S. 1981, \aj, 86, 1627, \dodoi{10.1086/113047}

\bibitem[{{Hodge} {et~al.}(1999){Hodge}, {Dolphin}, {Smith}, \& {Mateo}}]{Hodge1999}
{Hodge}, P.~W., {Dolphin}, A.~E., {Smith}, T.~R., \& {Mateo}, M. 1999, \apj, 521, 577, \dodoi{10.1086/307595}

\bibitem[{{Holmberg}(1950)}]{Holmberg1950}
{Holmberg}, E. 1950, Meddelanden fran Lunds Astronomiska Observatorium Serie II, 128, 1

\bibitem[{{Hughes} {et~al.}(2021){Hughes}, {Sand}, {Seth}, {Strader}, {Voggel}, {Dumont}, {Crnojevi{\'c}}, {Caldwell}, {Forbes}, {Simon}, {Guhathakurta}, \& {Toloba}}]{Hughes2021}
{Hughes}, A.~K., {Sand}, D.~J., {Seth}, A., {et~al.} 2021, \apj, 914, 16, \dodoi{10.3847/1538-4357/abf63c}

\bibitem[{{Kauffmann} \& {White}(1993)}]{Kauffmann1993}
{Kauffmann}, G., \& {White}, S.~D.~M. 1993, \mnras, 261, 921, \dodoi{10.1093/mnras/261.4.921}

\bibitem[{{Koleva} {et~al.}(2009){Koleva}, {de Rijcke}, {Prugniel}, {Zeilinger}, \& {Michielsen}}]{Koleva2009}
{Koleva}, M., {de Rijcke}, S., {Prugniel}, P., {Zeilinger}, W.~W., \& {Michielsen}, D. 2009, \mnras, 396, 2133, \dodoi{10.1111/j.1365-2966.2009.14820.x}

\bibitem[{{Krienke} \& {Hodge}(1974)}]{Krienke1974}
{Krienke}, Jr., O.~K., \& {Hodge}, P.~W. 1974, \aj, 79, 1242, \dodoi{10.1086/111666}

\bibitem[{{Kroupa}(2001)}]{Kroupa2001}
{Kroupa}, P. 2001, \mnras, 322, 231, \dodoi{10.1046/j.1365-8711.2001.04022.x}

\bibitem[{{Kruijssen} {et~al.}(2019){Kruijssen}, {Pfeffer}, {Crain}, \& {Bastian}}]{Kruijssen2019}
{Kruijssen}, J.~M.~D., {Pfeffer}, J.~L., {Crain}, R.~A., \& {Bastian}, N. 2019, \mnras, 486, 3134, \dodoi{10.1093/mnras/stz968}

\bibitem[{{Larsen} {et~al.}(2001){Larsen}, {Brodie}, {Forbes}, \& {Beasley}}]{Larsen2001}
{Larsen}, S.~S., {Brodie}, J.~P., {Forbes}, D.~A., \& {Beasley}, M.~A. 2001, in American Astronomical Society Meeting Abstracts, Vol. 199, American Astronomical Society Meeting Abstracts, 52.04

\bibitem[{{Lomel{\'\i}-N{\'u}{\~n}ez} {et~al.}(2022){Lomel{\'\i}-N{\'u}{\~n}ez}, {Mayya}, {Rodr{\'\i}guez-Merino}, {Ovando}, \& {Rosa-Gonz{\'a}lez}}]{Lomeli2022}
{Lomel{\'\i}-N{\'u}{\~n}ez}, L., {Mayya}, Y.~D., {Rodr{\'\i}guez-Merino}, L.~H., {Ovando}, P.~A., \& {Rosa-Gonz{\'a}lez}, D. 2022, \mnras, 509, 180, \dodoi{10.1093/mnras/stab2890}

\bibitem[{{Lomel{\'\i}-N{\'u}{\~n}ez} {et~al.}(2024){Lomel{\'\i}-N{\'u}{\~n}ez}, {Mayya}, {Rodr{\'\i}guez-Merino}, {Ovando}, {Alzate}, {Rosa-Gonz{\'a}lez}, {Cuevas-Otahola}, {Bruzual}, {Cortesi}, {G{\'o}mez-Gonz{\'a}lez}, \& {Escudero}}]{Luis2024}
{Lomel{\'\i}-N{\'u}{\~n}ez}, L., {Mayya}, Y.~D., {Rodr{\'\i}guez-Merino}, L.~H., {et~al.} 2024, \mnras, 528, 1445, \dodoi{10.1093/mnras/stae051}

\bibitem[{{Mayya} {et~al.}(2006){Mayya}, {Bressan}, {Carrasco}, \& {Hernandez-Martinez}}]{Mayya2006}
{Mayya}, Y.~D., {Bressan}, A., {Carrasco}, L., \& {Hernandez-Martinez}, L. 2006, \apj, 649, 172, \dodoi{10.1086/506270}

\bibitem[{{Mayya} {et~al.}(2008){Mayya}, {Romano}, {Rodr{\'{\i}}guez-Merino}, {Luna}, {Carrasco}, \& {Rosa-Gonz{\'a}lez}}]{Mayya2008}
{Mayya}, Y.~D., {Romano}, R., {Rodr{\'{\i}}guez-Merino}, L.~H., {et~al.} 2008, \apj, 679, 404, \dodoi{10.1086/587541}

\bibitem[{{Mayya} {et~al.}(2013){Mayya}, {Rosa-Gonz{\'a}lez}, {Santiago-Cort{\'e}s}, {Rodr{\'{\i}}guez-Merino}, {Vega}, {Torres-Papaqui}, {Bressan}, \& {Carrasco}}]{Mayya2013}
{Mayya}, Y.~D., {Rosa-Gonz{\'a}lez}, D., {Santiago-Cort{\'e}s}, M., {et~al.} 2013, \mnras, 436, 2763, \dodoi{10.1093/mnras/stt1784}

\bibitem[{{McLaughlin} \& {van der Marel}(2005)}]{McLaughlin2005}
{McLaughlin}, D.~E., \& {van der Marel}, R.~P. 2005, \apjs, 161, 304, \dodoi{10.1086/497429}

\bibitem[{{Minchev} {et~al.}(2013){Minchev}, {Chiappini}, \& {Martig}}]{Minchev2013}
{Minchev}, I., {Chiappini}, C., \& {Martig}, M. 2013, \aap, 558, A9, \dodoi{10.1051/0004-6361/201220189}

\bibitem[{{Mu{\~n}oz} {et~al.}(2014){Mu{\~n}oz}, {Puzia}, {Lan{\c{c}}on}, {Peng}, {C{\^o}t{\'e}}, {Ferrarese}, {Blakeslee}, {Mei}, {Cuillandre}, {Hudelot}, {Courteau}, {Duc}, {Balogh}, {Boselli}, {Bournaud}, {Carlberg}, {Chapman}, {Durrell}, {Eigenthaler}, {Emsellem}, {Gavazzi}, {Gwyn}, {Huertas-Company}, {Ilbert}, {Jord{\'a}n}, {L{\"a}sker}, {Licitra}, {Liu}, {MacArthur}, {McConnachie}, {McCracken}, {Mellier}, {Peng}, {Raichoor}, {Taylor}, {Tonry}, {Tully}, \& {Zhang}}]{Munoz2014}
{Mu{\~n}oz}, R.~P., {Puzia}, T.~H., {Lan{\c{c}}on}, A., {et~al.} 2014, \apjs, 210, 4, \dodoi{10.1088/0067-0049/210/1/4}

\bibitem[{{Mucciarelli} {et~al.}(2010){Mucciarelli}, {Origlia}, \& {Ferraro}}]{Mucciarelli2010}
{Mucciarelli}, A., {Origlia}, L., \& {Ferraro}, F.~R. 2010, \apj, 717, 277, \dodoi{10.1088/0004-637X/717/1/277}

\bibitem[{{Nantais} \& {Huchra}(2010)}]{Nantais2010}
{Nantais}, J.~B., \& {Huchra}, J.~P. 2010, \aj, 139, 2620, \dodoi{10.1088/0004-6256/139/6/2620}

\bibitem[{{Narloch} {et~al.}(2021){Narloch}, {Pietrzy{\'n}ski}, {Gieren}, {Piatti}, {G{\'o}rski}, {Karczmarek}, {Graczyk}, {Suchomska}, {Zgirski}, {Wielg{\'o}rski}, {Pilecki}, {Taormina}, {Ka{\l}uszy{\'n}ski}, {Pych}, {Hajdu}, \& {Rojas Garc{\'\i}a}}]{Narloch2021}
{Narloch}, W., {Pietrzy{\'n}ski}, G., {Gieren}, W., {et~al.} 2021, \aap, 647, A135, \dodoi{10.1051/0004-6361/202039623}

\bibitem[{{Narloch} {et~al.}(2022){Narloch}, {Pietrzy{\'n}ski}, {Gieren}, {Piatti}, {Karczmarek}, {G{\'o}rski}, {Graczyk}, {Smolec}, {Hajdu}, {Suchomska}, {Zgirski}, {Wielg{\'o}rski}, {Pilecki}, {Taormina}, {Ka{\l}uszy{\'n}ski}, {Pych}, {Rojas Garc{\'\i}a}, \& {Lewis}}]{Narloch2022}
---. 2022, \aap, 666, A80, \dodoi{10.1051/0004-6361/202243378}

\bibitem[{{Notni} {et~al.}(2004){Notni}, {Karachentsev}, \& {Makarova}}]{Notni2004}
{Notni}, P., {Karachentsev}, I.~D., \& {Makarova}, L.~N. 2004, Astronomische Nachrichten, 325, 307, \dodoi{10.1002/asna.200310175}

\bibitem[{{Okamoto} {et~al.}(2023){Okamoto}, {Arimoto}, {Ferguson}, {Irwin}, \& {{\v{Z}}emaitis}}]{Okamoto2023}
{Okamoto}, S., {Arimoto}, N., {Ferguson}, A. M.~N., {Irwin}, M.~J., \& {{\v{Z}}emaitis}, R. 2023, \apj, 952, 77, \dodoi{10.3847/1538-4357/acdad1}

\bibitem[{{Oser} {et~al.}(2010){Oser}, {Ostriker}, {Naab}, {Johansson}, \& {Burkert}}]{Oser2010}
{Oser}, L., {Ostriker}, J.~P., {Naab}, T., {Johansson}, P.~H., \& {Burkert}, A. 2010, \apj, 725, 2312, \dodoi{10.1088/0004-637X/725/2/2312}

\bibitem[{{Pan} {et~al.}(2022){Pan}, {Bell}, {Smercina}, {Price}, {Slater}, {Bailin}, {de Jong}, {D'Souza}, {Jang}, \& {Monachesi}}]{Pan2022}
{Pan}, J., {Bell}, E.~F., {Smercina}, A., {et~al.} 2022, \mnras, 515, 48, \dodoi{10.1093/mnras/stac1638}

\bibitem[{{Parisi} {et~al.}(2014){Parisi}, {Geisler}, {Carraro}, {Clari{\'a}}, {Costa}, {Grocholski}, {Sarajedini}, {Leiton}, \& {Piatti}}]{Parisi2014}
{Parisi}, M.~C., {Geisler}, D., {Carraro}, G., {et~al.} 2014, \aj, 147, 71, \dodoi{10.1088/0004-6256/147/4/71}

\bibitem[{{P{\'e}rez} {et~al.}(2013){P{\'e}rez}, {Cid Fernandes}, {Gonz{\'a}lez Delgado}, {Garc{\'\i}a-Benito}, {S{\'a}nchez}, {Husemann}, {Mast}, {Rod{\'o}n}, {Kupko}, {Backsmann}, {de Amorim}, {van de Ven}, {Walcher}, {Wisotzki}, {Cortijo-Ferrero}, \& {CALIFA Collaboration}}]{Perez2013}
{P{\'e}rez}, E., {Cid Fernandes}, R., {Gonz{\'a}lez Delgado}, R.~M., {et~al.} 2013, \apjl, 764, L1, \dodoi{10.1088/2041-8205/764/1/L1}

\bibitem[{{Pessev} {et~al.}(2006){Pessev}, {Goudfrooij}, {Puzia}, \& {Chandar}}]{Pessev2006}
{Pessev}, P.~M., {Goudfrooij}, P., {Puzia}, T.~H., \& {Chandar}, R. 2006, \aj, 132, 781, \dodoi{10.1086/505625}

\bibitem[{{Pfeffer} {et~al.}(2018){Pfeffer}, {Kruijssen}, {Crain}, \& {Bastian}}]{Pfeffer2018}
{Pfeffer}, J., {Kruijssen}, J.~M.~D., {Crain}, R.~A., \& {Bastian}, N. 2018, \mnras, 475, 4309, \dodoi{10.1093/mnras/stx3124}

\bibitem[{{Price} \& {Gullixson}(1989)}]{Price1989}
{Price}, J.~S., \& {Gullixson}, C.~A. 1989, \apj, 337, 658, \dodoi{10.1086/167137}

\bibitem[{{Rodr{\'{\i}}guez-Merino} {et~al.}(2011){Rodr{\'{\i}}guez-Merino}, {Rosa-Gonz{\'a}lez}, \& {Mayya}}]{Lino2011}
{Rodr{\'{\i}}guez-Merino}, L.~H., {Rosa-Gonz{\'a}lez}, D., \& {Mayya}, Y.~D. 2011, \apj, 726, 51, \dodoi{10.1088/0004-637X/726/1/51}

\bibitem[{{S{\'a}nchez-Bl{\'a}zquez} {et~al.}(2006){S{\'a}nchez-Bl{\'a}zquez}, {Peletier}, {Jim{\'e}nez-Vicente}, {Cardiel}, {Cenarro}, {Falc{\'o}n-Barroso}, {Gorgas}, {Selam}, \& {Vazdekis}}]{Sanchez2006}
{S{\'a}nchez-Bl{\'a}zquez}, P., {Peletier}, R.~F., {Jim{\'e}nez-Vicente}, J., {et~al.} 2006, \mnras, 371, 703, \dodoi{10.1111/j.1365-2966.2006.10699.x}

\bibitem[{{Sandage}(1961)}]{Sandage1961}
{Sandage}, A. 1961, {The Hubble Atlas of Galaxies}

\bibitem[{{Sandage} {et~al.}(1975){Sandage}, {Sandage}, \& {Kristian}}]{Sandage1975}
{Sandage}, A., {Sandage}, M., \& {Kristian}, J. 1975, {Galaxies and the universe}

\bibitem[{{Santiago-Cort{\'e}s} {et~al.}(2010){Santiago-Cort{\'e}s}, {Mayya}, \& {Rosa-Gonz{\'a}lez}}]{Mayra2010}
{Santiago-Cort{\'e}s}, M., {Mayya}, Y.~D., \& {Rosa-Gonz{\'a}lez}, D. 2010, \mnras, 405, 1293, \dodoi{10.1111/j.1365-2966.2010.16531.x}

\bibitem[{{Santos} \& {Piatti}(2004)}]{Santos2004}
{Santos}, J.~F.~C., J., \& {Piatti}, A.~E. 2004, \aap, 428, 79, \dodoi{10.1051/0004-6361:20041560}

\bibitem[{{Schiavon} {et~al.}(2005){Schiavon}, {Rose}, {Courteau}, \& {MacArthur}}]{Schiavon2005}
{Schiavon}, R.~P., {Rose}, J.~A., {Courteau}, S., \& {MacArthur}, L.~A. 2005, \apjs, 160, 163, \dodoi{10.1086/431148}

\bibitem[{{Schlegel} {et~al.}(1998){Schlegel}, {Finkbeiner}, \& {Davis}}]{Schlegel1998}
{Schlegel}, D.~J., {Finkbeiner}, D.~P., \& {Davis}, M. 1998, \apj, 500, 525, \dodoi{10.1086/305772}

\bibitem[{{Searle} \& {Zinn}(1978)}]{Searle1978}
{Searle}, L., \& {Zinn}, R. 1978, \apj, 225, 357, \dodoi{10.1086/156499}

\bibitem[{{Skrutskie} {et~al.}(2006){Skrutskie}, {Cutri}, {Stiening}, {Weinberg}, {Schneider}, {Carpenter}, {Beichman}, {Capps}, {Chester}, {Elias}, {Huchra}, {Liebert}, {Lonsdale}, {Monet}, {Price}, {Seitzer}, {Jarrett}, {Kirkpatrick}, {Gizis}, {Howard}, {Evans}, {Fowler}, {Fullmer}, {Hurt}, {Light}, {Kopan}, {Marsh}, {McCallon}, {Tam}, {Van Dyk}, \& {Wheelock}}]{Skrutskie2006}
{Skrutskie}, M.~F., {Cutri}, R.~M., {Stiening}, R., {et~al.} 2006, \aj, 131, 1163, \dodoi{10.1086/498708}

\bibitem[{{Strader} {et~al.}(2012){Strader}, {Seth}, \& {Caldwell}}]{Strader2012}
{Strader}, J., {Seth}, A.~C., \& {Caldwell}, N. 2012, \aj, 143, 52, \dodoi{10.1088/0004-6256/143/2/52}

\bibitem[{{Thomas} {et~al.}(2003){Thomas}, {Maraston}, \& {Bender}}]{Thomas2003}
{Thomas}, D., {Maraston}, C., \& {Bender}, R. 2003, \mnras, 339, 897, \dodoi{10.1046/j.1365-8711.2003.06248.x}

\bibitem[{{Tody}(1986)}]{Tody1986}
{Tody}, D. 1986, in Society of Photo-Optical Instrumentation Engineers (SPIE) Conference Series, Vol. 627, Instrumentation in astronomy VI, ed. D.~L. {Crawford}, 733, \dodoi{10.1117/12.968154}

\bibitem[{{Trager} {et~al.}(1998){Trager}, {Worthey}, {Faber}, {Burstein}, \& {Gonz{\'a}lez}}]{Trager1998}
{Trager}, S.~C., {Worthey}, G., {Faber}, S.~M., {Burstein}, D., \& {Gonz{\'a}lez}, J.~J. 1998, \apjs, 116, 1, \dodoi{10.1086/313099}

\bibitem[{{Usher} {et~al.}(2012){Usher}, {Forbes}, {Brodie}, {Foster}, {Spitler}, {Arnold}, {Romanowsky}, {Strader}, \& {Pota}}]{Usher2012}
{Usher}, C., {Forbes}, D.~A., {Brodie}, J.~P., {et~al.} 2012, \mnras, 426, 1475, \dodoi{10.1111/j.1365-2966.2012.21801.x}

\bibitem[{{van den Bergh}(1991)}]{vandenBergh1991}
{van den Bergh}, S. 1991, \apj, 369, 1, \dodoi{10.1086/169732}

\bibitem[{{van der Hulst}(1979)}]{vanderhulst1979}
{van der Hulst}, J.~M. 1979, \aap, 75, 97

\bibitem[{{VandenBerg} {et~al.}(2013){VandenBerg}, {Brogaard}, {Leaman}, \& {Casagrande}}]{VandenBerg2013}
{VandenBerg}, D.~A., {Brogaard}, K., {Leaman}, R., \& {Casagrande}, L. 2013, \apj, 775, 134, \dodoi{10.1088/0004-637X/775/2/134}

\bibitem[{{Vazdekis} {et~al.}(2010){Vazdekis}, {S{\'a}nchez-Bl{\'a}zquez}, {Falc{\'o}n-Barroso}, {Cenarro}, {Beasley}, {Cardiel}, {Gorgas}, \& {Peletier}}]{Vazdekis2010}
{Vazdekis}, A., {S{\'a}nchez-Bl{\'a}zquez}, P., {Falc{\'o}n-Barroso}, J., {et~al.} 2010, \mnras, 404, 1639, \dodoi{10.1111/j.1365-2966.2010.16407.x}

\bibitem[{{Walter} {et~al.}(2002){Walter}, {Weiss}, {Martin}, \& {Scoville}}]{Walter2002}
{Walter}, F., {Weiss}, A., {Martin}, C., \& {Scoville}, N. 2002, \aj, 123, 225, \dodoi{10.1086/324633}

\bibitem[{{Wang} {et~al.}(2019){Wang}, {Ma}, \& {Liu}}]{Wang2019}
{Wang}, S., {Ma}, J., \& {Liu}, J. 2019, \aap, 623, A65, \dodoi{10.1051/0004-6361/201834748}

\bibitem[{{Weisz} {et~al.}(2014){Weisz}, {Dolphin}, {Skillman}, {Holtzman}, {Gilbert}, {Dalcanton}, \& {Williams}}]{Weisz2014}
{Weisz}, D.~R., {Dolphin}, A.~E., {Skillman}, E.~D., {et~al.} 2014, \apj, 789, 147, \dodoi{10.1088/0004-637X/789/2/147}

\bibitem[{{Whitmore} {et~al.}(2014){Whitmore}, {Chandar}, {Bowers}, {Larsen}, {Lindsay}, {Ansari}, \& {Evans}}]{Whitmore2014}
{Whitmore}, B.~C., {Chandar}, R., {Bowers}, A.~S., {et~al.} 2014, \aj, 147, 78, \dodoi{10.1088/0004-6256/147/4/78}

\bibitem[{{Wisniewski} \& {Kleinmann}(1968)}]{Wisniewski1968}
{Wisniewski}, W.~Z., \& {Kleinmann}, D.~E. 1968, \aj, 73, 866, \dodoi{10.1086/110721}

\bibitem[{{Worthey} {et~al.}(1994){Worthey}, {Faber}, {Gonzalez}, \& {Burstein}}]{Worthey1994}
{Worthey}, G., {Faber}, S.~M., {Gonzalez}, J.~J., \& {Burstein}, D. 1994, \apjs, 94, 687, \dodoi{10.1086/192087}

\bibitem[{{Yun}(1999)}]{yun1999}
{Yun}, M.~S. 1999, in IAU Symposium, Vol. 186, Galaxy Interactions at Low and High Redshift, ed. J.~E. {Barnes} \& D.~B. {Sanders}, 81

\bibitem[{{Yun} {et~al.}(1994){Yun}, {Ho}, \& {Lo}}]{Yun1994}
{Yun}, M.~S., {Ho}, P.~T.~P., \& {Lo}, K.~Y. 1994, \nat, 372, 530, \dodoi{10.1038/372530a0}

\bibitem[{{Zepf} \& {Ashman}(1993)}]{Zepf1993}
{Zepf}, S.~E., \& {Ashman}, K.~M. 1993, \mnras, 264, 611, \dodoi{10.1093/mnras/264.3.611}

\end{thebibliography}
\bibliographystyle{aasjournal}



\end{document}